\documentclass[12pt,twoside,a4paper]{article}
\usepackage{epsfig}
\usepackage[hang,bf,small]{caption}
\DeclareGraphicsExtensions{.eps.gz,.eps,.ps,.ps.gz}
\parskip=\itemsep               
\newcommand{\beq}{\begin{equation}}
\newcommand{\eeq}{\end{equation}}
\newcommand{\beqar}[1]{\begin{eqnarray}\label{#1}}
\newcommand{\eeqar}{\end{eqnarray}}

\newcommand{\Le}{\left(}
\newcommand{\Ra}{\right)}
\newcommand{\uw}{\uparrow}
\newcommand{\dw}{\downarrow}

\newcommand{\A}{{\cal A}}

\newcommand{\as}{\alpha_S}

\def\eq#1{{Eq.~(\ref{#1})}}

\def\npb#1#2#3{    {\it Nucl. Phys. }{\bf B#1} (19#2) #3}
\def\plb#1#2#3{    {\it Phys. Lett. }{\bf B#1} (19#2) #3}
\def\prd#1#2#3{    {\it Phys. Rev. }{\bf D#1} (19#2) #3}

\def\zpc#1#2#3{    {\it Z. Phys. }{\bf C#1} (19#2) #3}

\relax

\begin{document}
\renewcommand{\baselinestretch}{0.9}
\title{
   \hfill {\bf\normalsize TAUP 2616-99}\\
   \hfill {\bf\normalsize BNL - NT-99/11}\\
   \hfill {\bf\normalsize \today}\\[1cm]
   {\bf \Large A Pomeron Approach to Hadron-Nucleus and Nucleus-Nucleus
``Soft"  Interactions at High Energy}}
\author{
{\large \bf  S. ~B o n d a r e n k o   ~\thanks{Email: 
serg@post.tau.ac.il} ~$\,\mathbf{^a}$,\quad E. ~G o t s m a n 
~\thanks{Email: gotsman@post.tau.ac.il} ~$\,\mathbf{^a}$}\\[1.5ex]
{\large \bf E. ~L e v i n
~\thanks{Email: leving@post.tau.ac.il} ~$\mathbf{\,^{a,\,\, b}}$ \quad and
\quad
U. ~M a o r ~\thanks{ Email: maor@post.tau.ac.il} ~$\mathbf{\,^a}$}
\\[3mm]
{\it\normalsize $^a$HEP Department}\\ 
 {\it\normalsize School of Physics and Astronomy,}\\
 {\it\normalsize Raymond and Beverly Sackler Faculty of Exact Science,}\\
 {\it\normalsize Tel-Aviv University, Ramat Aviv, 69978, Israel}\\[2mm]
{\it\normalsize $^b$ Physics  Department}\\
 {\it\normalsize Brookhaven National Laboratory}\\
 {\it\normalsize Upton, NY 11973-5000, USA}}
 \date{}
\maketitle
\thispagestyle{empty}
\vspace*{-1cm}
\begin{abstract}
We formulate a generalization of the Glauber formalism for hadron-nucleus
and nucleus-nucleus collisions based on the
Pomeron approach to high energy interactions. 
Our treatment is based on
two physical assumptions ( i.e. two small parameters )
: (i) that only sufficiently small
distances contribute to the Pomeron structure; and (ii)
the triple Pomeron vertex $G_{3P}/g_{P-N} \,\ll\,1$ (where
$g_{P-N}$ is the Pomeron-nucleon vertex) is small.
 A systematic method is developed
for calculating the 
total, elastic and diffractive dissociation cross sections as well as
the survival probability of large rapidity gap processes and inclusive
observables, both for hadron - nucleus and nucleus-nucleus collisions.
Our approach suggests saturation of the density
of the produced hadrons in nucleus-nucleus collisions,  the value of the
saturation density turns out to be large and depends on the number of
nucleons in the lightest nucleus.

\end{abstract}

\vspace*{.5cm}
\setcounter{page}{1}
\section{Introduction}
\setcounter{equation}{0}
Although  QCD is believed to be the microscopic theory of strong
interactions it has led  to little progress in the theoretical
understanding of  "soft" interactions at high energies. At present 
we only have a phenomenological description of such processes where the
main postulate  of this description is the Pomeron -   a Reggeon with 
a trajectory intercept $\alpha_P(0)$  close to unity,
\beq \label{POM}
\alpha_P(t)\, =\,
\alpha_P(0)\, +\,\alpha'_P(0) |t|\,\,;\,\,\,\,\,\,\, \alpha_P(0)= 1 +
\Delta,\,\,;\,\,\,\,\,\,
\Delta\,\,\ll\,\,1 \,\,.
\eeq
We have two comments on the Pomeron hypothesis:
\begin{enumerate}
\item\,\,\, The first one is negative. There is a paucity of 
ideas and/or examples of how a Reggeon such as the Pomeron can be justified
theoretically. As far as we know, the only theoretical model, where the
Pomeron appears naturally, is a 2 + 1 dimensional QCD \cite{2DQCD} which can
hardly be considered as a good approximation to reality. The entity closest 
to the Pomeron, which follows from perturbative QCD (pQCD) is the
BFKL Pomeron \cite{BFKL}. 
In the following, when dealing with "soft" interactions, we shall also refer to
the BFKL Pomeron, but only with regard to the
general structure of pQCD approach at high energy.
\item\,\,\, The second comment is positive. In spite of all
theoretical uncertainties the Pomeron  has been at the heart of high
energy phenomenology over the past three decades, providing a good
description of all available experimental data
\cite{DL},\cite{KAID}, \cite{GLMSOFT}. Consequently, we cannot ignore
the Pomeron hypothesis as well as the numerous attempts to find a
selfconsistent approach based on this postulate.
\end{enumerate}
Even if we accept the Pomeron hypothesis,
there still remains a lot 
of hard work to find
the high energy scattering amplitude that
includes the Pomeron exchanges, as well as the interactions between

Pomerons. The goal of this paper
is to find a solution to this problem for hadron-nucleus and
nucleus-nucleus high energy collisions.
Our approach is based on two main ideas: (i) a new
sufficiently hard scale for the Pomeron structure and (ii) a specific
parameter ($\kappa_A$)  for the Pomeron interaction with a nucleus suggested
by Schwimmer \cite{SCHW} many years ago,
\beq \label{PAR}
\kappa_A\,\,=\,\,A\,\cdot\,\frac{g_{PN}\,G_{3P}}{\pi
R^2_A}\,\,\cdot\,
\,\frac{1}{\Delta}\{\,(\,\frac{s}{s_0}\,)^{\Delta}\,\,-\,\,1\,\} \,\,,
\eeq
where $g_{PN}$ is the Pomeron-nucleon vertex, $G_{3P}$ is the triple
Pomeron vertex and $\Delta$ is defined in \eq{POM}. $A$ and $R_A$ are the
atomic number and the radius of a nucleus, which are
defined by
\beq
R_{A}=A^{\frac{2}{3}}\cdot 17 GeV^{-2}.
\eeq
\subsection{A new scale for the Pomeron structure}
Our key assumption is a new scale for the Pomeron structure, i.e. a
sufficiently large mean transverse momentum of partons in the Pomeron
( $< p_t > \,\geq\, 1 \,GeV$ ).  We  list, first, the experimental
and phenomenological observations  that support our point of view
(see also Ref.\cite{LEVSP}):
\begin{enumerate}
\item\,\,\,Experimental elasting scattering data can be fitted
using a very small value for the slope
of the Pomeron trajectory
( see Refs.~\cite{DL}, ~\cite{KAID}, ~\cite{GLMSOFT},~\cite{HERAGL} ):
$\alpha_{P}'=0.25 GeV^{-2}\,\ll\,\alpha_{R}'= 1 GeV^{-2}$, here
$\alpha_{R}'$ is the slope of the Reggeon trajectory ( see \eq{POM} );
\item \,\,\, The experimental $\frac{d\sigma}{dt}$
slope of single diffraction dissociation,
with final state 
secondary hadron system with a large mass, is approximately two times
smaller
than the slope for the elastic scattering. The description of the
diffractive dissociation processes is
closely connected to the triple Pomeron vertex, $G_{3P}$ ( see
Fig.~\ref{MF15} ).
A small slope leads to a small proper size
of this vertex. As a first approximation
we can assign a zero slope to the triple Pomeron vertex, as 
this provides a reasonable description of 
the experimental single diffractive dissociation data.
\begin{figure}[htbp]
\begin{center}
\epsfig{file=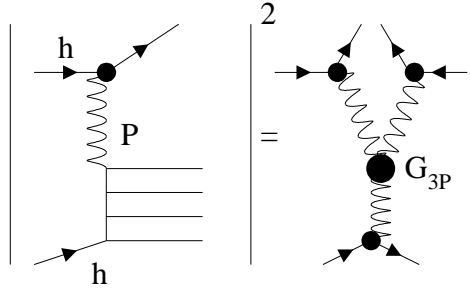,width=8cm}
\end{center}
\caption{\em $G_{3P}$ and the diffractive dissociation process.}
\label{MF15}
\end{figure}
\item\,\,\,The HERA data \cite{HERAPSI} on diffractive  photoproduction  of
J/$\Psi$ and deep inelastic scattering ( DIS),
show that the $t$- slope for elastic diffractive dissociation 
( $\gamma^* + p \,\,\rightarrow\,\,J/\Psi + p$ ) is larger than the
$t$-slope for the inelastic channel ( $\gamma^* + p \,\,\rightarrow\,\,J/\Psi
+ N^*$, where $N^*$ is the nucleon exitation ). These data provide  direct
experimental support for the idea that there are two typical scales for the
size inside a hadron, namely, the distance between quarks and
the proper size of the constituent quarks in the Additive Quark Model(
AQM )\cite{AQM}.  In the AQM the size of the constituent quark is
closely related to the typical transverse momentum of partons in the
Pomeron.

\item\,\,\, The main idea of the AQM that gluons inside a hadron are
confined in a volume of a smaller radius ( $R_G\,\approx\,0.1 fm \,\ll\,
R_h\,\approx\,1\,fm $ ), is a working hypothesis which is included in
the standard Pomeron phenomenology that describes the experimental data
\cite{DL}.
\item\,\,\,The HERA experimental data \cite{HERAGL} have confirmed the
theoretical expectation \cite{THEHDQCD} that the density of partons in a
hadron at low
$x$  reaches a high value.  In high density QCD we expect that a new
saturation  scale appears \cite{THEHDQCD}.
This   saturation scale means that the average
transverse momentum of a parton becomes large in the region of high density
and, therefore, it  confirms our assumption regarding a new scale
associated with the Pomeron.
\end{enumerate}

The consequence of assuming a new hard scale in the Pomeron leads to:

\begin{itemize}
\item\,\,\, The Pomeron exchange as well as all vertices for the 
Pomeron-Pomeron interaction,  can be viewed as delta functions in impact
parameter ( $b_t$ )  representation. Indeed, based on our assumption, we
can neglect both
$\alpha'_P(0)$ and the proper size of all Pomeron-Pomeron vertices, and
consider the Green's function for the Pomeron exchange, as well as the
vertices, as constants in momentum representation, or in other words, as delta
functions in $b_t$-representation:

\begin{eqnarray}
& P(s, b_t)\,\,=\,\,i\,\, \delta^{(2)}(\vec{b}_t
)\,(\,\frac{s}{s_0}\,)^{\Delta}\,\,;& \label{PB1}\\
& G_{3P}(b_t) \,\, =\,\,\, G_{3P}(0) \,\delta^{(2)}( \vec{b}_t)\,\,;
\,\,\,\,\,\,G_{4P}(b_t) \,\, =\,\,\, G_{4P}(0) \,\delta^{(2)}(
\vec{b}_t)&.
\label{PB2}
\end{eqnarray}

Consequently, in our approach, all the $b_t$ dependence is concentrated in
the Pomeron - hadron vertices which depend on the size of a hadron.
We use the simplest Gaussian parametrization
for the Pomeron-Nucleus vertex, i.e.:
\beq \label{POMA}
g_{PA}(b_t)\,\,\,=\,\,\,A\,\frac{g_{PN}}{\pi R^2_A}\,\,\exp\Le
-\frac{b^2_t}{R^2_A}\Ra\,\,,
\eeq
where the notation is the same as in \eq{PAR}.
Combining \eq{POMA} with \eq{PB1} one can obtain that for nucleus ($A_1$)
- nucleus ($A_2$) scattering, the single  Pomeron exchange has the form ( at
fixed $b_t$ ):
\beq \label{POMEXAA}
\int\,\,d^2\,b'_t\,\,\int\,d^2\,b"_t\, g_{PA_1} (b'_t)\,\,P(s,{\vec{b'}}_t-
{\vec{b"}}_t)\,\,
g_{PA_2} ({\vec{b}}_t - {\vec{b"}}_t ) \,\,=
\eeq
$$
\,\,\frac{ g^2_{PN}\,\cdot
A_1\,\cdot\,A_2 }{\pi R^2_{A_1}\,\cdot\,\pi R^2_{A_2}}\,\,\cdot\,\,
\int \,d^2\,b'_t
\,\,\exp\Le -\frac{b'^2_t}{R^2_{A_1}}\Ra\,\cdot\,
\left(\,\frac{s}{s_0}\,\right)^{\Delta}\,
\cdot\,\exp\Le -\frac{({\vec{b}}_t
-{\vec{b'}}_t
)^2}{R^2_{A_2}}\Ra\,\,
$$
$$
=\,\,\frac{ g^2_{PN}\,\cdot
A_1\,\cdot\,A_2 }{\pi ( R^2_{A_1}\,+\, R^2_{A_2})}\,\cdot\,
\left(\,\frac{s}{s_0}\,\right)^{\Delta}\,
\cdot\,
\exp\Le  -\frac{b^2_t}{R^2_{A_1}\,+\, R^2_{A_2}}\Ra\,\,.
$$

\item\,\,\, We can use pQCD to estimate the values for the vertices of the
Pomeron-Pomeron interactions, since the typical scale ( the Pomeron radius
) is small.  Simply counting the number of $\alpha_S$ in the pQCD diagrams
gives ( see Refs. \cite{BFKL} \cite{LEVSP} \cite{THEHDQCD} \cite{BAR} )
\begin{eqnarray}
& g_{PN}\,\,\propto \,\,\,\as\,\,;\,\,\,\,\,\,\,\,\,\Delta\,\,
\,\propto\,\,\,\as N_c\,\,;& \label{VE1}\\
& G_{3P}\,\,\,\propto\,\,\,\as^2 N_c\,\,; \,\,\,\,\,\,\,\,\,
G_{4p}\,\,\,\,\propto \,\,\,\as^2\,\,.\label{VE2}
\end{eqnarray}

To select the vertices of the Pomeron-Pomeron interaction we use the
main principle of the large $N_c$ ( number of colours ) approximation, that
has been formulated by Veneziano et al. \cite{VEN}. We sum 
separately in each topological configuration the leading
$N_c$ diagrams, considering $N_c \as \,\,\approx\,1$, while $\as
\,\,\ll\,\,1$. Applying
these rules to \eq{VE1} and \eq{VE2} we see that we can take into
account only the triple Pomeron interaction and neglect more complicated
vertices.
\end{itemize}
\subsection{The order parameter $\mathbf{\kappa_A}$}
The problem of summing all triple Pomeron interactions looks hopeless
without an additional small parameter that aids us in classifying the
numerous Reggeon diagrams, which describe  Pomeron interactions ( see
Ref, \cite{RFT} for details ).  Such a parameter is $\kappa_A$ given by
\eq{PAR} and suggested by Schwimmer\cite{SCHW}  more than thirty years
ago,  
\beq 
\kappa_A\,\,=\,\,A\,\cdot\,\frac{g_{PN}\,G_{3P}}{\pi
R^2_A}\,\,\cdot\,
\,\frac{1}{\Delta}\{\,(\,\frac{s}{s_0}\,)^{\Delta}\,\,-\,\,1\,\} \,\,.
\eeq
To clarify the meaning of $\kappa_A$ we calculate the
cross section for  diffractive dissociation of an incoming hadron  in the
region of large mass for the hadron-nucleus collision. This process is
illustrated in Fig.~\ref{MF15}, where the incoming 
hadron is shown at the bottom of the diagram.  
Using \eq{PB1} and \eq{PB2}, the cross section 
can be written as:
\begin{eqnarray}
\sigma^{SD}(h + A\, \rightarrow\, M + A )\,&=&\,\int^{Y}_0\, d y\,
\int\,d^2 \,b_t \,\int \,d^2\,b'_t g^2_{PA}(b^2_t) \,g_{PN}( (\vec{b}_t -
\vec{b'}_t )^2) \,\,G_{3P}\,\,e^{ 2\,\Delta ( Y - y)\, +
\,\Delta\,y}\,\nonumber\\
 &=&\,\frac{A^2}{\Le\pi
R^2_A\Ra^2} \, g^3_{PN} \,G_{3P} e^{ 2\,\Delta \,Y}
\,\int^{Y}_0 \,d\,y \,e^{- \,\Delta\,y}\,\nonumber\\
 &=&\,\frac{A \,g^2_{PN}}{\pi
R^2_A}\,e^{\Delta\,Y}\cdot\,\frac{A\,g_{PN}\,G_{3P}}{\pi
R^2_A}\,\frac{1}{\Delta}\,\{\,e^{\Delta \,Y} \,\,-\,\,1\,\} \nonumber \\
 &=& \,\frac{A \,g^2_{PN}}{\pi
R^2_A}\,e^{\Delta\,Y}\,\cdot\,\kappa_A\,\,=\,\,g_{PN}
\,g_{PA}\Le b_{t}=0\Ra
P(Y) \,\cdot\,\kappa_A\,\,,\label{DDXS}
\end{eqnarray}
where \eq{POMA} was used and we introduce the  notation: $Y =
ln(s/s_0)$,
$y=\,ln(M^2/s_0)$ and $P\Le Y\Ra=e^{\Delta\,Y}$.  From
\eq{DDXS} we see that the parameter $\kappa_A$ now
indicates   how large or small the cross section of 
diffractive dissociation is, in
comparison with the total cross section given by Pomeron
exchange ( see \eq{POMEXAA} ).
The principle idea of the selection rules for Reggeon diagrams is to sum
all diagrams of the order $\kappa^n_A$, neglecting the diagrams that are
proportional to $g^{2n}_{PN}$ or $\kappa^n_N$. In other words, we consider
the following set of parameters:
\beq \label{PSR}
\kappa_A\,\,\approx\,\,1\,\,;\,\,\,\,\,\,\,\,\,\,\kappa_N\,\,\ll\,\,1\,\,;
\,\,\,\,\,\,\,\,\,\,\frac{\gamma}{g_{PN}}\,\cdot\,\kappa_N\,\,\ll\,\,1\,\,.
\eeq
Below we show how this  set of parameters helps, and the type of 
selection rules it leads to, both for hadron-nucleus and nucleus-nucleus
interaction. \eq{PSR} only holds for the case of heavy
nuclei with $A \,\,\gg\,\,1$. We consider the
experimental fact that $G_{3P}/(g_{PN}\cdot \Delta)$ is rather small
($\approx$ 1/4 - 1/8), which follows from measurements of $\sigma_{el}(pp)$
and $\sigma^{SD}(pp)$, as supportive for our approach 
( see also section 2.6 for details ).

\subsection{The Glauber-Gribov Approach}
As we have mentioned, the main goal of this paper is to find a natural
generalization of Glauber-Gribov approach\cite{GG}\cite{GG1}  to hadron -
nucleus
and nucleus-nucleus interaction at high energies. In this subsection we
recall the  Glauber-Gribov approach for the nucleus -
nucleus interaction (see Fig. 2 ).
\begin{figure}[htbp]
\begin{center}
\begin{tabular}{c c }
\epsfig{file=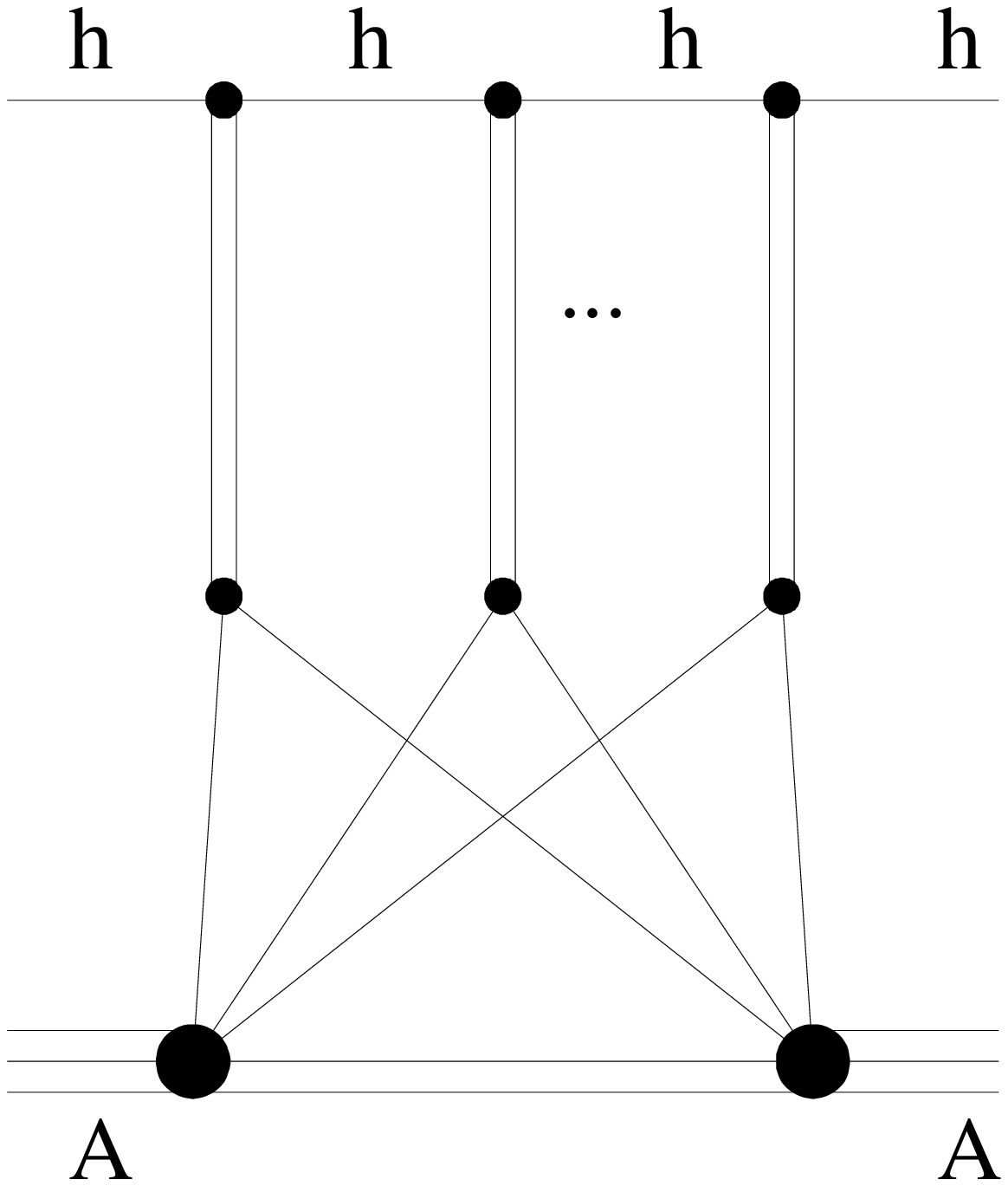,width=7cm,height=4cm} &
\epsfig{file=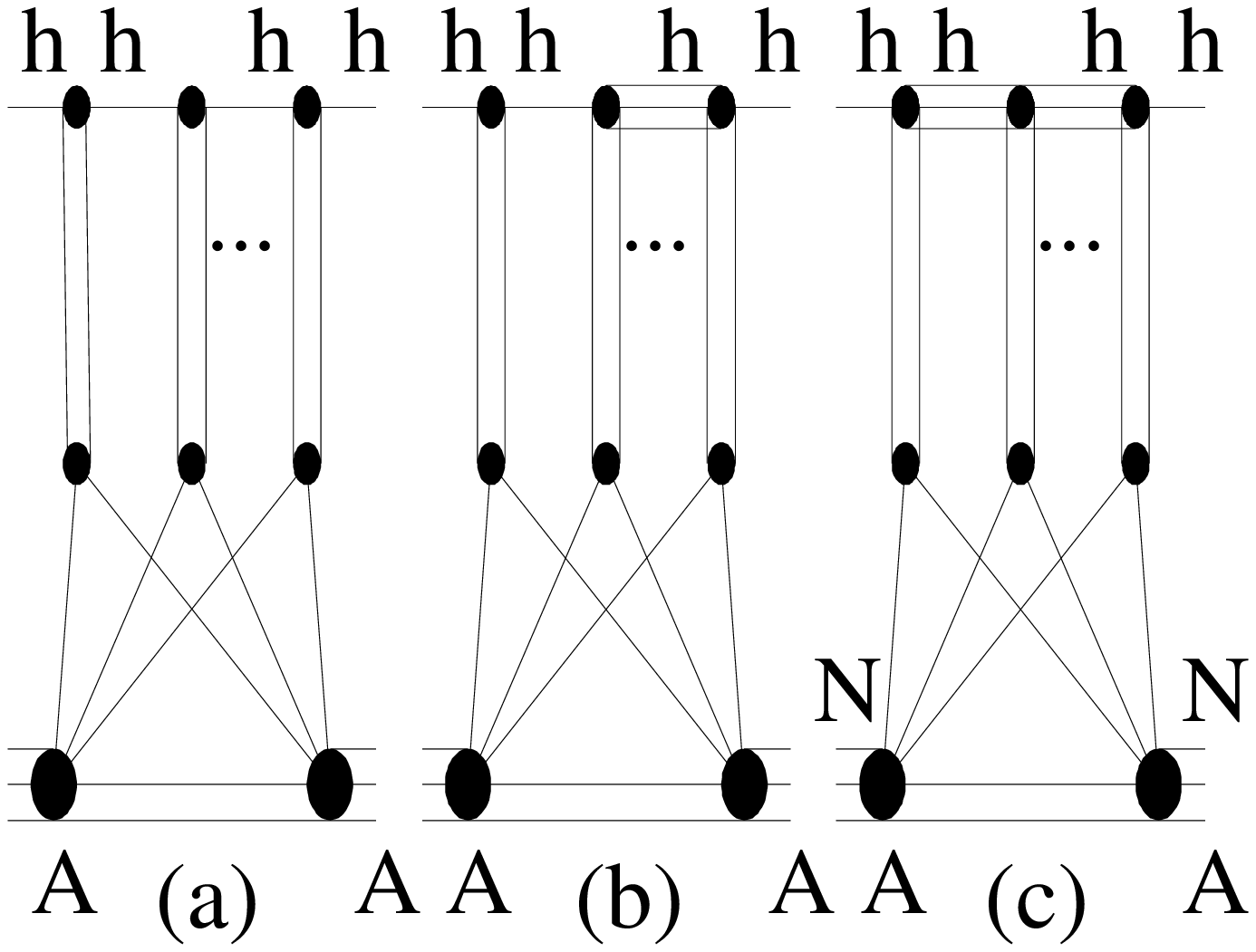,width=9cm,height=4.5cm}\\
Fig. 2-a & Fig. 2-b
\end{tabular}
\end{center}
\caption{\it  Nucleus - Nucleus scattering: The  Glauber-Gribov approach
(Fig.2-a) and  the first corrections (Fig.2-b) to the Glauber-Gribov
approach. Figures are taken fron Ref. \protect\cite{GG1}.}
\label{gg}
\end{figure}
Fig. 2-a shows that in the Glauber-Gribov approach we view a nucleus as a
collection of almost free nucleons which could interact elastically and/or
inelastically with a nucleon in another nucleus. In this approach the

unitarity constraint 
\beq \label{UNTRTY}
2\,Im\,a_{el}(s,b_t)\,\,\,=\,\,|a_{el}(s,b_t)|^2\,\,\,+\,\,\,G_{in}(s,b_t)
\,\,,
\eeq
has a simple solution
\begin{eqnarray}   
a_{el}(s,b_t)\,\,&=&\,\,i\,\left(\,1\,\,-\,\,\exp\Le-
\,\frac{\Omega(s,b_t)}{2}\Ra\,\right) \,\,;\label{UB1}\\
G_{in}(s,b_t)\,\,&=&\,\,1\,\,\,-\,\,\,\exp\Le
- \Omega(s,b_t)\Ra\,\,, \label{UB2}
\end{eqnarray}
assuming that the elastic amplitude $a_{el}(s,b_t)$
is pure imaginary at high energy. 
In \eq{UNTRTY}
$G_{in}(s,b_t)$ denotes the contribution of all inelastic processes.
The advantages
of the Glauber-Gribov approach is that we can 
express  the cross section of a nucleon-nucleon interaction
in terms of an arbitrary real function $\Omega(s,b_t)$, which is known as
opacity. Indeed, we can
use \eq{POMEXAA} for rewriting the opacity in in the form:
\beq \label{GG1}
\Omega^{A_1 - A_2}(s,b_t) \,\,=\,\,A_1\times A_2\times
\sigma^{N-N}_{tot}(s) \int 
S(| \vec{b}_t - \vec{b'}_t)\,\, \cdot\,\,S(b't)\,
d^{2}b \,d^{2}b'\,,
\eeq
where $\sigma^{N-N}_{tot}(s)$ is the total cross section of the 
nucleon-nucleon interaction and $S(b_t)$ is the distribution of nucleons
in a nucleus $S(b_t) = \int d z \rho(z,b_t)$ where $\rho(z,b_t)$ is the
nucleon density in a nucleus. In \eq{POMEXAA} we used a Gaussian
parametrization for  $S(b_t)$ to simplify our calculations. 
Formulae given by \eq{UB1} and \eq{UB2} with $\Omega$ defined in \eq{GG1}
have been studied in detail two decades ago
\cite{GG1,GGA1,GGA2,GGA3,GGA4,GGA5,GGA6,GGA7,LRN,REG1,REG2,REG3,REG4,REG5,AGK} 
and they provide the first natural estimates for all observables in
hadron-nucleus and nucleus - nucleus collisions. Using the AGK cutting rules
\cite{AGK}, we can often reduce the number of Reggeon diagrams and
simplify
our calculation. A well known example of such a simplification is the
inclusive cross section which can be described by a single diagram of Fig.3
instead of the whole series shown in Fig.2.  

\begin{figure}[htbp]
\begin{center}
\epsfig{file=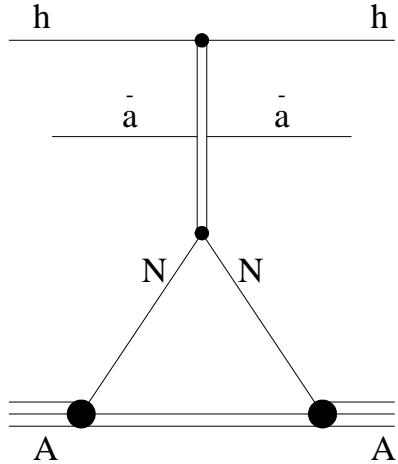,width=9cm} 
\end{center}
\caption{\it  	Single inclusive cross section for Nucleus - Nucleus
scattering in  the Glauber-Gribov approach.}
\label{gg1}
\end{figure}
The corresponding cross-section is 
\beq \label{GG2}
\frac{d \sigma(A_1 - A_2)}{d y} \,\,\,=\,\,\,A_1 \times A_2 \times 
\frac{d \sigma(N - N)}{d y}\,\,.
\eeq
\eq{GG2} is our point of reference for  all further calculations. We will
show that the interaction between Pomerons, that have been neglected in
the Glauber-Gribov approach,  change  the result of \eq{GG2} drastically
reducing the A-dependence to
\beq \label{GG3}
\frac{d \sigma(A_1 - A_2)}{d y} \,\,\,=\,\,\,A^{\frac{2}{3}}_1 \times
A^{\frac{2}{3}}_2 \times
\frac{d \sigma(N - N)}{d y}\,\,.
\eeq

Fig.2-a shows that  in the Glauber-Gribov approach we have neglected the
intermediate processes
of diffraction dissociation ( see Fig.2-b). In this paper we will take
them into account replacing diffractive processes by large mass
diffraction
described by the triple Pomeron vertex ( see Fig.4 ). 

\subsection{Motivation and structure of the paper}
Our primary goal in this paper is to solve 
the problem of high energy behaviour within the Pomeron 
framework. We are hopeful that our phenomenological approach
will provide information as to what one can expect to learn from
high energy soft interactions. We will concern ourselves here only with 
the problem of  hadron-nucleus and nucleus-nucleus scattering. In these
cases we have an additional simplification as we are dealing with systems 
with a high density of partons. We trust that matching our approach
to that of QCD 
for high parton density systems \cite{THEHDQCD} will provide a bridge
between the microscopic theory and the physics of large
cross sections.

  Our second  motivation is a practical one. RHIC is due to start operating
shortly and we believe that it will be much easier 
to understand new
phenomena such as quark-gluon plasma production, or the observation of the
saturation scale if a reliable phenomenological model of the "soft'
nucleus-nucleus interaction is available. Our approach is just a natural
generalization of the Glauber-Gribov \cite{GG}\cite{GG1}  approach for
the nucleus-nucleus interaction, where we
take into account not only the rescattering of the fastest partons, as was
done in the Glauber-Gribov approach, but also the interaction of all
partons with  the target and the projectile.
 
  The structure of the paper is the following: In the next section we
consider the hadron-nucleus interaction. The high energy amplitude for
this reaction has been calculated by Schwimmer \cite{SCHW} but we recall
the result of this calculation which we will
need  for the nucleus-nucleus collision. We
then calculate the diffraction dissociation cross section as well as the 
survival probability for the hadron-nucleus reaction. The main properties
of
the inclusive production are discussed. 
At the end of the section we present a phenomenological
application to exemplify the values 
that we are dealing with. The content of this section is not completely
new and many properties of hadron-nucleus interaction have been studied
for decades ( see
Refs.\cite{GG1,GGA1,GGA2,GGA3,GGA4,GGA5,GGA6,GGA7,LRN,REG1,REG2,
REG3,REG4,REG5} ). However, the equation for the diffraction dissociation
processes as well as the calculation of the survival probability of the
large rapidity gap processes are described for the first time.
 
  The third section is the key section of this paper. In it  we develop our
technique for dealing with
nucleus-nucleus scattering  at high energy. We present 
the analytic formulas, as well as 
phenomenological estimates, for the total ion-ion cross sections at high
energy, their elastic and diffractive dissociation cross sections and
an estimate of the  survival probability for large
rapidity gap processes. 
In section four our results are summarized and discussed in light
of the new experiments at RHIC.

\section{The Hadron - nucleus interaction }
\setcounter{equation}{0}
\subsection{Selection rules for  Reggeon diagrams}
\begin{figure}[htbp]
\begin{center}
\epsfig{file=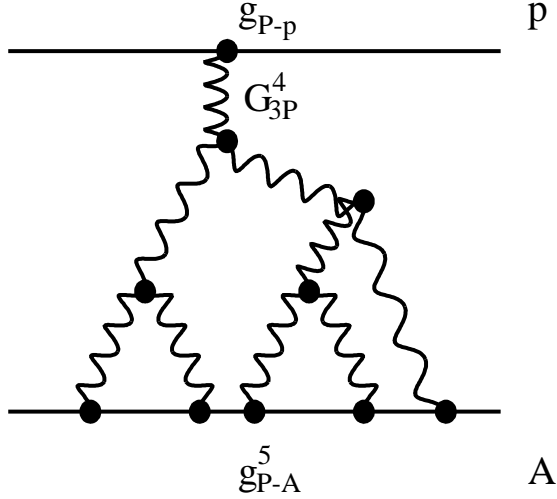,width=10cm}
\end{center}
\caption{\it  "Fan"  Pomeron diagrams.}
\label{MF40}
\end{figure}

In this subsection we demonstrate that \eq{PSR} leads to a selection
rule: the main contribution to the
hadron-nucleus amplitude given by the "fan" diagrams of Fig.~\ref{MF40}
with the incoming hadron in the handle of this fan.

As has been shown in Eqs.(~\ref{UNTRTY}),
(~\ref{UB1}) and (~\ref{UB2}) we can introduce an arbitrary real function 
for the opacity
$\Omega(s, b_t)$. This function is very convenient since we  can  use it
to rewrite all observables, i.e.:
\begin{eqnarray}
\sigma_{tot}\,\,&=&\,\,2\,\int\,d^2\,b_t\,\,\left(\,1\,\,-\,\,
\exp\Le
-\,\frac{\Omega(s,b_t)}{2}\Ra\,\right)\,\,;\label{O1}\\
\sigma_{el}\,\,&=&\,\,\int\,d^2b_t
\,\left(\,1\,\,-\,\exp\Le
-\,\frac{\Omega(s,b_t)}{2}\Ra\,\right)^2\,\,;\label{O2}\\
\sigma_{in}\,\,&=&\,\,\int\,d^2\,b_t\,\left(\,1\,\,\,-\,\,\,\exp\Le
-\Omega(s,b_t)\Ra\,\right)\,\,; \label{O3}\\
\frac{d \sigma_{el}}{dt}\,\,&=&\,\,\pi\,|f(s,t)|^2\,\,;\,\,\,\,\,\,
\sigma_{tot}\,\,=\,\,4\,\pi\,\,f(s,0)\,\,;
\label{O4}\\
f(s,t= - q^2)\,\,&=&\,\,\frac{1}{2 \pi}\,\,\int\,\,d^2b_t\,\,e^{i
{\mathbf{\vec{q}\,\cdot\,\vec{b}_t}}}\,\,a_{el}(s,b_t)\,\,.\label{O5}
\end{eqnarray}

\begin{figure}[htbp]
\begin{center}
\epsfig{file=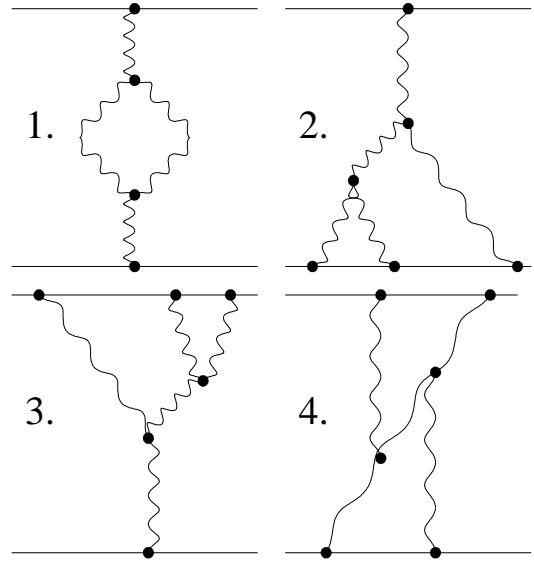,width=10cm}
\end{center}
\caption{\it  "Fan" diagrams of order $G^2_{3P}$  with and without
loops.}
\label{nf1}
\end{figure}

We now calculate the contributions of the Reggeon diagrams to the opacity
$\Omega$. Neglecting the triple Pomeron vertex $G_{3p}$ we have only a
single
Pomeron exchange given by \eq{POMEXAA} with $A_1 = 1$.
\beq \label{OPHA}
\Omega(Y,b_t)\,\,=\,\,\frac{g^2_{PN} \,\cdot\,A}{\pi
\,R^2_A}\,\cdot\,e^{\Delta\,Y}\,\exp\Le
-\,\frac{b^2_t}{R^2_A}\Ra\,\,,
\eeq
where we take the integral over $b'_t$ in \eq{POMEXAA} and neglect the
value
of $R_N$ as it is much smaller then $R_A$. Substituting \eq{OPHA}
into \eq{UB1} and \eq{UB2} we obtain the Glauber formula for 
the hadron-nucleus
interaction \cite{GG,GG1} .
At first order with respect to $G_{3P}$ we have two "fan" diagrams of the
type of
Fig.~\ref{MF15} with diffraction of the incoming hadron or nucleus.
It is easy to evaluate both these diagrams using \eq{DDXS}. It turns out
that their contributions to $\Omega$ are 
\begin{eqnarray} \label{FOFD}
\Omega( h + A \,\rightarrow\,M + A)\,\,&=&\,\,-\,\,g_{PN}(b_t-b'_{t})
\,g_{PA}(b_t)
e^{\Delta\,Y}\,\,\kappa_A(b_t)\,\,;\nonumber\\
\Omega( h + A \,\rightarrow\,h +  M)\,\,&=&
\,\,-\,\,g_{PN}(b_t-b'_{t})\,g_{PA}(b_t)
e^{\Delta\,Y}\,\,\kappa_N(b_t-b'_{t})\,\,,
\end{eqnarray}
where $\kappa_{A,N}(b_t)\,\,=\,\,\kappa_{A,N}\,exp( - \frac{-
b^2_t}{R^2_{A,N}})$. Accordingly to the rules of \eq{PSR} we neglect
$\Omega( h + A \,\rightarrow\,h +  M)$ and, therefore, at this order we
have only a "fan" diagram with two Pomerons attached to the  nucleus.
We have more diagrams of order $G^2_{3P}$ ( see Fig.\ref{nf1}
where both up and bottom lines could be incoming nucleus ).

The "fan" diagram, in which three Pomerons
attach to the nucleus ( the second diagram of Fig.\ref{nf1} )
has a contribution to $\Omega$ which is equal to
\beq \label{SOFD}
\Omega_{fan} (G^2_{3P})\,\,=\,\,g_{PN}\,g_{PA}(b_t)
e^{\Delta\,Y}\,\kappa^2_A(b_t)\,\,;
\eeq
while all other diagrams are much smaller since:

\begin{enumerate}
\item\,\,\, The "fan" diagrams, where three Pomerons are attached to the
hadron, are proportional to
$g_{PN}\,g_{PA}(b_t)e^{\Delta\,Y}\kappa^2_N(b_t-b'_{t})$ and, therefore, this
diagram gives a much smaller contribution than \eq{SOFD}.
\item\,\,\, The diagrams, in which two Pomerons are attached to the hadron and
two Pomerons are attached to the nucleus, also have  a tree-like structure.
However, its contribution is of the order of
$g_{PN}\,g_{PA}(b_t)e^{\Delta\,Y}\kappa_N\,\kappa_A$ which is smaller than
the contribution of \eq{SOFD}.
\item\,\,\, The loop diagram ( the first in Fig.\ref{nf1} ) is
proportional to  $g_{PN}\,g_{PA}(b_t)e^{\Delta\,Y}\kappa^2_N$
and can be neglected.
\end{enumerate}

Repeating this procedure for higher order of $G_{3P}$ we
will obtain the selection rules that have been mentioned above, namely,
only "fan" diagrams of Fig. ~\ref{MF40} contribute to the value of the opacity
$\Omega$.

\subsection{The Hadron-nucleus amplitude at high energy}
\begin{figure}[htbp]
\begin{center}
\epsfig{file=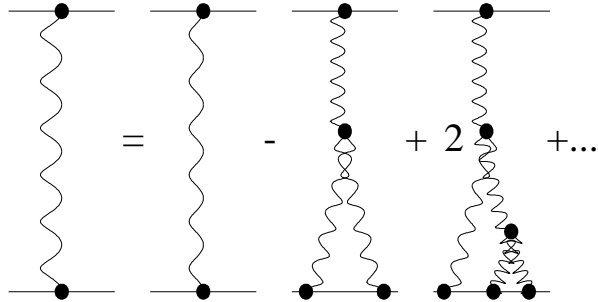,width=10cm}
\end{center}
\caption{ \it Hadron-Nucleus amplitude at high energy.}
\label{nf3}
\end{figure}

To calculate the hadron-nucleus amplitude we have to sum the
"fan" diagrams of Fig.~\ref{nf3}.

\begin{figure}[htbp]
\begin{center}
\epsfig{file=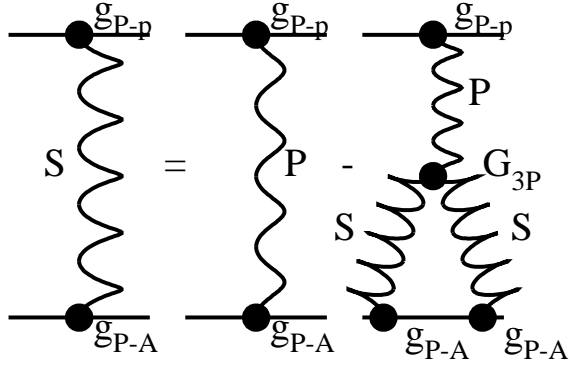,width=10cm}
\end{center}
\caption{\it The graphic equation for  the hadron-nucleus
amplitude.}
\label{nf4}
\end{figure}

It is easy to sum ``fan"  diagrams using the equation shown in
Fig.~\ref{nf4}.

This equation has a simple analytic form:
\beq \label{FANEQ}
S\Le Y, b_t \Ra\,\, =\,\,g_{PA}\Le b_t \Ra \,e^{\Delta Y}\,-\,G_{3P}\cdot
\int_{0}^{Y}
e^{\Delta\Le Y-y\Ra}\cdot S^{2}\Le y, b_t\Ra dy \,\,,
\eeq
where $\Omega(Y,b_t)\,\,=\,\,g_{PN}\,S(Y,b_t)$.
Taking the derivative with respect to $Y$ and  using \eq{FANEQ}
we obtain a simple differential equation
\beq \label{DEQFAN}
\frac{dS\Le y, b, b'\Ra}{dY}=\Delta\cdot S\Le y, b_t\Ra -
G_{3P}\,\cdot\, S^{2}\Le Y, b_t\Ra\,\,,
\eeq
with initial conditions which follows directly from \eq{FANEQ}:
\beq \label{INCONFEQ}
S\Le Y=0,b_t \Ra\,\,\,=\,\,\, g_{PA}\Le b_t \Ra\,\,.
\eeq
Rewriting \eq{DEQFAN} in the form ($\gamma = G_{3P}/\Delta$):
\beq \label{FFAN}
\frac{dS\Le Y, b_t\Ra}{\Delta\cdot dY}=S\Le y, b_t\Ra -
\gamma \cdot S^{2}\Le Y,b_t\Ra ,
\eeq
we obtain the solution to \eq{FFAN} and \eq{INCONFEQ}
\beq \label{SFD}
S\Le Y,b_t\Ra =
\frac{e^{\Delta Y}\,g_{PA}\Le b_t \Ra }
{g_{PA}\Le b_t\Ra\cdot\gamma\cdot\Le e^{\Delta Y}-1\Ra +1}\,\,=\,\,
\frac{e^{\Delta Y}\,g_{PA}\Le b_t \Ra }{\kappa_A\Le Y, b_t\Ra \,\,+\,\,1}.
\eeq
Finally,
\beq \label{FFOM}
\Omega_{fan} \Le Y, b_t \Ra\,\,=\,\,g_{PN}\,\,S\Le Y,b_t
\Ra\,\,=\,\,\frac{e^{\Delta Y}\,g_{PN}\,\,g_{PA}\Le b_t \Ra }{\kappa_A\Le
Y, b_t\Ra
\,\,+\,\,1}.
\eeq

\subsection{Single diffraction dissociation}

\begin{figure}[htbp]
\begin{center}
\epsfig{file=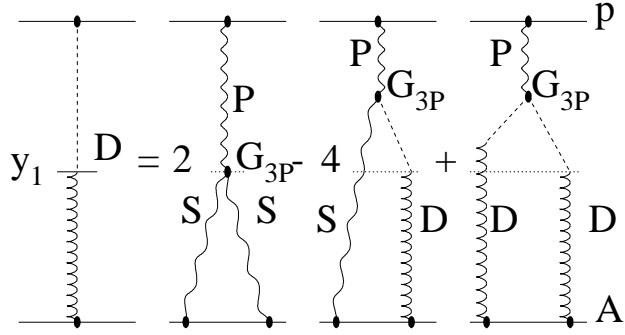,width=10cm}
\end{center}
\caption{\it Pictorial form of the equation for function
$D(Y,y,b_t)$.}
\label{fig6}
\end{figure}

In this subsection we derive the equation for the 
diffractive dissociation cross section
of the incoming hadron.  The general formula for
the
diffractive cross section can be obtained directly from \eq{UB1} and
\eq{UB2} and it is
\beq \label{DF1}
\sigma^{SD} ( h + A \,\rightarrow\,M + A )\,\,=\,\,\int\,d^2b_t\,e^{ -
\Omega(s,b_t)}\,D(Y,y,b_t)\,\,,
\eeq
where $D(Y,y,b_t) $ is the cross section  for diffraction dissociation
of the "fan" diagram at fixed $b_t$. Here, $ Y = ln(s/s_0)$ and $Y - y =
ln(M^2/s_0)$.  For the function $D(Y,y,b_t)$ we can write the equation which
is shown in Fig.\ref{fig6}

This equation generates all diagrams that
contribute to the function $D$, but before doing so we comment on each
term of the equation separately:

\begin{enumerate}
\item \,\,\, We want to  calculate the
function $D\Le Y, y,b_t\Ra $
on  the l.h.s. of the equation.
This function is a cross-section
for the single diffractive dissociation generated by "fan" diagrams , with
the gap in rapidity from
0 to $y$ and production of particles in the rapidity interval (Y, y)
 with $Y - y = ln M^2 $.

\item\,\, The first term on the r.h.s. of the equation, is our simplest
diagram of Fig.\ref{MF15} where we substitute the full amplitude
$S(y,b_t)$ for the "fan" diagrams (see \eq{FANEQ} and Fig.\ref{nf3} )
instead of the two Pomerons which are attached to the nucleus.
The coefficient 2 comes from the
AGK cutting rules \cite{AGK} since in the upper part of the
diagram we have the cut Pomeron. In our equation this term plays
the  role
of the initial condition and we obtain the full series of the
diagrams for the diffractive cross section by iterating this term in the
equation.

\item\,\, The second term on the r.h.s. is written with the help of
our function, $D\Le y', y,b_t\Ra $. The coefficient 4 arises due to
(i) we have a cut Pomeron in the upper part of diagram
and (ii) there are two possible types of this diagram,
which we can get by interchanging
$S\Le y,b_t\Ra$ and $D\Le Y, y,b_t\Ra $.

\item \,\,The third term on the r.h.s.  describes the possibility of
having
a diffractive production from more than one Pomeron.
The coefficient  2 from the cut Pomeron is cancelled
by the
coefficient 1/2, due to the symmetric form of the diagram ( we have
$D^2\Le
y',y,b_t \Ra$ in it ) giving an overall coefficient of 1.
\end{enumerate}

\begin{figure}[htbp]
\begin{center}
\epsfig{file=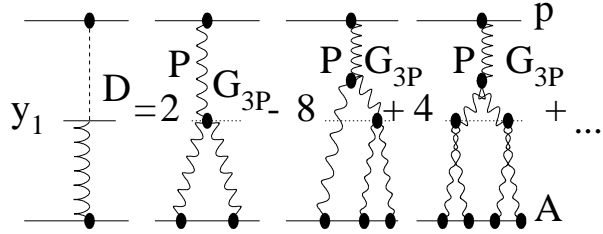,width=10cm}
\end{center}
\caption{\it The first terms obtained from the iteration of the equation of
Fig.8.}
\label{fig7}
\end{figure}

In writing the equation we have used the AGK cutting rules.
We follow 
Refs.\cite{AGK} and \cite{BR} 
which establish the relation between different
production processes including diffractive dissociation. 
It is easy to see that the iteration of the equation of
Fig.\ref{fig6} yields a series for the diagram for diffractive production
shown in Fig.\ref{fig7}.

This equation can be written in the form:
\begin{eqnarray}
D\Le Y, y, b_t \Ra&=& 2\cdot g_{PN}\cdot G_{3P}\cdot
e^{\Delta\Le Y-y\Ra}\cdot S_{\downarrow}^{2}\Le y, b_t\Ra \nonumber\\
 &-& -4\cdot g_{P-p}\cdot G_{3P}\int^{Y}_{y} d y'\,e^{\Delta\Le y-y'
\Ra}\cdot
 S\Le y',b_t\Ra \cdot D \Le y', y, b_t\Ra \nonumber\\
&+& g_{PN}\cdot G_{3P}\int_{y}^{Y}\,d\,y'\, e^{\Delta\Le Y-y' \Ra}\cdot
D\Le y', y,b_t\Ra\cdot\int_{y}^{y'}\,d\,y"\, D\Le y', y"\Ra\,\,.
\label{DF2}
\end{eqnarray}
We introduce a new function, $SD\Le Y, y,b_t\Ra$, which is the cross
section of the
single diffraction  of the incoming hadron in the hadron system with mass
(M) smaller than $\ln (M^2/s_0) \,\leq\, Y - y $
\beq \label{DF3}
SD\Le Y, y, b_t \Ra \,\,=\,\,\int_{y}^{Y}\,d\,y' S\Le Y, y', b_t\Ra \,\,,
\eeq
with the initial condition
\beq \label{DF4}
SD\Le Y, Y,b_t\Ra =0\,\,.
\eeq

Integrating \eq{DF2} we obtain an equation for $SD\Le Y, y, b_t \Ra $
\begin{eqnarray}
SD\Le Y, y, b_t\Ra &=&
2\cdot g_{PN}\cdot G_{3P}\cdot\int_{y}^{Y}\,d\,y'\,\, S^{2}\Le
y', b_t\Ra\cdot
e^{\Delta\Le Y-y'\Ra}  \nonumber\\
&-& 4\cdot g_{PN}\cdot G_{3P}\int_{y}^{Y} \,d\,y'\,\,
e^{\Delta\Le Y-y' \Ra}\cdot S\Le y',b_t\Ra\cdot
SD\Le y', y, b_t\Ra \nonumber\\
&+& g_{PN}\cdot G_{3P}\int_{y}^{Y}\,d\,y'\,\, e^{\Delta\Le Y-y' \Ra}\cdot
SD\Le y', y, b_t\Ra^{2} \,\,.\label{DF5}
\end{eqnarray}
The derivative of \eq{DF5} with respect to $Y$ gives
\begin{eqnarray}
\frac{dSD\Le Y, y,b_t\Ra}{d\Delta Y} &=&
SD\Le Y, y,b_t\Ra\,\,+
g_{PN}\cdot\gamma\cdot\Le\, 2\,S\Le Y, b_t\Ra\,\, -\,\,
SD\Le Y, y,b_t\Ra\,\Ra^{2} \nonumber\\
&-&  2\,\cdot g_{PN}\cdot\gamma\cdot S^2\Le Y,b_t\Ra\,\, ,
\label{DF6}
\end{eqnarray}
where $\gamma = G_{3P}/\Delta$.

Subtracting  \eq{FFAN} for $S$ in  \eq{DF6} we  obtain
\beq \label{DF7}
\frac{dF\Le Y, y, b_t \Ra}
{\Delta\cdot d Y}\,\,= \,\,F\Le Y, y, b_t \Ra\,\, -
\,\,\gamma F^2\Le Y, y, b_t \Ra\,\, ,
\eeq
where the function $F$ is defined as
\beq \label{DF8}
F\Le Y,y,b_t \Ra
\,\,=\,\,g_{PN}\,T\Le Y,y,b_t
\Ra\,\,=\,\,g_{PN}\,\,\cdot\,\,\{\,2\,S\Le Y,y,b_t
\Ra\,\,\,-\,\,SD\Le Y,y,b_t\Ra\,\}\,\,.
\eeq
We obtain for $F$ the same equation as for $S$, however there is a
difference in the initial and boundary conditions, which are
\begin{eqnarray}
F(Y,Y,b_t)\,\,&=&\,\,2
\,g_{PN}\,S(Y,b_t)\,\,=\,\,2\,\,\frac{g_{PN}\,g_{PA}(b_t)\,\,e^{\Delta
\,Y}}{\kappa_A(Y,b_t) \,+\,1}\,\,;\label{DFIN1}\\
F(Y,0)\,\,&=&\,\,2\,\frac{g_{PN}\,g_{PA}(b_t)\,\,e^{\Delta
\,Y}}{2\,\kappa_A(Y,b_t) \,+\,1}\,\,.
\label{DFIN2}\,\,
\end{eqnarray}

To understand \eq{DFIN2} we need to recall that $F(Y,0)$ denotes the total
inelastic cross section $F(Y,0,b_t) = \sigma_{in}(Y,b_t)$ due to the 
unitarity constraint of \eq{UNTRTY}. On the
other hand we can calculate the inelastic cross section using the AGK
cutting rules which lead to $\sigma_{in}(Y,b_t) =  2 S(Y, 2\,\kappa_A)$.

The solution of \eq{DF8} with the initial and boundary conditions of
\eq{DFIN1} and \eq{DFIN2} is
\beq \label{DF9}
F(Y,y.b_t)\,\,\,=\,\,\,2\,\frac{
g_{PN}\,g_{PA}(b_t)e^{\Delta\,Y}}{2\,\kappa_A(Y,b_t) - \kappa_A(y,b_t)
\,\,+\,\,1}\,\,.
\eeq
Using \eq{DF8} we can find the  function $ SD \Le Y,y, b_t \Ra $ defined
in
\eq{DF3}
\beq \label{SD}
 SD \Le Y,y, b_t \Ra\,\,=\,\, \frac{g_{PN}
\,e^{\Delta Y}\,\,\cdot\,\{\,\kappa_A(Y,b_t)
\,\,-\,\,\kappa_A(y,b_t)\,\}}
{\{\,2\,\kappa_A(Y,b_t)\,\,-\,\,\kappa_A(y,b_t)\,\}\,\,
\{\,\kappa_A(Y,b_t)\,\,+\,\,1\,\}}
\,\,.
\eeq
>From \eq{DF9} and \eq{DF1} we can obtain the total ( integrated over mass
) cross section for single diffraction production:
\begin{eqnarray} \label{DF10}
\sigma^{SD}\Le Y\Ra
\,&=&\\
\,& &
\int\,d^2\,b_t\,e^{-
\Omega_{fan}\Le Y.b_t\Ra}\,\,SD\Le Y,0,b_t
\Ra\,\,=\,\,\int\,d^2\,b_t\,\,\{\,2\,g_{PN}
\,S\Le Y,b_t\Ra\,\,-\,\,F\Le Y,0,b_t\Ra\,\}\,e^{- \Omega_{fan}\Le Y,b_t\Ra} .
\nonumber
\end{eqnarray}

Using \eq{DF9} we can also calculate $D\Le Y,y,b_t\Ra$ in \eq{DF1},
which is equal to
\beq \label{DF11}
D\Le Y,y,b_t\Ra\,\,=\,\,2\,\frac{g^2_{PA}\Le b_t \Ra
\,\cdot\,g_{PN}\,\cdot\,G_{3P}\,\cdot\,e^{\Delta\,(Y +
y)}}{\{\,2\,\kappa_A\Le
Y,
b_t \Ra \,\,-\,\,\kappa_A\Le y, b_t \Ra\,\,+\,\,1\,\}^2 }\,\,.
\eeq
\eq{DF10} and \eq{DF11} solve the problem of single diffractive
cross section in our approach.

\subsection{Survival Probability of Large Rapidity Gaps}

A large rapidity gap (LRG)  process, is any process in which no particles are
produced in sufficiently large rapidity region \cite{BJ}. The simplest
example of such process is the production of two large transverse momentum
jets with LRG between them ($\Delta y = | y_1 - y_2 | \,\,\gg\,\,1$ )
in back-to-back kinematics ( $\vec{p}_{t1} \,\sim\, - \vec{p}_{t2} $ ):
\beq \label{SP1}
p \,\,+\,\,p\,\,\,\longrightarrow
\eeq
$$
M_1\{ hadrons\,\,+\,\,jet_1 \Le(y_1,\vec{p}_{t1} \Ra
\,\}\,\,+\,\,[LRG(\Delta y)]\,\,+\,\,M_2\{ hadrons\,\,+\,\,jet_2
\Le(y_1,\vec{p}_{t1} \Ra\,\}.
$$

We believe that the exchange of colourless "hard" Pomeron is the only way
to describe  this production making it a unique source for 
experimental information regarding
pQCD at high energies. Assuming factorization \cite{FT},
the cross section for this reaction can be written as
\beq \label{SP2}
\sigma_{jet}\,\,=\,\,f\Le\Delta y, y_{1}+y_{2}, p_{t1}, p_{t2}\Ra=F^{\Le
1\Ra}_{p}\Le
x^{1},p_{t1}\Ra\cdot  F^{\Le 2\Ra}_{p}\Le x^{2},p_{t2}\Ra\cdot
\sigma_{hard}\Le p_{t1}, x^{1}x^{2}s\Ra\,\,,
\eeq
where $F^{\Le i\Ra}_{p}$ is the probability to find a parton with
$x^{i}= \frac{2p_{ti}}{\sqrt{s}}e^{y_{i}}
$ in the proton and
$\sigma_{hard}$ denotes the cross-section
of the "hard"parton-parton interaction at sufficiently high  energies.
This "hard"
process is due  to the exchange of a "hard" Pomeron.
We need to multiply \eq{SP2} by the damping factor $ <|S|^2>$ to
obtain the correct answer for the LRG cross section. This factor
$<|S|^{2}>$, gives a probability that no partons with $x>x^{1}$ from
one
proton, and no partons with $x<x^{2}$ from the other proton will
interact with each other inelastically. Therefore \eq{SP2}
should  be rewritten in the following form:
\beq \label{SP3}
f\Le\Delta y, y=y_{1}+y_{2}, p_{t1}, p_{t2}\Ra=<|S|^{2}>\cdot f\Le
\eq{SP2} \Ra\,\,.
\eeq
This damping factor $<|S|^{2}>$, which is called survival probability of the LRG
gap processes \cite{BJ}, has been calculated in  eikonal-type
models \cite{EM} for hadron - hadron collisions.  We proceed to calculate
$<|S|^{2}>$ for a hadron - nucleus collision in our approach, which allows us
to go beyond the eikonal - type models.

\begin{figure}[htbp]
\begin{center}
\epsfig{file=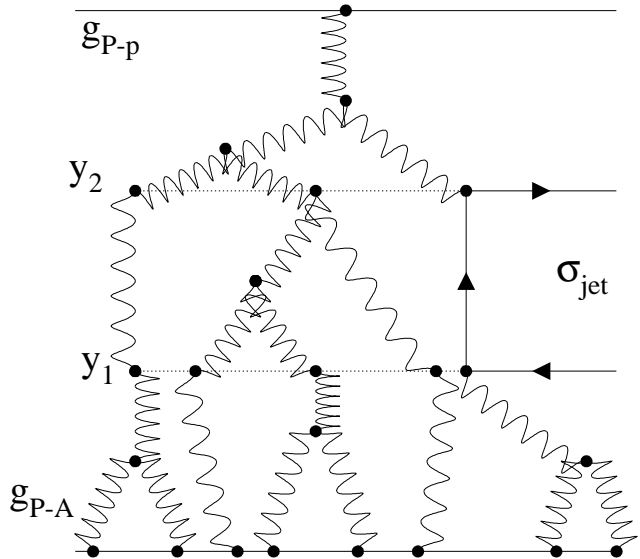,width=10cm}
\end{center}
\caption{The Survival probability diagram.}
\label{fig8}
\end{figure}

The general formula for the cross section  of LRG processes  is
\beq \label{SP4}
\sigma^{LRG}\,\,\,=\,\,\,\int\,d^2\,b_t \,e^{- \Omega_{fan}(Y,b_t)}\,\,
             L\Le Y,y_1,y_2,b_t\Ra\,\,,
\eeq
where $ L\Le Y,y_1,y_2,b_t\Ra$ is the cross section of LRG processes
induced by
"fan" diagrams. We need to calculate the "fan" diagrams of
Fig.\ref{fig8}  to estimate  $ L\Le Y,b_t\Ra$.

We now consider  Fig. \ref{fig8} . The upper part of this diagram,
from rapidity $y_2$ to rapidity $Y$, is our usual function $G_{PN}\cdot S(Y -
y_2 )$ but without $g_{PA}(b_t)$. We introduce the function $R(Y - y_2,z)
\,\,\equiv \,\,S(Y - y_2,g_{PA}(b_t)\,\rightarrow\,z )$ , which is a
generating function for the number of Pomerons at rapidity $y_2$  in the
diagrams of Fig.\ref{fig8}. This means that the functions $C(Y - y_2) $ in
series
\beq \label{SP5}
R(Y - y_2,z)\,\,=\,\,\sum \,C_n(Y - y_2)\,z^n
\eeq
can be interpreted as the probability to have $n$ Pomerons at rapidity
$y_2$.  To obtain the cross section of LRG processes from the function $R(Y -
y_2,z)$ we need to do four things:
\begin{enumerate}
\item\,\,\, We have to insert the "hard" cross section of \eq{SP2} in all
Pomeron lines at rapidity  $y_2$. That means that we have to change
$z^n\,\rightarrow\,n\,z^{n -1}\,\sigma_{jet}$.

\item\,\,\, To describe the absence of particle production in the
interval
$\Delta y  = y_1  -  y_2 $ we should replace $\gamma \,\rightarrow\,\,2
\gamma$ in the function $R$. As we have mentioned above, these rules are a
direct
consequence of the AGK cutting rules. Recall, that the factor $\Omega$ in
\eq{SP1} can also be derived from the eikonal
formula for the total cross section, making this substitution.

\item\,\,\, To generate Pomerons at rapidity $y_1$ we should replace $z $
in \eq{SP5}  by
\beq \label{SP6}
z\,\,\longrightarrow\,\,S\Le y_2 - y_1, 2 \gamma,z\Ra\,\,.
\eeq
This replacement produces the generating function at the rapidity 
$y_1$ and the argument
$2 \gamma $ in \eq{SP6}, reflecting  the fact that none of the 
Pomerons
created in the interval from $y_2$ to $y_1$,  can be the source of
the particle production.

\item\,\,\, Finally, we  substitute $S\Le y_1,g_{PA}(b_t) \Ra $ for
$z$, since below rapidity $y_1$ we have  normal ``fan" diagrams for every
Pomeron at  rapidity level $y_1$.
\end{enumerate}

The resulting formula is
\begin{eqnarray}
 L\Le Y,y_1,y_2,b_t\Ra\,\,&=& \label{SP7} \\
 & & \Le
\Le\frac{dR\Le y-y_{2},z, 2\gamma\Ra}
{d z}\Ra_{z=R\Le  y_{2}- y_{1},z, 2\gamma\Ra}
\Ra_{z=S\Le  y_{1},b_t \gamma\Ra}\cdot
\sigma_{jet}\cdot S\Le y_{1},b_t, \gamma\Ra\,\,,\nonumber
\end{eqnarray}
which can be rewritten in the form
\begin{eqnarray}
 L\Le Y,y_1,y_2,b_t\Ra\,\,&=& \label{SP8} \\
& &
\sigma_{jet}\frac{
g_{PN}\cdot g_{PA}(b_t) e^{\Delta\Le Y-y_{2}+y_{1}\Ra }}
{\Le \kappa_A(y_1,b_t) +1 \Ra}\cdot
\frac{\Le 2\, \kappa_A(y_2,b_t) \,-\,\kappa_A(y_1,b_t)  +1 \Ra^{2}}
{\Le 2\,\kappa_A(Y,b_t) - \kappa_A(y_1,b_t)  +1 \Ra^{2}}\,\,.\nonumber
\end{eqnarray}
To find the survival probability of the LRG we need to divide \eq{SP1} by the
inclusive cross section for di-jet production. We will show in the next
subsection that this cross section is
\beq \label{SP9}
\sigma_{incl} \Le Y,y_1,y_2,b_t \Ra = \sigma_{jet}\frac{
g_{PN}\cdot g_{PA}(b_t)  e^{\Delta\Le y-y_{2}+y_{1}\Ra }}
{\Le \kappa_A(y_1,b_t) +1 \Ra}\,\,.
\eeq

Finally, we obtain for the survival probability of a LRG process the
following formula
\beq \label{SP10}
<|S^2\Le Y,y_1,y_2 \Ra |>\,\,=\,\,
\frac{\int\,d^2\,b_t \,e^{- \Omega_{fan}(Y,b_t)}\,\,
L\Le Y,y_1,y_2,b_t\Ra\,\,,}{\int\,d^2\,b'_{t}\,
d^2\,b_t\,\,
\sigma_{incl} \Le Y,y_1,y_2,b_t,b'_{t} \Ra }\,\,.
\eeq

\subsection{Inclusive production}

In this subsection  we discuss the inclusive production of
particles for hadron-nucleus interactions at high energy. It is well known
that for inclusive production we have a number of very useful sum rules
which follow from the AGK cutting rules \cite{AGK}\cite{BR}. These sum
rules simplify our calculations, reducing them to estimates of several
Mueller diagrams \cite{MUDI}.

\subsubsection{Single inclusive cross section}
\begin{figure}[htbp]
\begin{center}
\epsfig{file=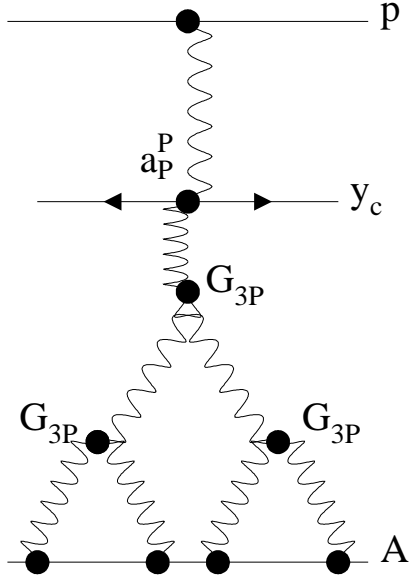,width=10cm}
\end{center}
\caption{\it The Mueller diagram for single inclusive cross section.}
\label{fig9}
\end{figure}

The single inclusive cross section can be described by one Mueller
diagram \cite{MUDI} shown in Fig.\ref{fig9}.

Recall, that the  diagram of Fig.\ref{fig9} is what remains after the AGK
cancellation  in a system of Reggeon diagrams which includes "fan"
diagrams in the entire region of rapidity ($Y$) and their eikonal
repetition.
The final result is
\beq \label{IP1}
\frac{d\sigma( h + A)}{d y}\,\,\,=\,\,\,\int\,d^2b_t\,g_{PN}\,a^P_P\,
e^{\Delta (Y - y)}\,S\Le y,b_t \Ra\,\,=\,\,\int\,d^2b_t\,g_{PN}\,a^P_P\,
g_{PA}(b_t) \frac{ e^{\Delta Y }}
{\Le \kappa_A(y,b_t) +1 \Ra}\,\,,
\eeq
where $a^P_P$ denotes the vertex of emission from the Pomeron.

\subsubsection{Rapidity correlations}

\begin{figure}[htbp]
\begin{center}
\epsfig{file=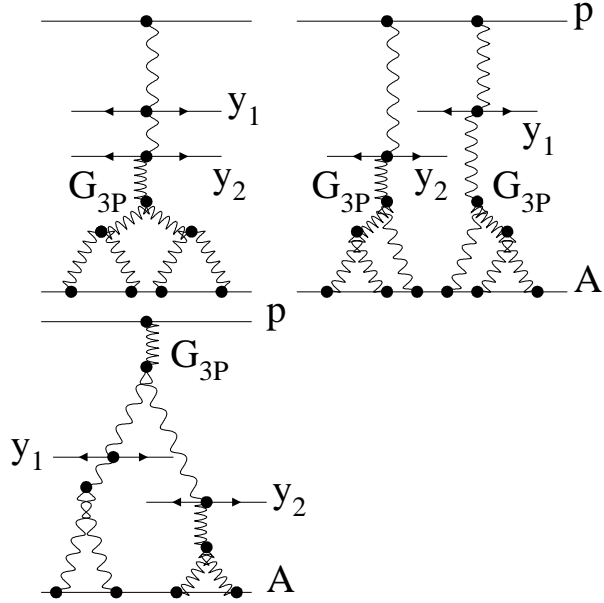,width=10cm}
\end{center}
\caption{\it The Mueller diagram for double  inclusive cross section.}
\label{fig10}
\end{figure}

For the double inclusive cross section, the only diagrams that survive the
AGK
cancellation are shown in Fig.\ref{fig10} and their contributions are
\begin{eqnarray}
\frac{d^2 \sigma}{d y_1 \,d y_2}\,\,\,&=&  
\int\,\,d^2\,b_t \label{IPCOR}\cdot
\\
 & & \cdot 
g^2_{PN} \,( a^P_P )^2 \,\,e^{ \,\Delta \,( \,2\,Y - y_1 - y_2
\,)} S\Le y_1,b_t \Ra \,\,S\Le y_2,b_t
\Ra  \label{IP2}\\
&+ & g_{PN} \,\,( a^P_P )^2\,\,e^{ \Delta (\, Y -  y_2 )}\,S\Le y_2,b_t
\Ra \,\label{IP3} \\
 & + & 2g_{PN} \,\,( a^P_P )^2\,\,\gamma \{\,e^{ \Delta (\,2 Y - y_1 -
y_2\,
)}\,\,-\,\,e^{ \Delta (\,Y - y_2\,)}\,\}  S\Le y_1,b_t \Ra \,\,S\Le
y_2,b_t \Ra .
\label{IP4}
\end{eqnarray}
 The rapidity correlation function is defined as
\beq \label{IP5}
R(y_1,y_2)\,\,\,=\,\,\, \frac{\frac{1}{\sigma_{tot}}\frac{d^2 \sigma}{d
y_1 \,d y_2}}{\frac{1}{\sigma_{tot}} \frac{d \sigma}{d
y_1}\,\,\,\frac{1}{\sigma_{tot}} \frac{d \sigma}{d
y_2}}\,\,\,-\,\,\,1\,\,.
\eeq
Note that correlation function $R$ is not equal to zero. In the
case of the Glauber approach the second term in \eq{IPCOR} ( see \eq{IP3} )
cancels with 1 in \eq{IP5} and only the first term  ( \eq{IP2} ) remains to
generate a correlation function. In our approach we do not have such a
cancellation and there are many correlations in the "fan" diagrams
alone, without Glauber rescatterings. We present our numerical estimates for the
correlation function in the next  subsection.

\subsection{Numerical estimates}

{\bf Proton - proton interaction:} We first estimate phenomenologically
 the vertices of the Pomeron-Pomeron 
and the Pomeron-proton interactions using data from total, elastic and
diffractive cross sections for high energy proton-proton collisions.
Our goal is to describe proton-proton scattering using a phenomenological model
as
close as possible, to one Pomeron exchange which successfully described the
hadron data \cite{DL}. However, we also want to include the experimental
data on   diffractive production  which cannot be described in 
the one Pomeron
exchange model.  This is the reason why we decided to include 
"fan" Pomeron diagrams of \eq{FFOM} in the proton-proton interaction. 
It is difficult to justify apriori such an approach, but
developing such a model will allows us to extract the values of
parameters for nucleon-nucleon interaction which we need to estimate the 
hadron-nucleus interaction.  For the Pomeron-proton vertex we use \eq{POMA}
with $A\,=\,1$ and we describe the the total 
cross section using the following generalization of \eq{FFOM}:
\beq \label{NE1}
\sigma_{tot}\,\,\,=\,\,\,\,\,\int\,d^2\,b_t \,\int\,\,d^2\,b'_t\,\,
\frac{ g_{PN}(b_t) \,g_{PN}(b'_t)\,e^{\Delta\,Y}}{\kappa_N(Y,b_t)
\,\,+\,\,1}\,\,.
\eeq
We introduce similar modifications  in \eq{O1} - \eq{O3}, as well as in
\eq{DF10} and \eq{DF11}. Fitting the experimental data we obtain the
following set of parameters:

\begin{enumerate}
\item\,\,$R^2_N\,\,=\,\,25\,\,GeV^{-2}$\,\,.
\item\,\,$g^2_{PN}(b_t = 0 )\,\,=\,\,70\,\,GeV^{-2}$\,\,.
\item\,\,$\gamma\,\,=\,\,0.14\,g_{PN}(b_t =
0)\,\,=\,\,1.19\,\,GeV^{1}$\,\,.
\item\,\,\,$\Delta\,\,=\,\,0.07$.
\end{enumerate}

\begin{figure}[htbp]
\begin{tabular}{ c c}
\epsfig{file=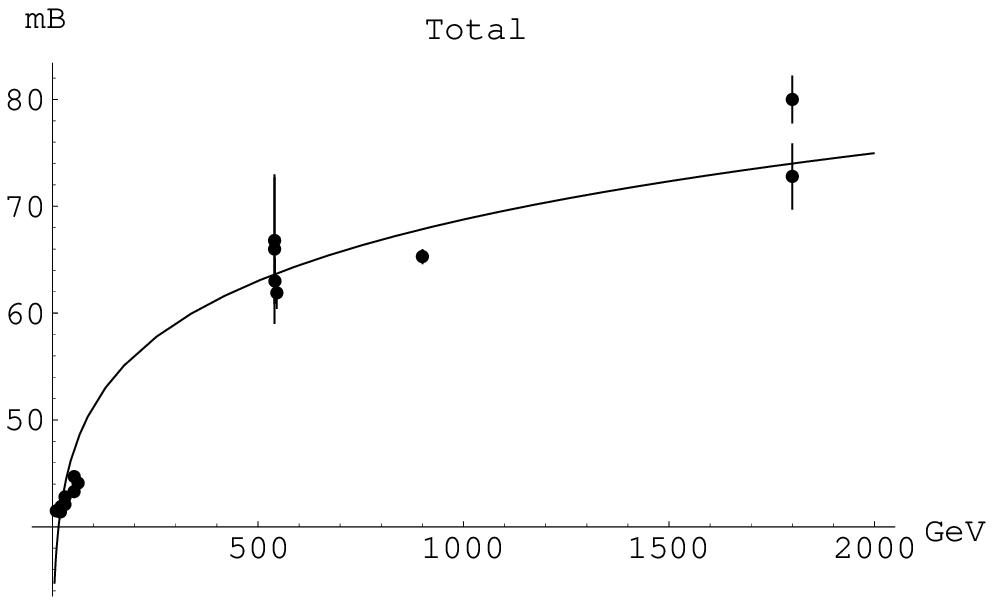,width=7cm} &
\epsfig{file=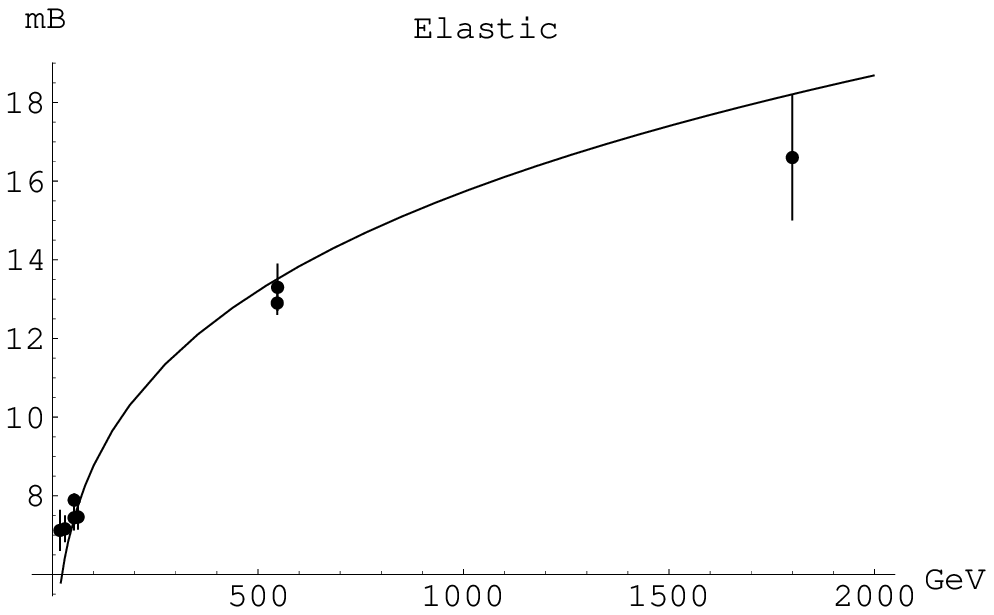,width=7cm} \\
Fig. 13-a &
Fig.13-b\\
\epsfig{file=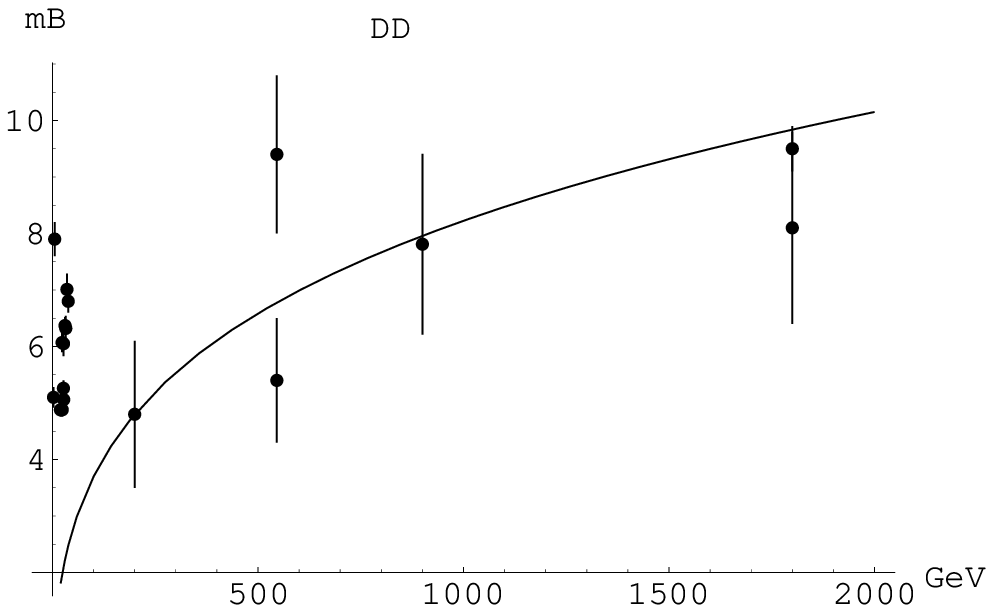,width=7cm} &
\epsfig{file=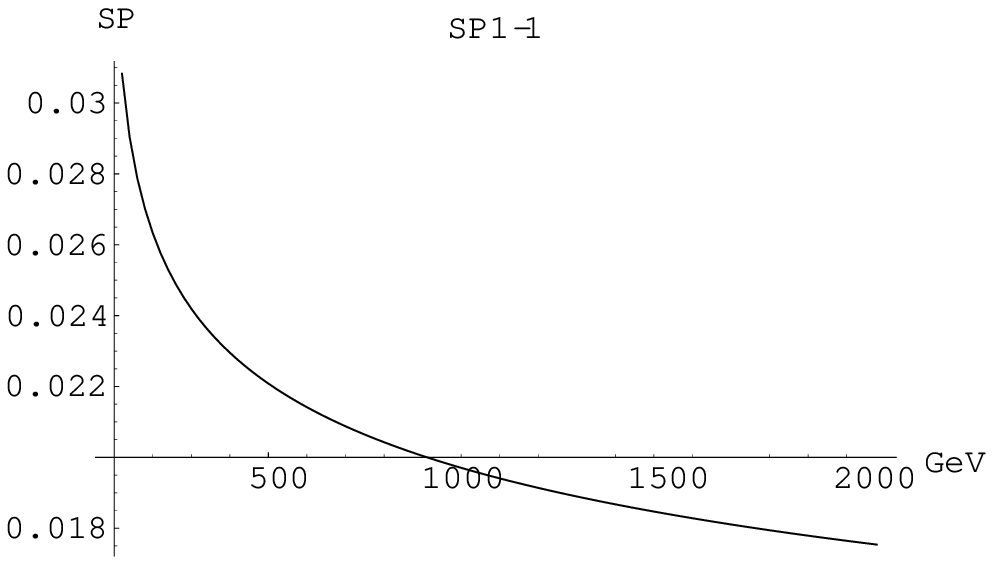,width=7cm}\\
Fig. 13-c &
Fig.13-d\\
\end{tabular}
\caption{\it Total  ( Fig.13-a ), elastic ( Fig.13-b ), diffractive
dissociation ( Fig.13-c )  cross sections and the survival probability of
LRG processes ( Fig.13-d ) for proton - proton scattering.}
\label{fig11}
\end{figure}

Fig. \ref{fig11} shows  our fit compared to
the experimental data,
where the survival probability for the case of  p-p interaction
was calculated using the eikonal approach.
The results are reasonable for a first attempt.

\begin{figure}[htbp]
\begin{tabular}{ c c}
\epsfig{file=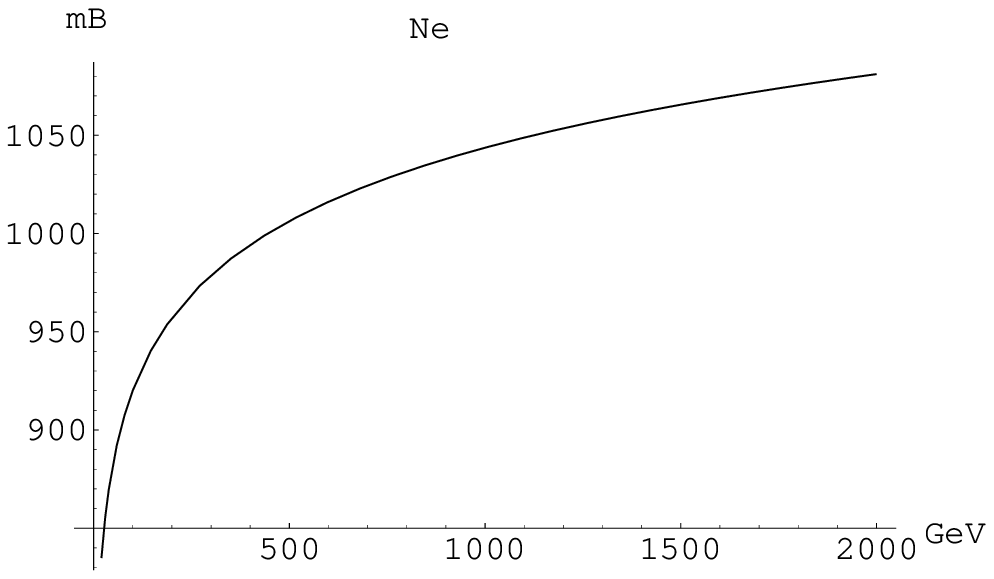,width=7cm} &
\epsfig{file=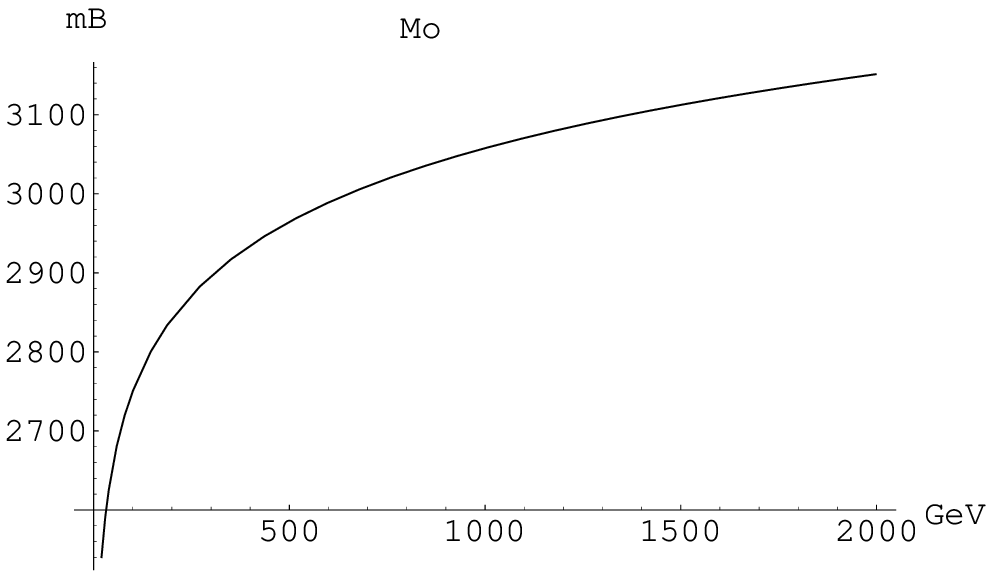,width=7cm} \\
Fig. 14-a &
Fig.14-b\\
\epsfig{file=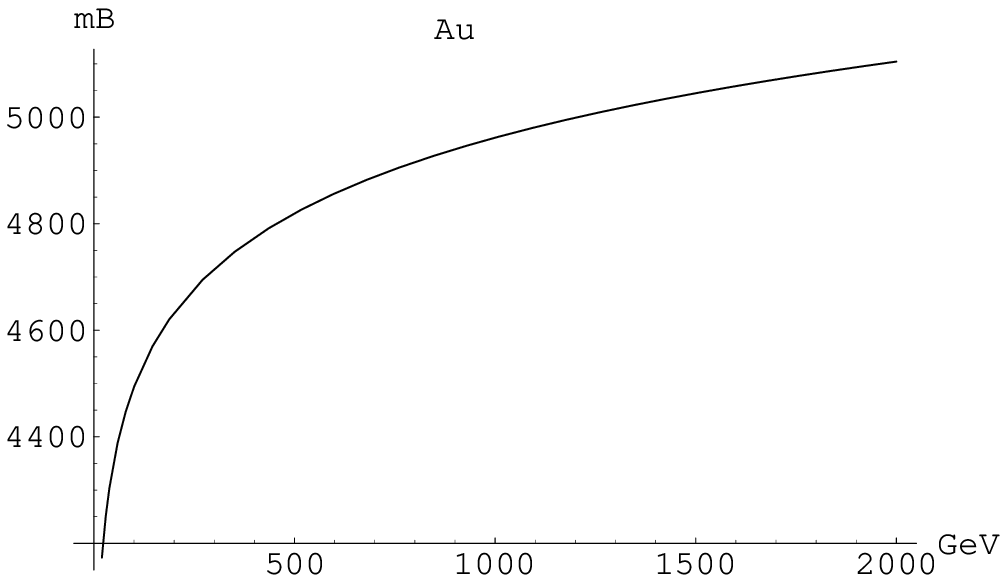,width=7cm} &
\epsfig{file=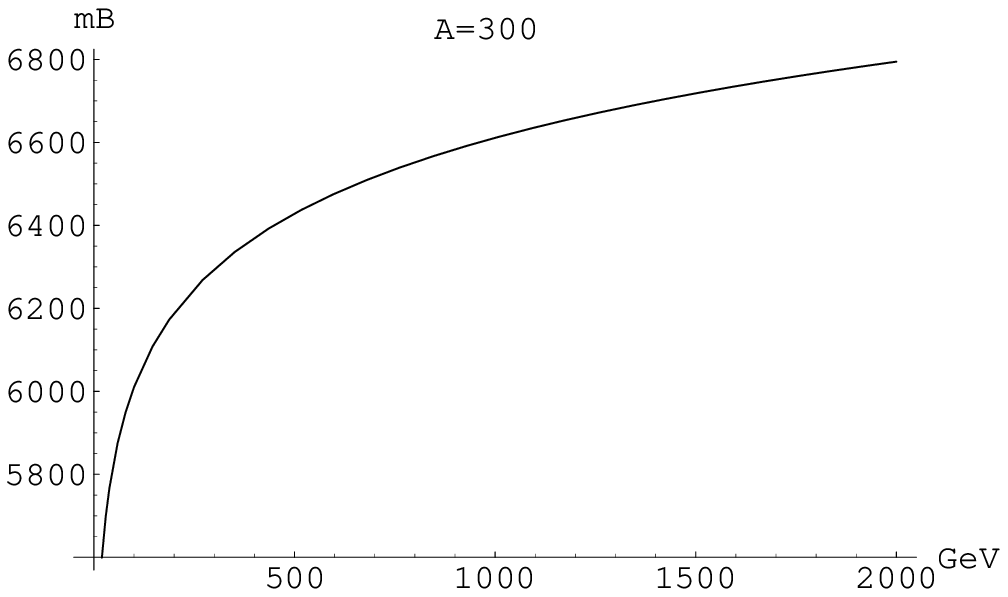,width=7cm}\\
Fig. 14-c &
Fig.14-d\\
\end{tabular}
\caption{\it Total proton - Nucleus 
cross sections 
for Ne ( Fig.14-a ), Mo ( Fig.14-b ), 
Au ( Fig.14-c ) and A=300( Fig.14-d ).}
\label{fig12}
\end{figure}

{\bf Proton - nucleus interaction:}
Using the parameters mentioned above, we calculate the 
total cross sections for
the hadron-nucleus interaction at high energies.
(see Fig.\ref{fig12} ).

\begin{figure}[htbp]
\begin{tabular}{ c c}
\epsfig{file=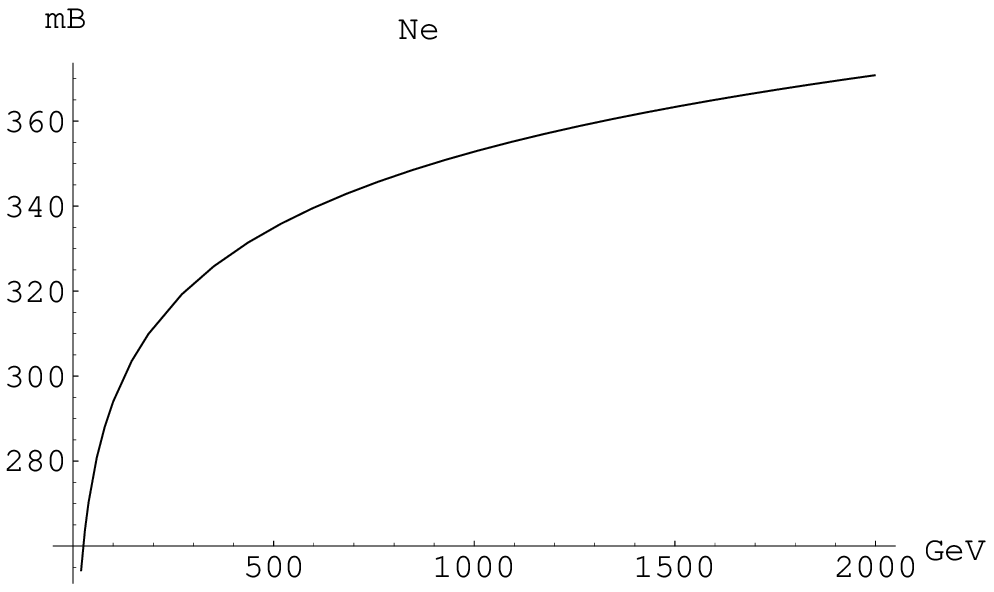,width=7cm} &
\epsfig{file=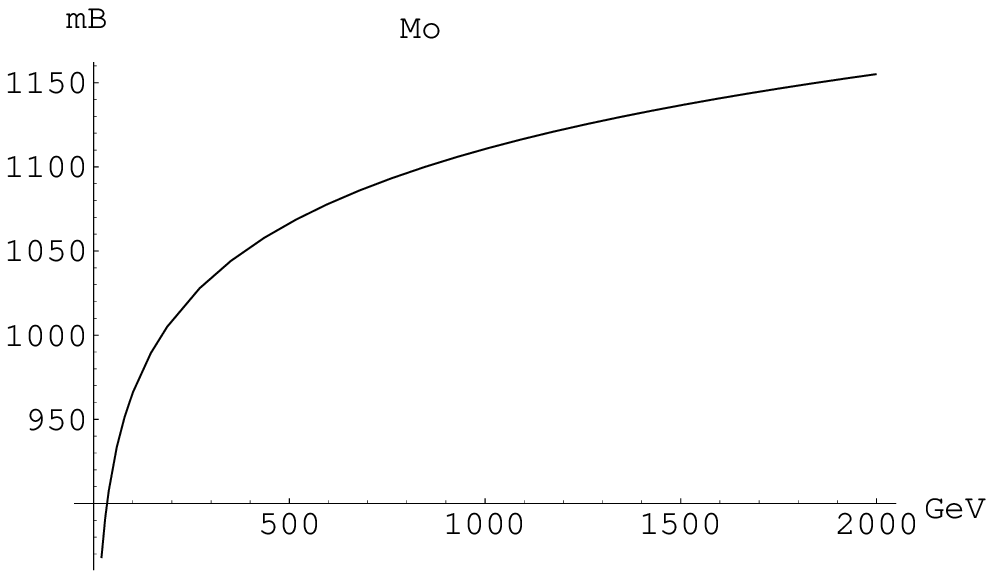,width=7cm} \\
Fig. 15-a &
Fig.15-b\\
\epsfig{file=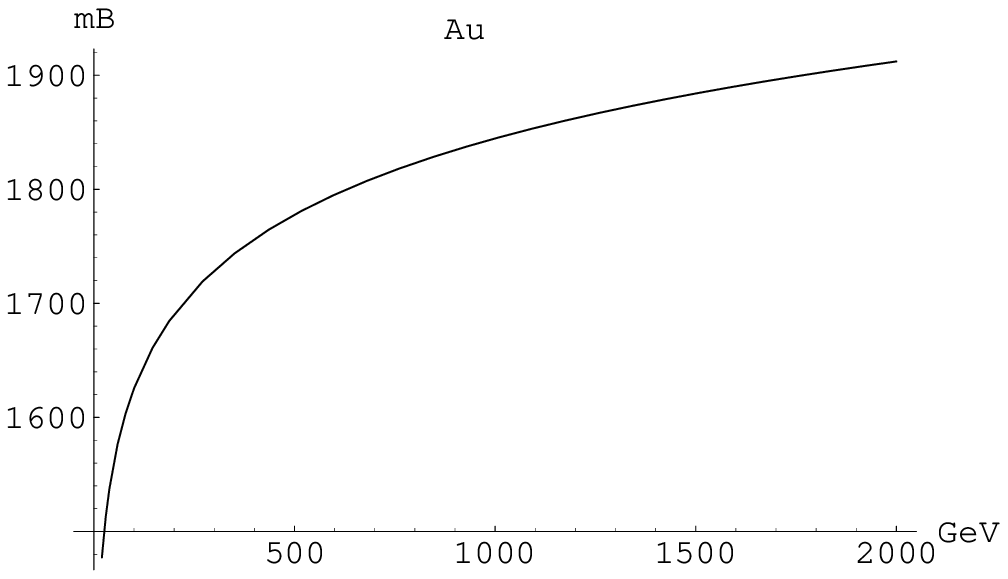,width=7cm} &
\epsfig{file=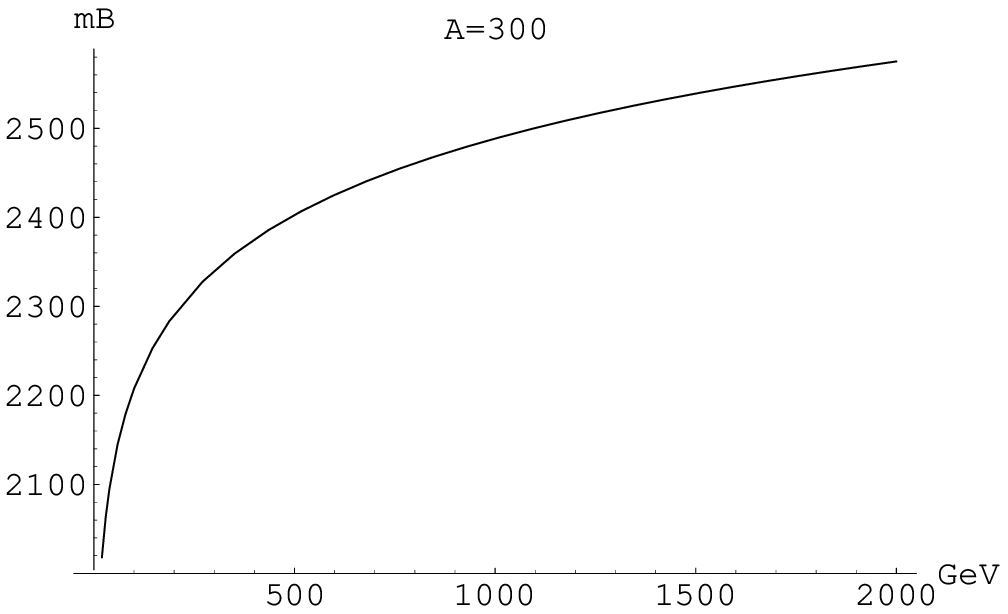,width=7cm}\\
Fig. 15-c &
Fig.15-d\\
\end{tabular}
\caption{\it Elastic  cross sections 
for Ne ( Fig.15-a ), Mo ( Fig.15-b ), 
Au ( Fig.15-c ) and A=300( Fig.15-d ) for proton - Nucleus scattering.}
\label{fig13}
\end{figure}

The corresponding elastic cross sections are
shown in Fig.\ref{fig13}. 

\begin{figure}[htbp]
\begin{tabular}{ c c}
\epsfig{file=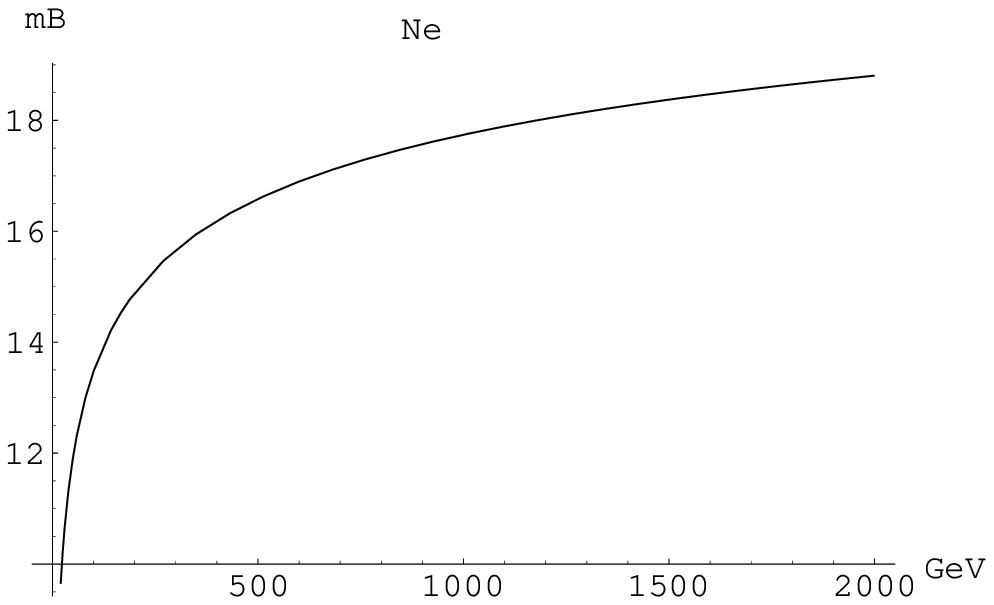,width=7cm} &
\epsfig{file=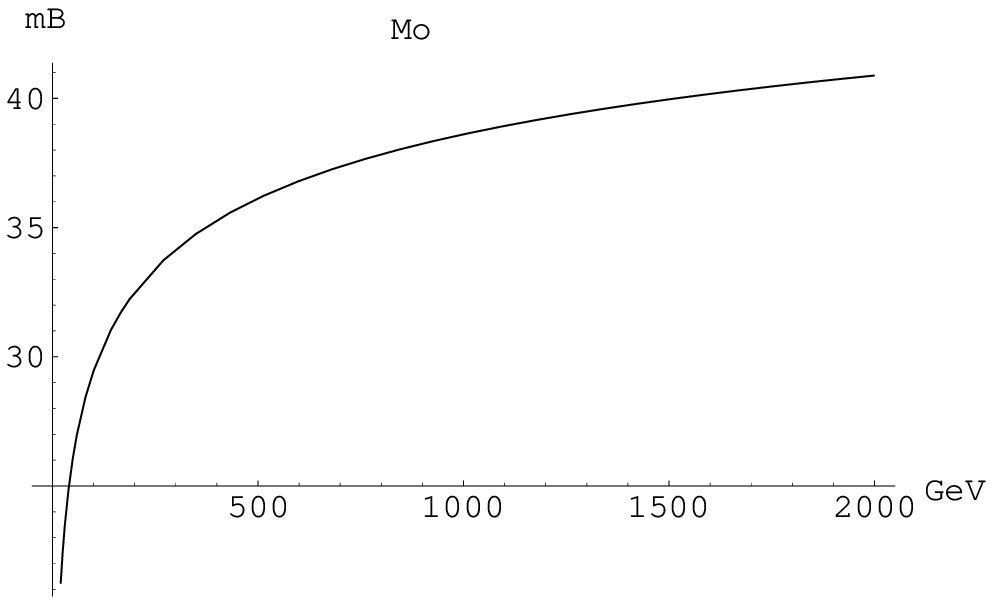,width=7cm} \\
Fig. 16-a &
Fig.16-b\\
\epsfig{file=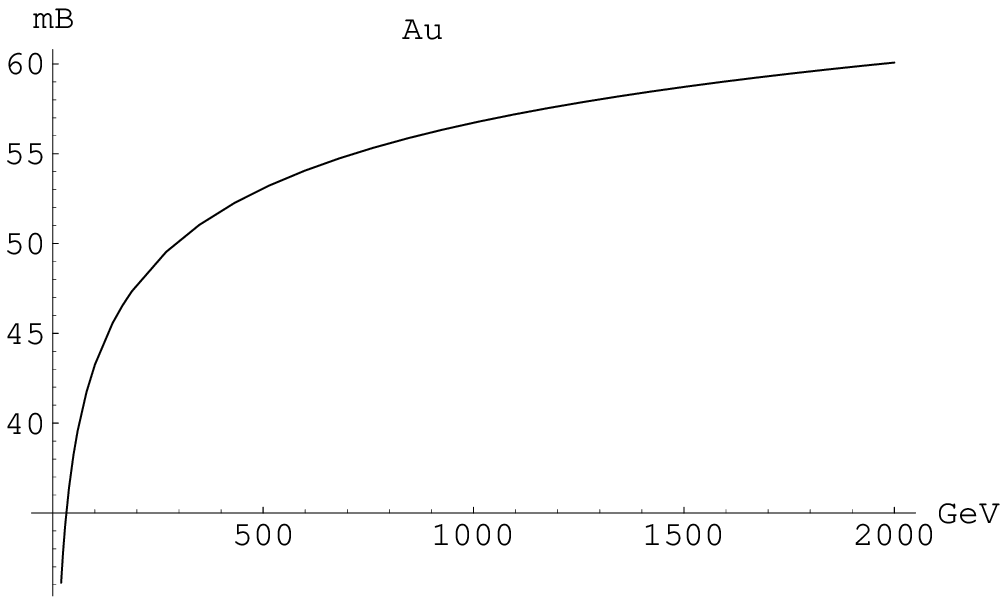,width=7cm} &
\epsfig{file=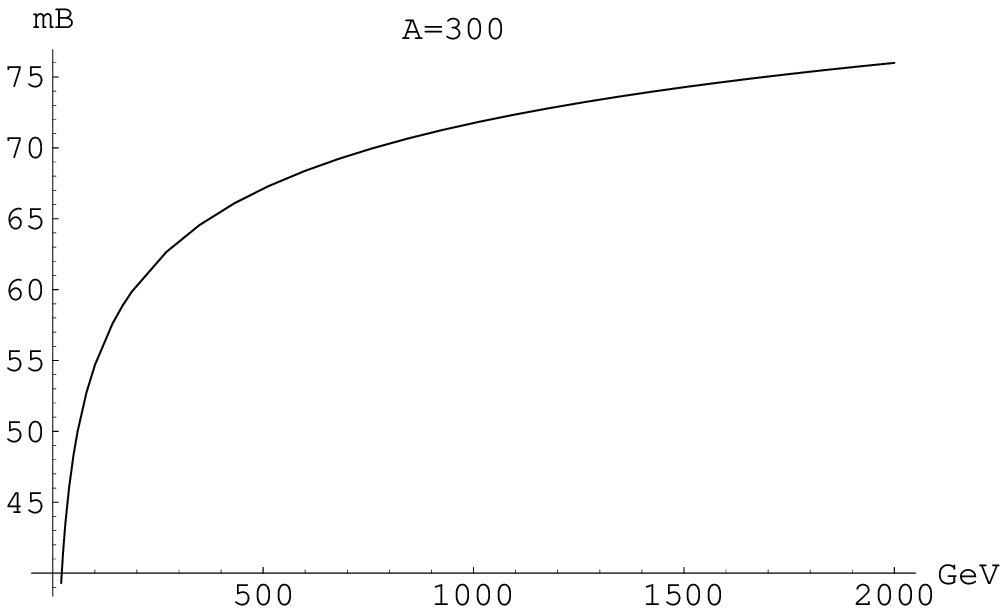,width=7cm}\\
Fig. 16-c &
Fig.16-d\\
\end{tabular}
\caption{\it Total diffractive production cross 
section as a function of energy
for Ne ( Fig.16-a ), Mo ( Fig.16-b ), 
Au ( Fig.16-c ) and A=300( Fig.16-d ) for proton - Nucleus scattering.}
\label{fig14}
\end{figure}

\begin{figure}[htbp]
\begin{tabular}{ c c}
\epsfig{file=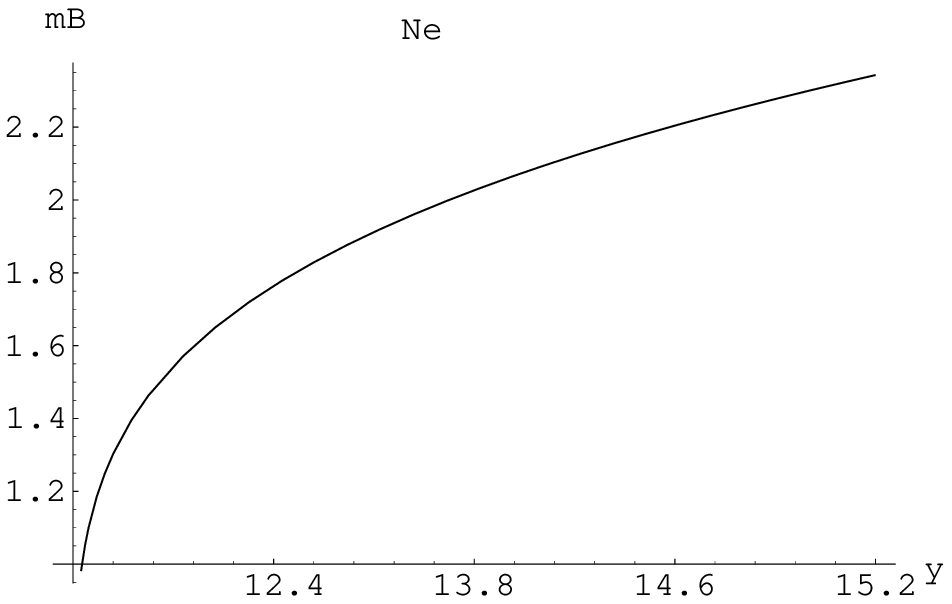,width=7cm} &
\epsfig{file=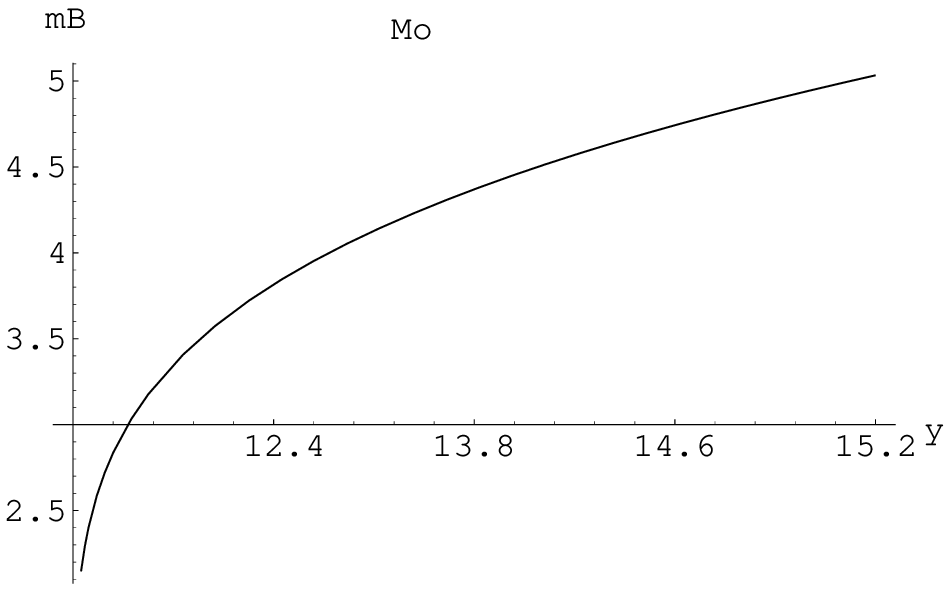,width=7cm} \\
Fig. 17-a &
Fig.17-b\\
\epsfig{file=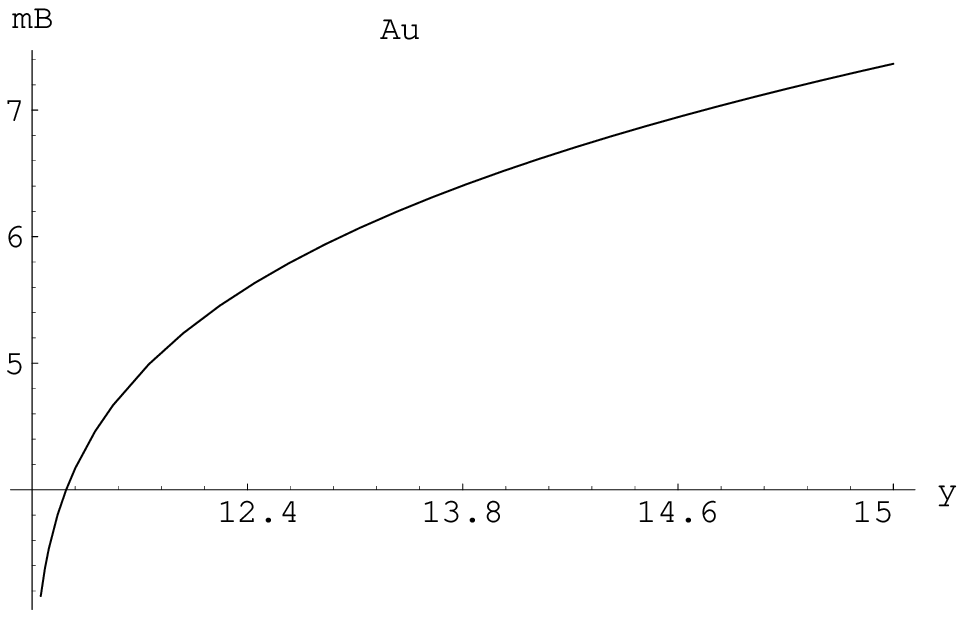,width=7cm} &
\epsfig{file=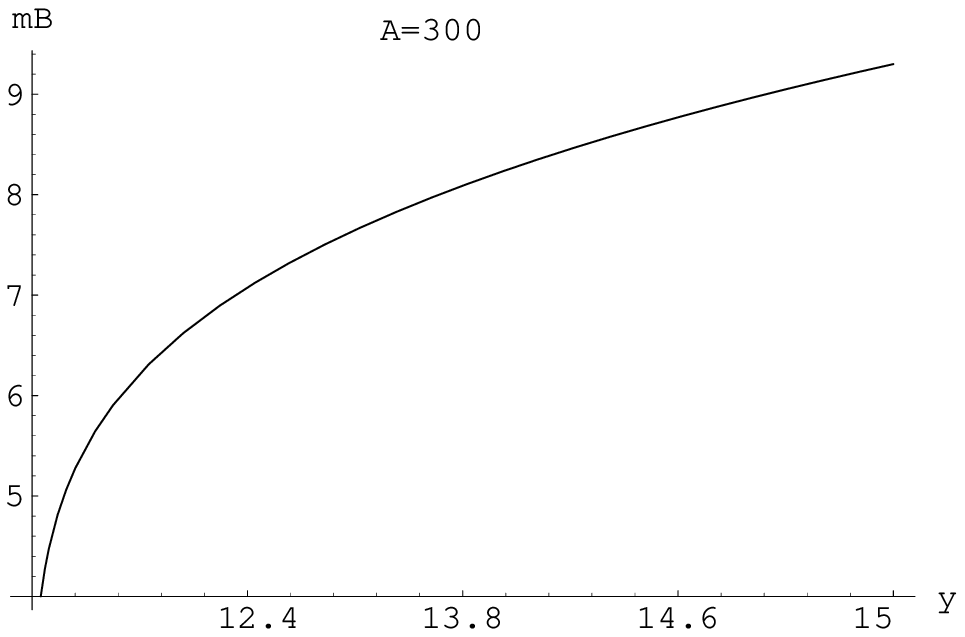,width=7cm}\\
Fig. 17-c &
Fig.17-d\\
\end{tabular}
\caption{\it Single diffractive dissociation cross section
at fixed rapidity, 
$Y=\ln\Le S/S_{0}\Ra$, $\sqrt{S}=2000$ $GeV$
and $\sqrt{S_{0}}=1$ $GeV$ ,
as a function of  rapidity gap, 
$y=\ln\Le s/s_{0}\Ra$, from $0$ to $15.2$
for $\sqrt{s}=1-2000$ $GeV$ and $\sqrt{s_{0}}=1$  $GeV$,
for Ne ( Fig.17-a ), Mo ( Fig.17-b ), 
Au ( Fig.17-c ) and A=300( Fig.17-d ) for proton - Nucleus scattering.}
\label{fig15}
\end{figure}

In Fig.\ref{fig14} we present 
the total diffractive production cross section as a function of
energy. The
single diffractive dissociation cross section,
\eq{DF11},
is 
presented in Fig.\ref{fig15}
at fixed rapidity, $Y=\ln\Le S/S_{0}\Ra$, $\sqrt{S}=2000$ $GeV$
and $\sqrt{S_{0}}=1$ $GeV$ , where $S$ denotes the center mass
energy squared,
as a function of the
rapidity gap, $y=\ln\Le s/s_{0}\Ra$, from $0$ to $15.2$,
which corresponds to
$\sqrt{s}=1-2000$ $GeV$ and $\sqrt{s_{0}}=1$ $GeV$.

\begin{figure}[htbp]
\begin{tabular}{ c c}
\epsfig{file=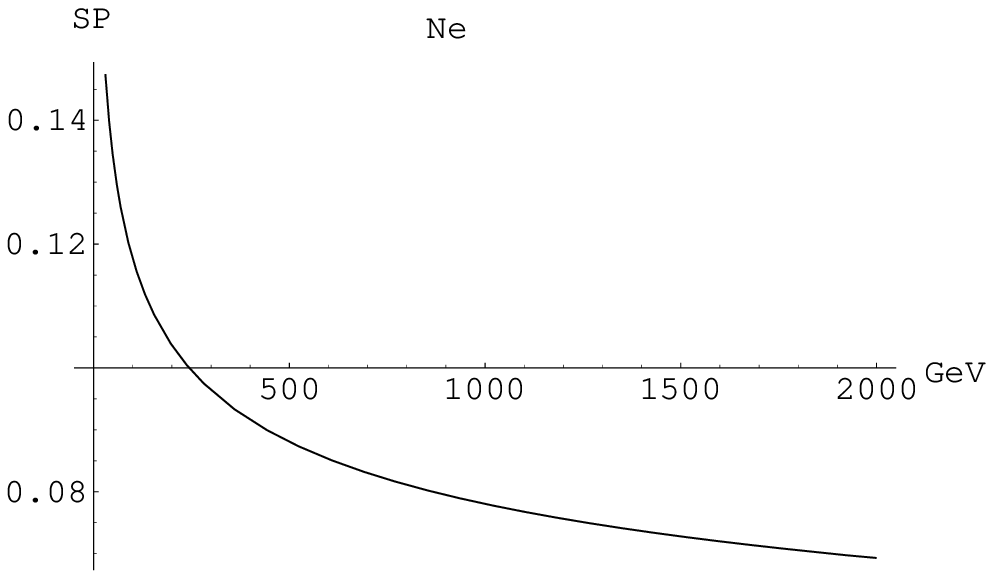,width=7cm} &
\epsfig{file=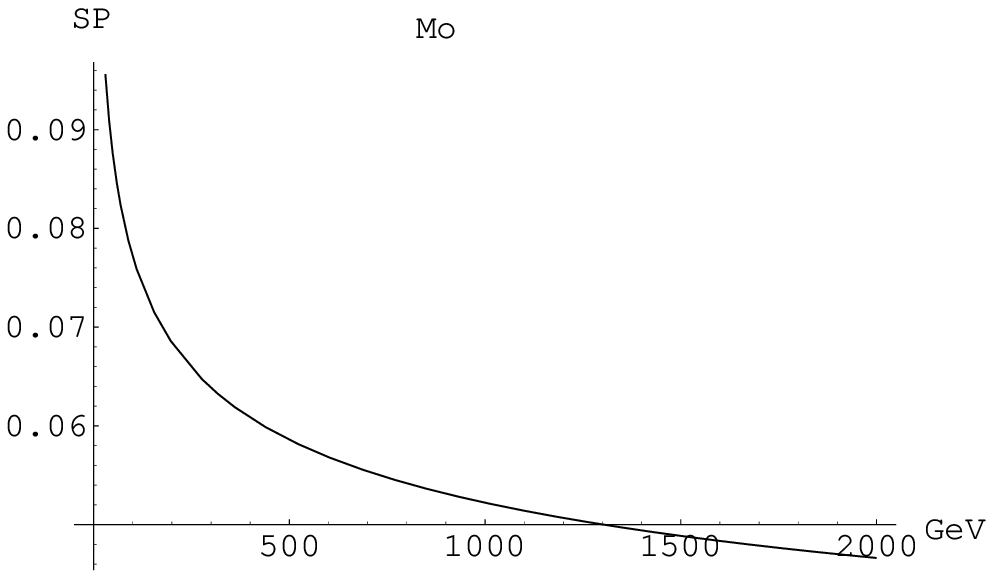,width=7cm} \\
Fig. 18-a &
Fig.18-b\\
\epsfig{file=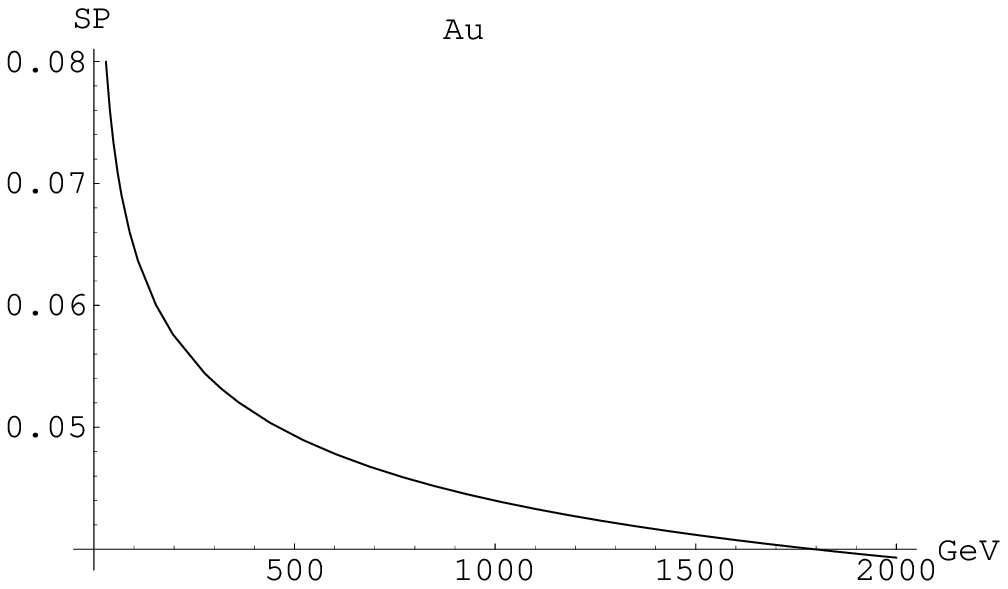,width=7cm} &
\epsfig{file=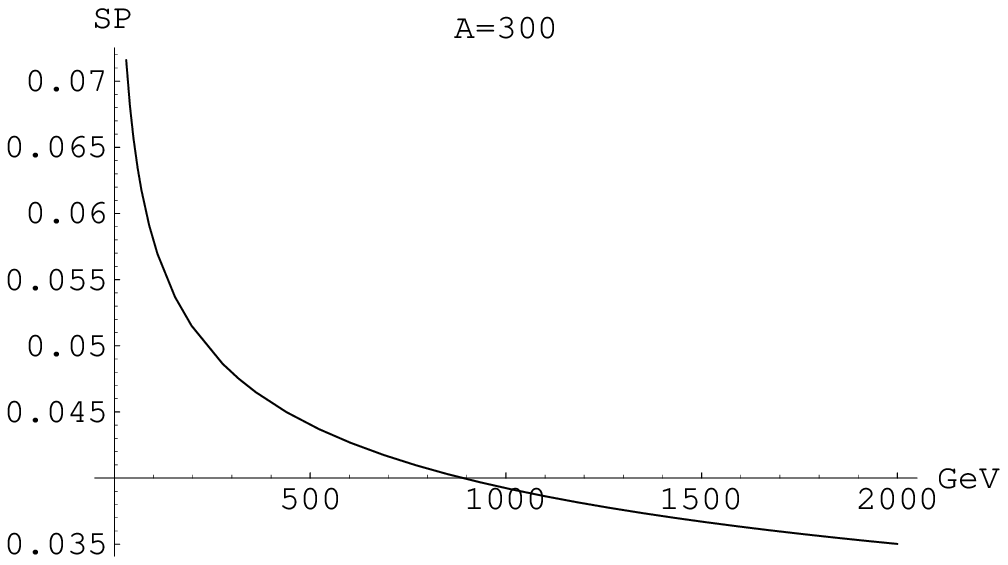,width=7cm}\\
Fig. 18-c &
Fig.18-d\\
\end{tabular}
\caption{\it Survival probability
for Ne ( Fig.18-a ), Mo ( Fig.18-b ), 
Au ( Fig.18-c )and A=300( Fig.18-d )
for the fixed jet produced in the interval of rapidity
10-12.5 for proton - Nucleus scattering.}
\label{fig16}
\end{figure}

In Fig.~\ref{fig16} we show our
estimates for survival probabilities of large rapidity gaps 
for a jet produced in the interval of rapidity
10-12.5, as a function of energy.

\begin{figure}[htbp]
\begin{tabular}{ c c}
\epsfig{file=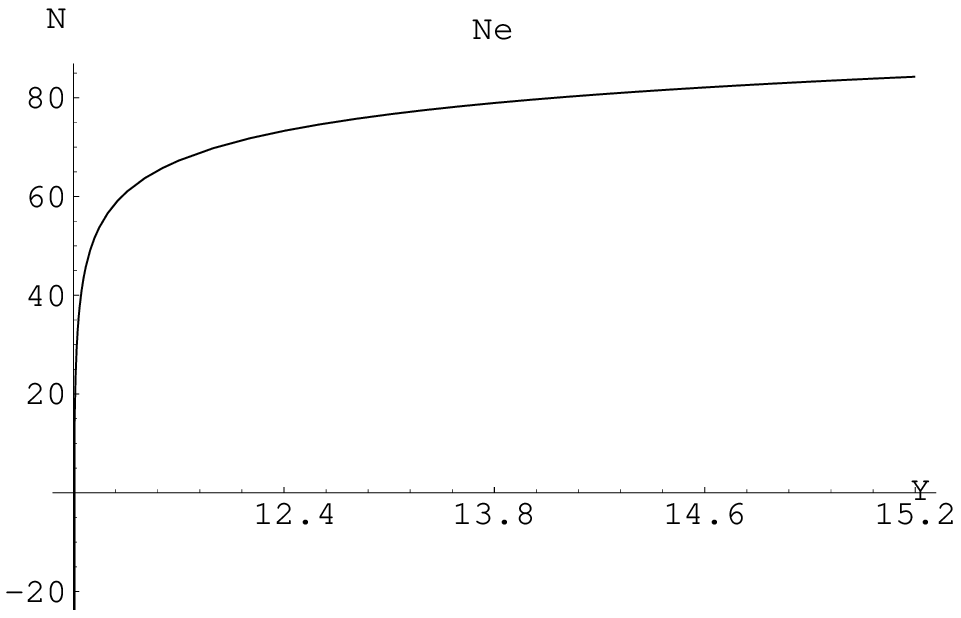,width=7cm} &
\epsfig{file=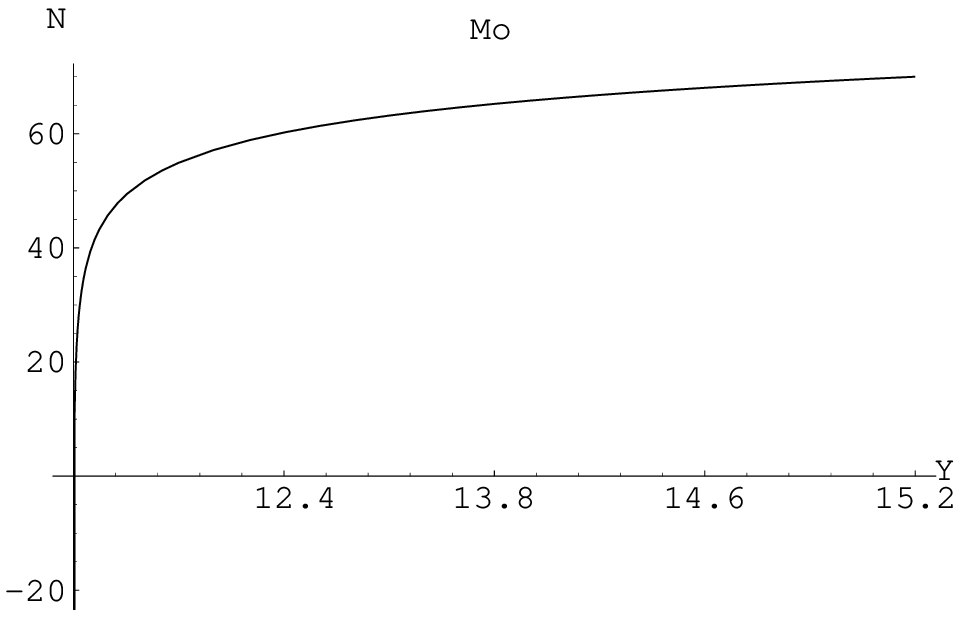,width=7cm} \\
Fig. 19-a &
Fig.19-b\\
\epsfig{file=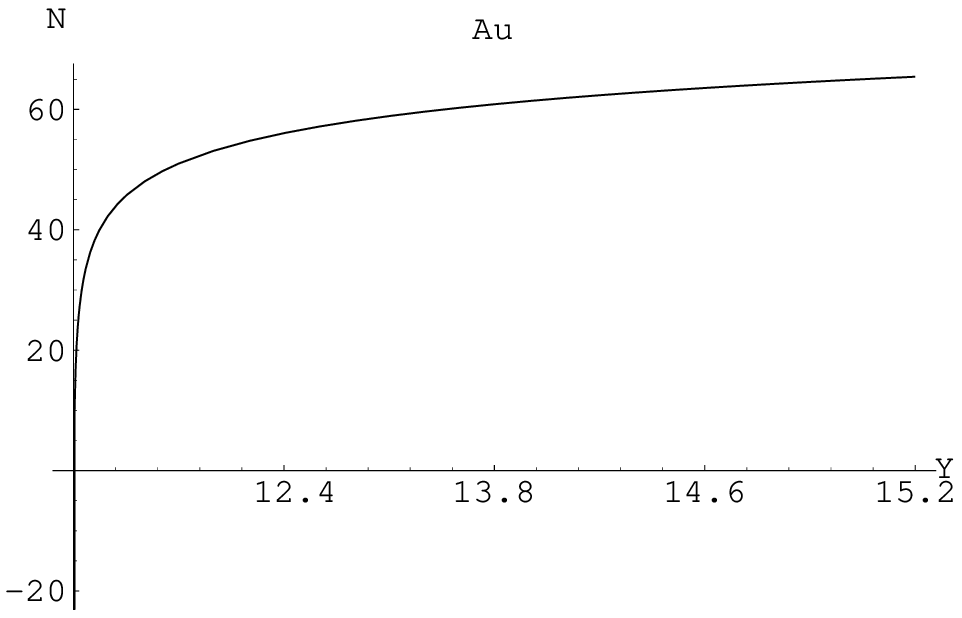,width=7cm} &
\epsfig{file=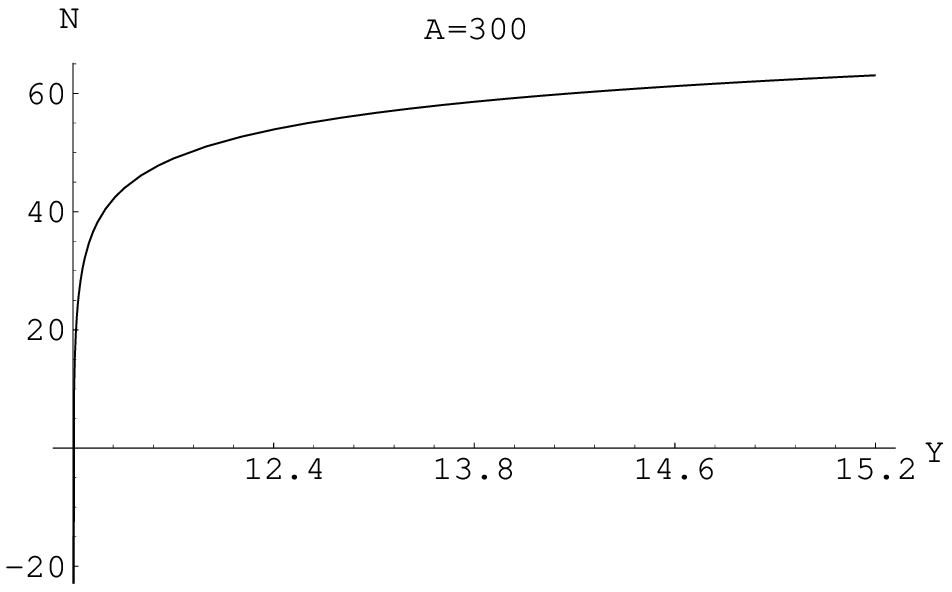,width=7cm}\\
Fig. 19-c &
Fig.19-d\\
\end{tabular}
\caption{\it Density of produced hadrons
for Ne ( Fig.19-a ), Mo ( Fig.19-b ), 
Au ( Fig.19-c ) and A=300( Fig.19-d ) 
as a function of rapidity, 
$Y$, from $0$ to $15.2$,
for proton - Nucleus scattering.}
\label{fig17}
\end{figure}

\begin{figure}[htbp]
\begin{tabular}{ c c}
\epsfig{file=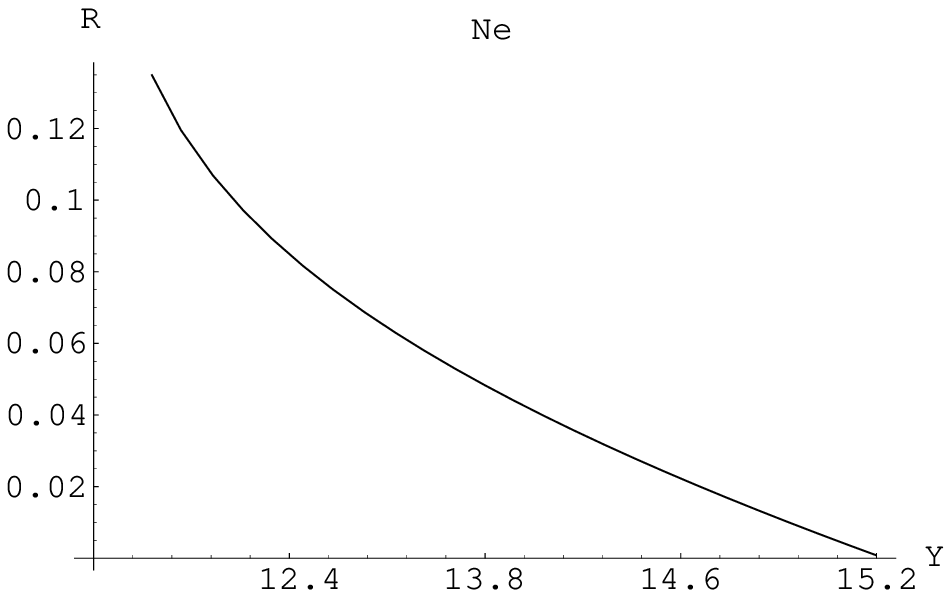,width=7cm} &
\epsfig{file=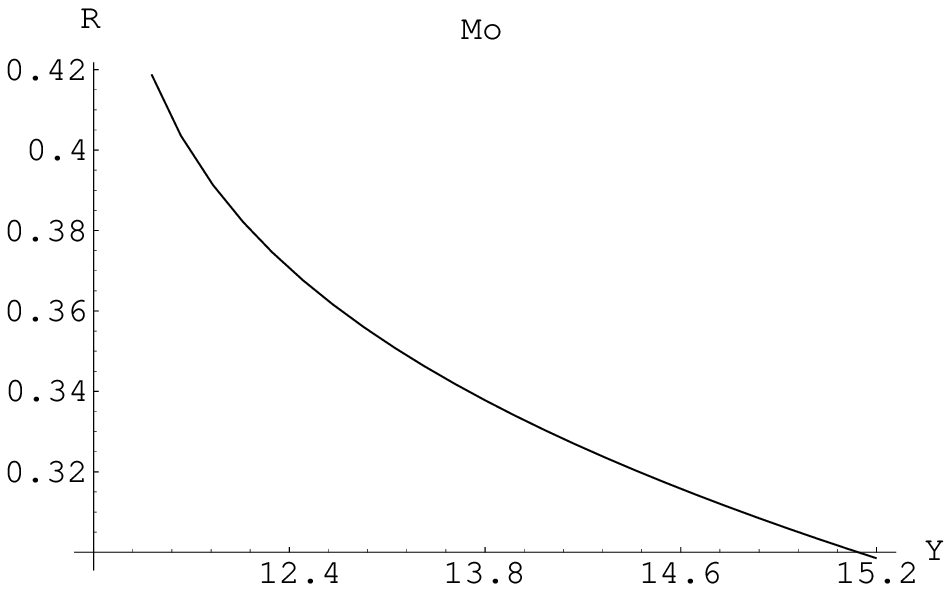,width=7cm} \\
Fig. 20-a &
Fig.20-b\\
\epsfig{file=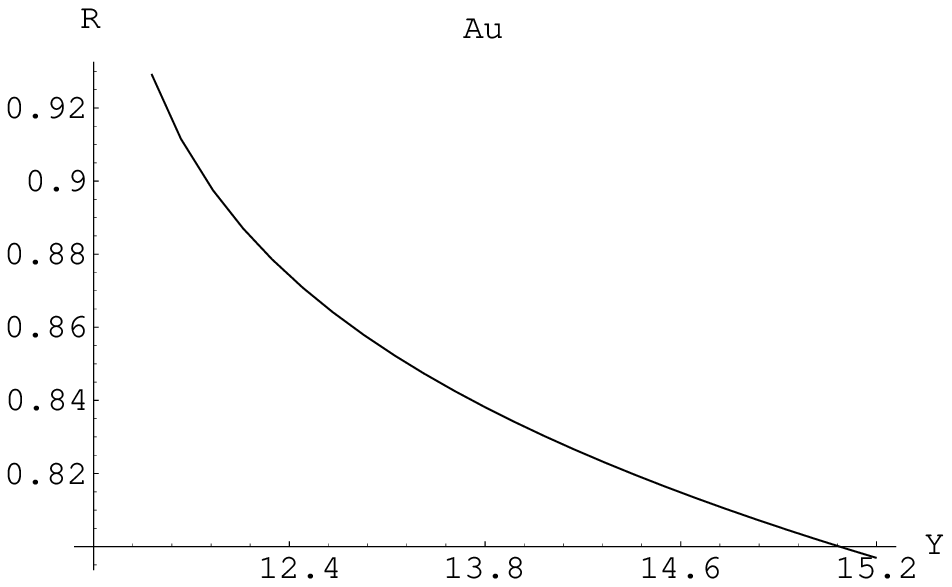,width=7cm} &
\epsfig{file=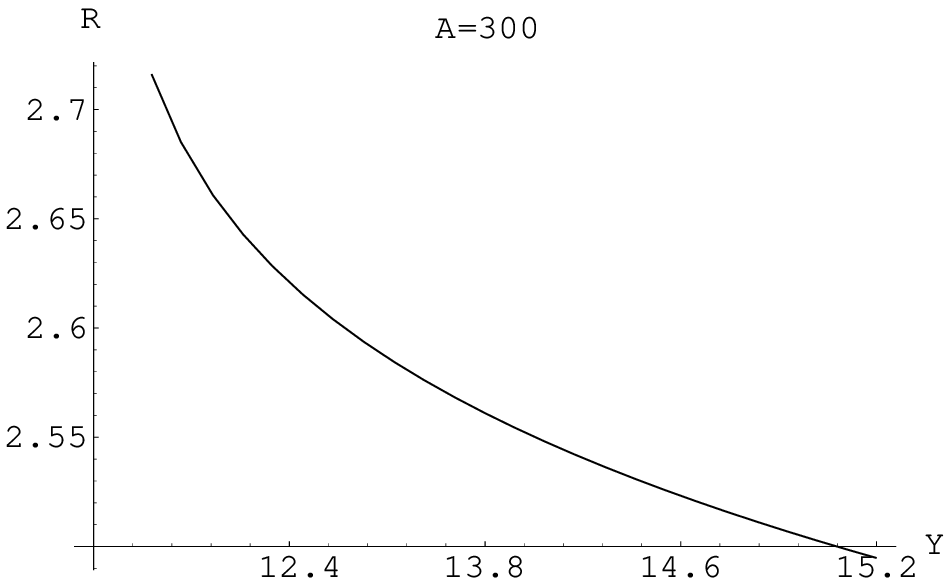,width=7cm}\\
Fig. 20-c &
Fig.20-d\\
\end{tabular}
\caption{\it Correlation function
for Ne ( Fig.20-a ), Mo ( Fig.20-b ), 
Au ( Fig.20-c ) and A=300( Fig.20-d )
for the
fixed value of rapidity $y_{2}=10$ as a function of $y_{1}$ from
$10$ to $15.2$,
for the fixed rapidity
$Y=\ln\Le S/S_{0}\Ra$, $S=2000$ $GeV$
and $S_{0}=1$ $GeV$  
for proton - Nucleus scattering.}
\label{fig18}
\end{figure}

Estimates for single
inclusive cross section, $N=\sigma_{incl}$/$\sigma_{total}$
are given in Fig.~\ref{fig17} as a function of rapidity, 
$y$=$0$-$15.2$.
The value of
the correlation function in Fig.~\ref{fig18}, is calculated for the
fixed value of rapidity $y_{2}=10$ as a function of $y_{1}$ from
$10$ to $15.2$
for the fixed energy
$\sqrt{s}=2000$ $GeV$.

\section{Nucleus\,\,-\,\, Nucleus interaction}
\setcounter{equation}{0}
\subsection{Classification and selection rules for Reggeon diagrams}
 
\begin{figure}[htbp]
\begin{center}
\epsfig{file=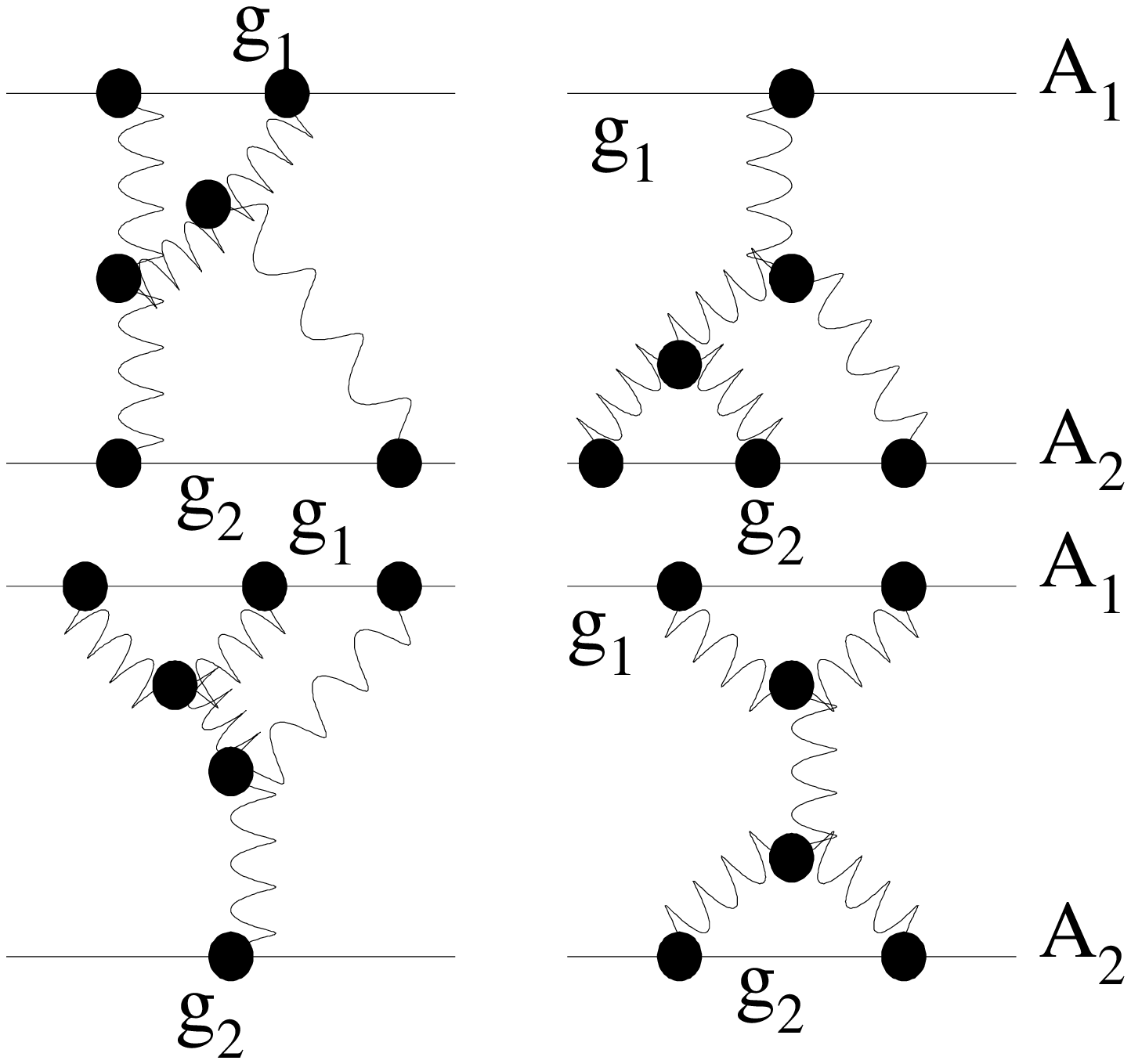,width=10cm}
\end{center}
\caption{\it The Pomeron diagrams for nucleus-nucleus interaction  of
order $G^2_{3p}$.}
\label{fig19}
\end{figure}

Describing the nucleus-nucleus interaction is much more difficult than 
the description of
hadron-nucleus scattering.  In Fig.~\ref{fig19} one can see four
topological classes of Reggeon diagrams, which are
second order 
in $G_{3p}$, the triple Pomeron coupling. In general, they cannot be reduced 
to "fan" diagrams, as can be seen for the first diagram, which we shall call
a "net" diagram.

\begin{figure}[htbp]
\begin{center}
\epsfig{file=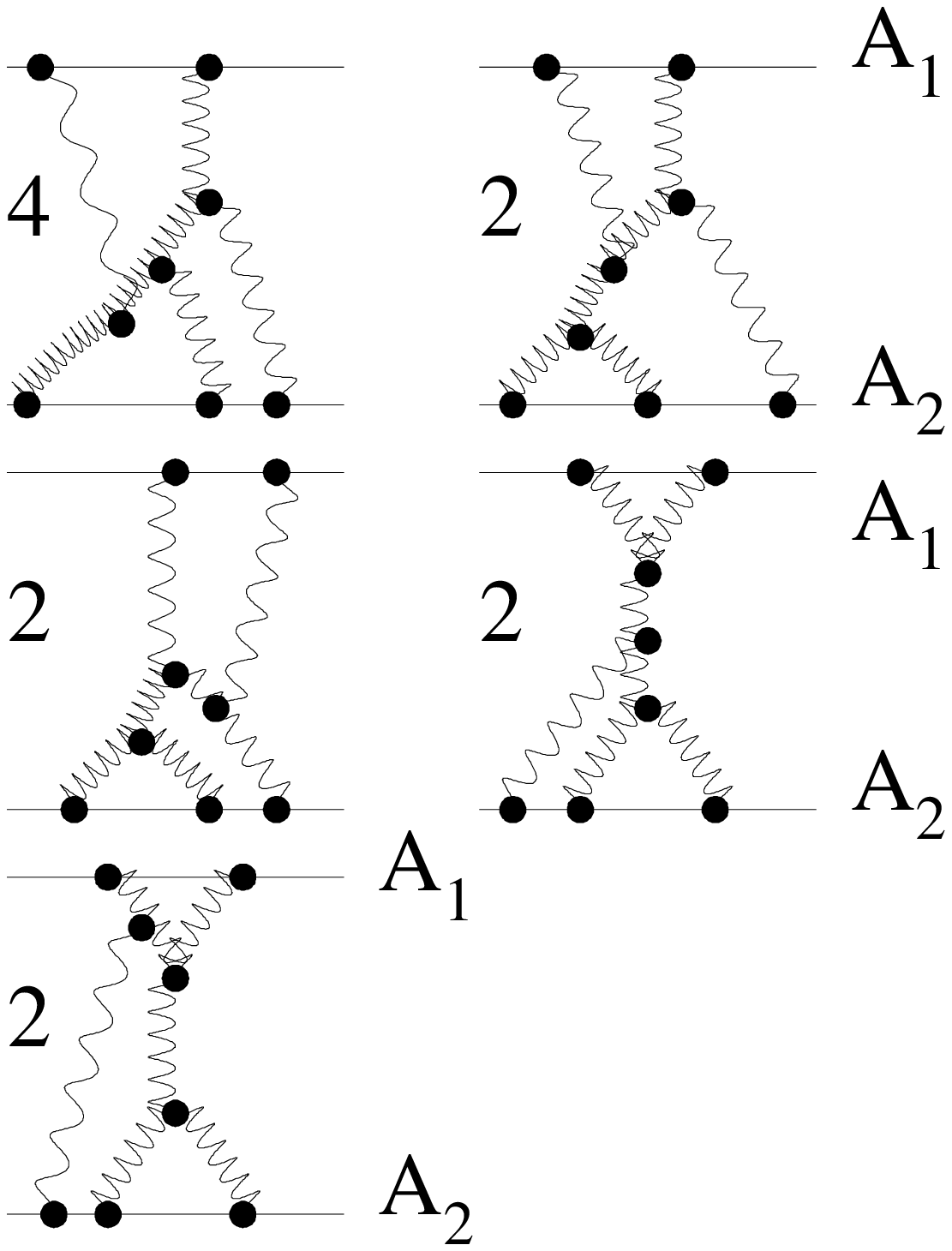,width=10cm}
\end{center}
\caption{\it The ``net"  Pomeron diagrams for nucleus-nucleus intereaction
of order
$G^3_{3p}$.}
\label{fig20}
\end{figure}

Using our main parameter $\kappa_{A} $, 
the rules of our selection are  well defined as
we have discussed in section 2.1. 
Complications arise as we have two
parameters:$\kappa_{A_1} $ and $\kappa_{A_2} $.  The general type of 
diagram that we select is the diagram which gives a contribution of the
order of $g_{P-A_1}g_{P-A_2} \cdot e^{\Delta Y} \cdot \kappa^n_{A_1}
\cdot \kappa^m_{A_2} $. This means that we do not consider only   
diagrams
with Pomeron loops. The whole structure of the diagrams looks rather
complicated and, we hope, that summing all of them  will yield a new
example of high energy behaviour which is a natural generalization of the
Glauber approach. Unfortunately, our new approach
is quite different and more difficult than a
simple summation of "fan" diagrams. As an instructive example, of how rich
the system of the diagrams is, we show in Fig.~\ref{fig20} the "net" diagrams
of order $G^3_{3p}$.

\begin{figure}[htbp]
\begin{center}
\epsfig{file=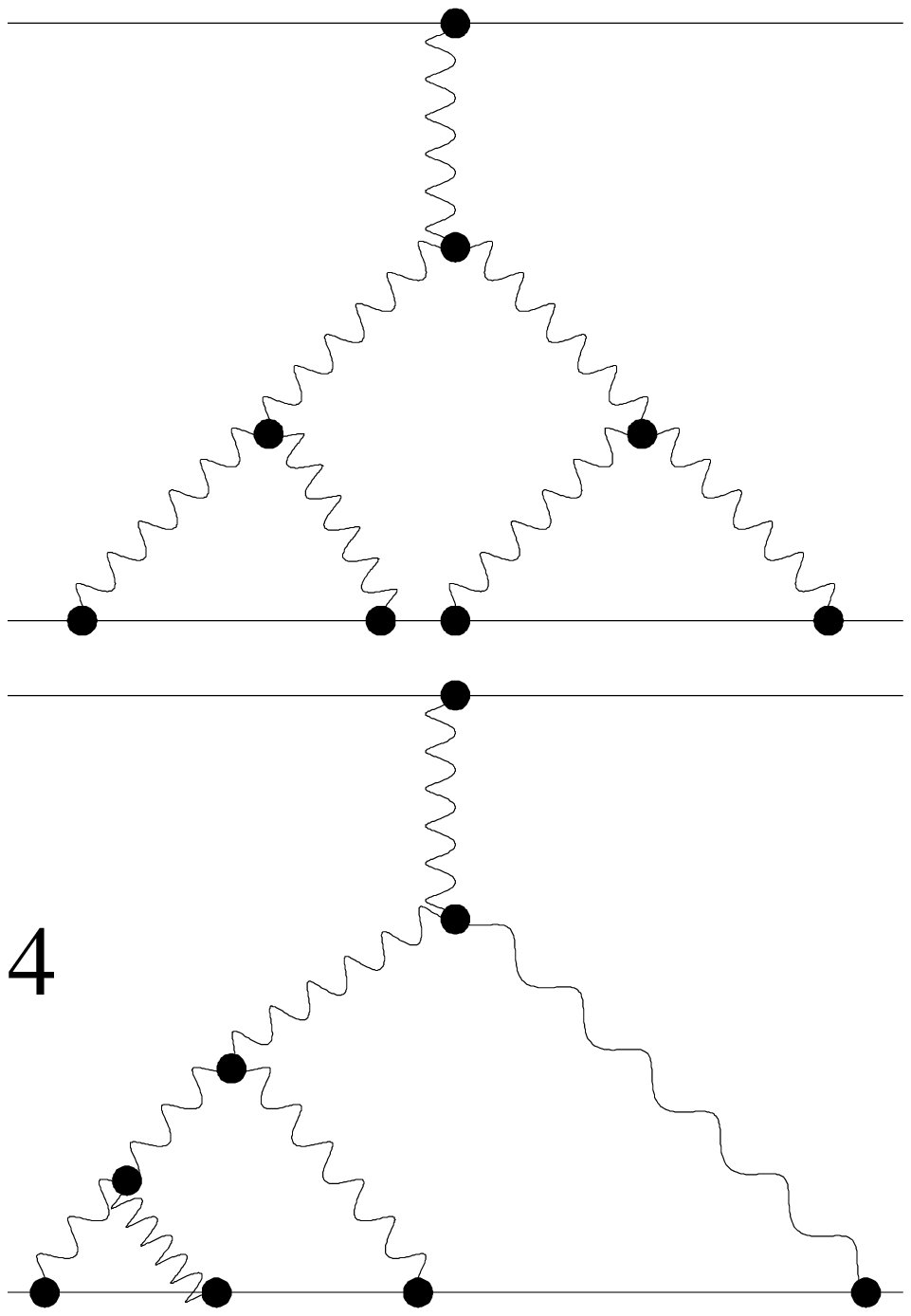,width=10cm}
\end{center}
\caption{\it The ``fan"  Pomeron diagrams for nucleus-nucleus intereaction
of order
$G^3_{3p}$.}
\label{fig21}
\end{figure}

\subsection{ An instructive example: the $\mathbf{G^2_{3P}}$ - order
Pomeron diagrams}

In order to demonstrate both the technique of calculation and the
classification of the diagrams, using our general approach, we consider the
"net" diagrams of order $G^3_{3P}$, shown in Fig.~\ref{fig20}.
In Fig.~\ref{fig20} we have \footnote{To abreviate we use the notation $g_1
\equiv
g_{P-A_1}$ and $g_2
\equiv g_{P-A_2}$.}:

\begin{enumerate}

\item Four  diagrams
with the same structure as the first diagram
and their total contribution to the amplitude is
\beq \label{AS1}
4g_{1}^{2}\cdot g_{2}^{3}\cdot e^{\Delta y}\cdot\gamma^{3}\cdot
\Le \frac{e^{3\Delta y}-2e^{2\Delta y}+e^{\Delta
y}}{2}-\Delta^{2}e^{\Delta y}
\int_{0}^{y} e^{\Delta y'}y' dy'\Ra\,\,.
\eeq
\item Two  diagrams
with the same structure as the second diagram,
their total contribution to the amplitude is
\beq \label{AS2}
2g_{1}^{2}\cdot g_{2}^{3}\cdot e^{\Delta y}\cdot\gamma^{3}\cdot
\Le e^{\Delta y}-e^{2\Delta y}+\Delta^{2}e^{\Delta y}
\int_{0}^{y} e^{\Delta y'}y' dy'+e^{\Delta y}\Delta y\Ra\,\,.
\eeq

\item Two  diagrams
with the same structure as the third diagram
and their total contribution to the amplitude is
\beq  \label{AS3}
2g_{1}^{2}\cdot g_{2}^{3}\cdot e^{\Delta y}\cdot\gamma^{3}\cdot
\Le \frac{e^{3\Delta y}-4e^{2\Delta y}+3e^{\Delta y}}{2}+
e^{\Delta y}\Delta y\Ra \,\,.
\eeq

\item Two  diagrams
with the same structure as the fourth diagram,
their total contribution to the amplitude is
\beq \label{AS4}
2g_{1}^{2}\cdot g_{2}^{3}\cdot e^{\Delta y}\cdot\gamma^{3}\cdot
\Le e^{\Delta y}-e^{2\Delta y}+\Delta^{2}e^{\Delta y}
\int_{0}^{y} e^{\Delta y'}y' dy'+e^{\Delta y}\Delta y\Ra\,\,.
\eeq

\item Two  diagrams
with the same structure as the fifth diagram,
their total contribution to the amplitude is
\beq \label{AS5}
2g_{1}^{2}\cdot g_{2}^{3}\cdot e^{\Delta y}\cdot\gamma^{3}\cdot
\Le\frac{e^{2\Delta y}-1}{2}-e^{\Delta y}\Delta y\Ra\,\,.
\eeq

\end{enumerate}

\begin{figure}[htbp]
\begin{center}
\epsfig{file=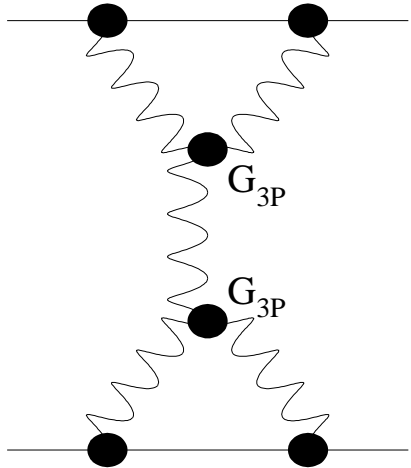,width=10cm}
\end{center}
\caption{\it The general structure of the "fan tree" diagrams.}
\label{fig22}
\end{figure}

Summing all contributions given by \eq{AS1} - \eq{AS5} we obtain
\beq \label{AS6}
-g_{1}^{2}\cdot g_{2}^{3}\cdot e^{\Delta y}\cdot\gamma^{3}\cdot
\Le 3e^{3\Delta y}-11e^{2\Delta y}+9e^{\Delta y}-1+
4\Delta e^{\Delta y} y\Ra\,\,,
\eeq
where the  minus sign in front reflects the odd number of Pomeron loops in
the
diagrams. \eq{AS6} looks rather complicated. The key observation is that
adding four digrams of "fan tree" type ( see the fourth diagram in Fig.20
 for the $G^3_{3P}$- order and Fig.~\ref{fig22} for the general structure of these
diagrams ) we obtain a very simple formula. Indeed, the four diagrams of
the "fan
tree" type give the contribution:
\beq \label{AG1}
-g_{1}^{2}\cdot g_{2}^{3}\cdot e^{\Delta y}\cdot\gamma^{3}\cdot
\Le 2e^{2\Delta y}-2-4e^{\Delta y}\Delta y\Ra\,\,,
\eeq
which leads to sum
\beq \label{AG2}
\sum \left( \,Fig.20-1 \,\,-\,\,Fig.20-5 \,\right)\,\,+\,\,4 \cdot
Fig.20-4\,\,=\,\,3\cdot g_{1}^{2}\cdot g_{2}^{3}\cdot e^{\Delta
y}\cdot\gamma^{3}\cdot
\Le e^{\Delta y}-1\Ra^{3}\,\,.
\eeq
We, therefore, obtain the result in a very compact form. On the other hand
the "fan tree" diagrams have a very simple form and their contribution
can be calculated easily  just by using the simple formula ( see \eq{SFD} )
for the "fan" diagrams.

To obtain the complete contribution of order $G^2_{3P}$ we have
to add to \eq{AG2} the contributions of the "fan" diagrams of the type
given in
Fig.~\ref{fig21} and subtract the contribution of Fig.20-4 diagrams. The
contribution of the "fan" diagrams with "fan" looking down
consists of two terms:
\begin{enumerate}
\item
One diagram of Fig.21-1
which leads to
\beq \label{AG3}
\frac{1}{3}g_{1}\cdot g_{2}^{4}\cdot e^{\Delta y}\cdot\gamma^{3}\cdot
\Le e^{\Delta y}-1\Ra^{3}\,\,.
\eeq

\item
Four diagrams of Fig.21-2, which give
\beq \label{AG4}
\frac{4}{3!}g_{1}\cdot g_{2}^{4}\cdot e^{\Delta y}\cdot\gamma^{3}\cdot
\Le e^{\Delta y}-1\Ra^{3}\,\,.
\eeq

\end{enumerate}

The total contribution of these diagrams is
\beq \label{AG5}
-g_{1}\cdot g_{2}^{4}\cdot e^{\Delta y}\cdot\gamma^{3}\cdot
\Le e^{\Delta y}-1\Ra^{3}\,\,.
\eeq

Summing \eq{AG2} and \eq{AG5} we obtains a beautiful result
\footnote{ We add the ''fan" diagrams with the ``fan" looking down to get
\eq{AG6}.}:
\begin{eqnarray}
All\,\,\,\,contributions\,\,\,of\,\,\,\,G^3_{3P}\,\,-\,\,order\,\,+\,\,4
\cdot Fig.\,\,20-4\,\,\,&=&  \label{AG6} \\
 &-& g_{1}\cdot g_{2}\cdot e^{\Delta y}\cdot\gamma^{3}\cdot
\Le e^{\Delta y}-1\Ra^{3}\Le g_{1}+g_{2}\Ra^{3}\,\,.\nonumber
\end{eqnarray}
 From \eq{AG6} one can see that we can easily obtain the answer if we
calculate the "fan tree" diagrams of Fig.~\ref{fig22} separately.

\subsection{Sum of the Pomeron diagrams:  general approach}

Based on the above example we have formulated a general
algorithm: to the order of $G^n_{3P}$ the answer is
\begin{eqnarray}
(- 1 )^n\,\,g_{1}\cdot g_{2}\cdot e^{\Delta y}\cdot\gamma^{3}\cdot
\Le e^{\Delta y}-1\Ra^{3}\Le g_{1}+g_{2}\Ra^{n}\,\,&- &
\label{AGA1}\\
  &- &\sum^{n}_{k=1}
\,\,C_k \Le``fan\,\,\,tree"\,\,diagrams; G^n_{3P}, Y \Ra \,\,g^k_1\,g^{n -
k}_{2}\,\,,\nonumber
\end{eqnarray}
where the contributions of the "fan tree" diagrams have to be evaluated.
First, we discuss why we get  such a  simple first term in \eq{AGA1}.
In the simplest example of the diagrams in Fig.~\ref{fig19}
one can see that if we
integrate the  "fan tree" diagram of Fig.~\ref{fig19}-5
over the position of
the upper $G_{3P}$ ( $y_1$ ) we will get the contribution of the  "net"
diagram of Fig.~\ref{fig19}-4, for the region of integration $y_1 < y_2$.
However, this
construction yields only one of the diagrams
in Fig.~\ref{fig19}-4.
We can put it
differently, that by integrating the diagrams in
Fig.~\ref{fig19}-4 over $y_1$ and $y_2$
without any restriction, we obtain a factor $ \Le \frac{1}{\Delta} \{
\,e^{\Delta\,\,Y}\,\,-\,\,1\,\}\,\Ra^2$. i.e., we include two diagrams
of Fig.~\ref{fig19}-5 instead of one. Therefore, to obtain the correct answer we
have to subtract one diagram of 
Fig.~\ref{fig19}-4.

Before formulating the rules as to what number of "fan tree" diagrams we need
to subtract, let us consider in more detail the contribution of these
diagrams (see Fig.~\ref{fig22} ). Replacing each line in the 
upper and low "fan" in
Fig.~\ref{fig22} by the full "fan" diagram function of \eq{SFD}, we obtain the
explicit expression for the sum of these diagrams:
\begin{eqnarray}
\Phi\Le y, b, b'\Ra& =&G_{3P}^{2}\int_{0}^{y}
\frac{g_{1}^{2}\cdot e^{2\Delta\Le y-y_{1}\Ra}\cdot
e^{\Delta\Le y_{1}-y_{2}\Ra}}{\Le g_{1}\gamma\Le e^{\Delta\Le y-y_{1}\Ra}
-1\Ra +1\Ra^{2}}dy_{1}\cdot
\int_{0}^{y_{1}}
\frac{g_{2}^{2}\cdot e^{2\Delta y_{2}}}
{\Le g_{2}\gamma\Le e^{\Delta y_{2}}+1\Ra +1\Ra^{2}}dy_{2} \nonumber\\
 &=&
g_{1}\cdot g_{2}\sum_{n=1, m=1} A_{n+1,m+1}\Le G_{3P}, y\Ra
\cdot g_{1}^{n}\cdot g_{2}^{m} ,
\label{AGA2}
\end{eqnarray}
where
\beq \label{AGA3}
g_{1} =\frac{g_{0,1}}{\pi R_{1}^{2}}
exp\Le
\frac{-b^{2}}{R_{1}^{2}}\Ra\,\,\,;\,\,\,\,\,\,g_{2}=\frac{g_{0,2}}{\pi
R_{2}^{2}}
exp\Le \frac{-b'^{2}}{R_{2}^{2}}\Ra\,\,.
\eeq
Integrating over  $b$ and $b'$
we have
\begin{eqnarray}
\Phi\Le y\Ra & =& G_{3P}^{2}\cdot R_{1}^{2}\cdot
e^{\Delta y}\cdot \pi^{2}\cdot
 R_{2}^{2}\cdot\int_{0}^{y} e^{\Delta\Le y-y_{1}\Ra} dy_{1}\cdot
\int_{0}^{y_{1}} e^{\Delta y_{2}}\cdot dy_{2}\cdot \nonumber \\
 & & \cdot\int_{0}^{\hat{g_{1}}}\frac{x\cdot dx}
{\Le x\gamma\Le e^{\Delta\Le y-y_{1}\Ra}
-1\Ra +1\Ra^{2}}\cdot
\int_{0}^{\hat{g_{2}}}\frac{y\cdot dy}
{\Le y\gamma\Le e^{\Delta y_{2}}
-1\Ra +1\Ra^{2}}\,\,.
\label{AGA4}
\end{eqnarray}
The final expression for this diagram is
\begin{eqnarray}
\Phi\Le y\Ra &=& G_{3P}^{2}\cdot R_{1}^{2}\cdot
e^{\Delta y}\cdot \pi^{2}\cdot
R_{2}^{2}\cdot\int_{0}^{y} e^{\Delta\Le y-y_{1}\Ra} dy_{1}\cdot
\int_{0}^{y_{1}} e^{\Delta y_{2}}\cdot dy_{2}\cdot \nonumber \\
& & \cdot\Le -\frac{\hat{g_{1}}\gamma\Le e^{\Delta\Le y-y_{1}\Ra}
-1\Ra}{\hat{g_{1}}\gamma\Le e^{\Delta\Le y-y_{1}\Ra}
-1\Ra +1}+Ln\left[ \hat{g_{1}}\gamma\Le e^{\Delta\Le y-y_{1}\Ra}
-1\Ra +1\right]\Ra\cdot \nonumber \\
& & \cdot\Le -\frac{\hat{g_{2}}\gamma\Le e^{\Delta y_{2}}
-1\Ra}{\hat{g_{2}}\gamma\Le e^{\Delta y_{2}}
-1\Ra +1}+Ln\left[ \hat{g_{2}}\gamma\Le e^{\Delta y_{2}}
-1\Ra +1\right]\Ra \,\,,
\label{AGA5}
\end{eqnarray}
where
$$
\hat{g_{i}}=\frac{g_{0,i}}{\pi R_{2}^{i}}\,\,.
$$
There are several important properties of the "fan tree"  diagrams, which
we will use below:
\begin{enumerate}

\item These diagrams, in contrast to the "net" diagrams, yield a
term, which is
proportional to $e^{\Delta y}$. The coefficient
in front of this term is $1$ to all orders of $G_{3P}$.

\item The first term in the expansion of these diagrams
is proportional to the $G_{3P}^{2}\cdot g_{1}^{2}\cdot g_{2}^{2}$.
"Fan" diagrams also have a term  $e^{\Delta y}$, but 
only have one of the vertices  $g_{1}$ or $g_{2}$  
to the first power, for all orders of $G_{3P}$.

\item As has been stated above, 
to obtain a compact expression for
the sum of the diagrams in the nucleus-nucleus 
amplitude at high energy
to all orders of $G_{3P}$ 
we need to add a definite number of 
"fan tree" diagrams of the same order.
\end{enumerate}

We now begin our calculation of the amplitude
using these properties of the
"fan tree" diagrams.

First of all, we introduce a function, which contains all
the  "fan" diagrams, all  the "net" diagrams  and part of the "fan tree"
diagrams, namely, those contained in the first term in \eq{AGA1}:
\beq \label{AGA6}
F_{1}\Le y, b, b'\Ra= \frac{g_{1}\cdot g_{2}\cdot e^{\Delta y}}
{\Le g_{1}+g_{2}\Ra \gamma\Le e^{\Delta y}
-1\Ra +1}\,\,.
\eeq
the term of  order $\gamma^3$ in
\eq{AGA6}  is \eq{AG6}. To find the second term in \eq{AGA1}
we expand \eq{AGA6} with respect to the powers of $\gamma$:
\begin{enumerate}
\item To first order in $\gamma$ we have:
\beq \label{AGAE1}
-g_{1}\cdot g_{2}\cdot e^{\Delta y}\cdot\gamma\cdot
\Le g_{1}+g_{2}\Ra\Le e^{\Delta y}
-1\Ra\,\,.
\eeq
This is the first contribution from the "fan" diagrams,
Fig.~\ref{fig19}.
\item Second order in $\gamma$ has the form
\beq \label{AGAE2}
2\cdot g_{1}\cdot g_{2}\cdot e^{\Delta y}\cdot\gamma^{2}\cdot
\Le g_{1}+g_{2}\Ra^{2}\Le e^{\Delta y}
-1\Ra ^{2}\,\,.
\eeq
 One can see that this term which is proportional to  $2\gamma^{2}
g_{1}^{2}\cdot g_{2}^{2}\cdot e^{\Delta y}$ in \eq{AGAE2}
has the same structure as a term of the same order
in  the "fan tree" diagram in the total amplitude.
The difference is only in the value of the numerical
coefficient: in the amplitude this term has  coefficient
$1$ instead of the $2$ in \eq{AGAE2}. Therefore,
using the properties of the "tree" diagrams we can conclude, that
to obtain the exact coefficient $1$
we must subtract from this term the first term of the expansion
\eq{AGA2}, namely, $g_{1}^{2}\cdot g_{2}^{2}\cdot
A_{2,2}\Le G_{3P}, y\Ra$, which also has a coefficient $1$.

\item The third order in $\gamma$ is
\beq \label{AGAE3}
-g_{1}\cdot g_{2}\cdot e^{\Delta y}\cdot\gamma^{3}\cdot
\Le e^{\Delta y}-1\Ra^{3}\Le g_{1}+g_{2}\Ra^{3}.
\eeq
In order to obtain a coefficient $1$, we
subtract from this term the second term of the expansion
\eq{AGA1}  multiplied by 2,
\beq \label{AGAE4}
2\cdot\Le g_{1}^{2}\cdot g_{2}^{3}\cdot
A_{2,3}+
g_{1}^{3}\cdot g_{2}^{2}\cdot
A_{3,2}\Ra\,\,.
\eeq
This gives us four diagrams of the required order.

\end{enumerate}

Conforming this expansion we find that the number of ``fan
tree" diagrams of order $g_{1}^{n}\cdot g_{2}^{m}\cdot \gamma^{n+m-2}$
 which we have to subtract is equal to
\beq  \label{AGA7}
\frac{\Le n+m-2\Ra !}{\Le n-1\Ra !\Le m-1\Ra !}-1
\,\,.
\eeq
This means that the function  to be subtracted is equal to
\beq \label{AGA8}
F_{2}\Le y, b, b'\Ra=
g_{1}\cdot g_{2}\sum_{n=1, m=1}\Le
\frac{\Le n+m\Ra !}{n! m!}
-1\Ra\cdot A_{n+1,m+1}\cdot
g_{1}^{n}\cdot g_{2}^{m}\,\,.
\eeq
Using \eq{AGA2},  we obtain
\begin{eqnarray}
F_{2}\Le y, b, b'\Ra  &=& 2\cdot G_{3P}^{2}\cdot g_{1}^{2}\cdot
g_{2}^{2}\cdot
e^{\Delta y}\cdot
\int_{0}^{y} e^{\Delta\Le y-y_{1}\Ra} dy_{1}\cdot
\int_{0}^{y_{1}} e^{\Delta y_{2}}\cdot dy_{2}\cdot \nonumber \\
 & & \frac{1}{\Le  g_{1}\gamma\Le e^{\Delta\Le y-y_{1}\Ra}-1\Ra
+g_{2}\gamma\Le e^{\Delta y_{2}}-1\Ra +1\Ra^{3}}-
\Phi\Le y, b, b'\Ra \,\,. \label{AGA9}
\end{eqnarray}
Finally, the two particle irreducible set of Pomeron diagrams which gives
us the opacity $\Omega(s,b_t)$ ( see \eq{UB1} and \eq{UB2} ) is equal to
\beq \label{AGA10}
\Omega(s,b_t)\,\,=\,\,\int \,d^2 b \,\,\,F \Le y, b, | \vec{b}_t -
\vec{b}|
 \Ra \,\,,
\eeq
where
\begin{eqnarray}
F\Le y, b, b' \Ra\,&=&\,F_1\Le y, b, b' \Ra\,\,-\,\,F_2\Le y, b, b'
\Ra\,\,= \nonumber \\
 & =& \Phi_1\Le y, b, b' \Ra\,\,+\,\,\Phi_2\Le y, b, b'
\Ra\,\,-\,\,\Phi_3\Le y, b, b' \Ra \,\,,
\label{AGA11}\\
\Phi_1\Le y, b, b' \Ra &=& F_1\Le y, b, b' \Ra = \frac{g_{1}\cdot
g_{2}\cdot e^{\Delta y}}
{\Le g_{1}+g_{2}\Ra \gamma\Le e^{\Delta y}
-1\Ra +1}\,\,,
\label{AGA12}\\
\Phi_2\Le y, b, b' \Ra &=& \Phi \Le y, b, b' \Ra \,\,\label{AGA13} \\
&=& +G_{3P}^{2}\int_{0}^{y}
\frac{g_{1}^{2}\cdot e^{2\Delta\Le y-y_{1}\Ra}\cdot
e^{\Delta\Le y_{1}-y_{2}\Ra}}{\Le g_{1}
\gamma\Le e^{\Delta\Le y-y_{1}\Ra}
-1\Ra +1\Ra^{2}}dy_{1}\cdot
\int_{0}^{y_{1}}
\frac{g_{2}^{2}\cdot e^{2\Delta y_{2}}}
{\Le g_{2}\gamma\Le e^{\Delta y_{2}}+1\Ra +1\Ra^{2}}dy_{2} \,\,,
\nonumber\\
\Phi_3\Le y, b, b' \Ra &=& 2\cdot G_{3P}^{2}\cdot g_{1}^{2}\cdot
g_{2}^{2}\cdot
e^{\Delta y}\cdot
\int_{0}^{y} e^{\Delta\Le y-y_{1}\Ra} dy_{1}\cdot
\int_{0}^{y_{1}} e^{\Delta y_{2}}\cdot dy_{2}\cdot
\label{AGA14} \\
& & \frac{1}{\Le  g_{1}\gamma\Le e^{\Delta\Le y-y_{1}\Ra}-1\Ra
+g_{2}\gamma\Le e^{\Delta y_{2}}-1\Ra +1\Ra^{3}}\,\,.
\nonumber
\end{eqnarray}

\subsection{The Ultra high energy asymptote}

The asymptotic form at
ultra high energy of the opacity $\Omega $ originates in
\eq{AGA11} from the difference between the second and the third terms in this
equation. To our surprise this asymptotic behaviour is just the
exchange of the Pomeron but with an intercept which 
is two times smaller than the
intercept of the input Pomeron:
\beq \label{AU1}
F\Le y, b, b'\Ra_{asymp}=
\frac{\sqrt{2}}{2^{3}}\cdot\frac{\sqrt{g_1\,\,g_2}}{\gamma}\cdot
e^{\frac{\Delta}{2}\cdot y}\,\,.
\eeq

The opacity $\Omega(s,b_t)$ is equal to
\beq \label{AU2}
\Omega_{asymp} \Le s,b_t \Ra \,\,=\,\,\int \,d^2  b  F_{asymp}\Le y, b,
b'\Ra \,\,=\,\,\frac{\sqrt{2}}{2^{2}}\cdot
\frac{\,g_0}{\gamma}\cdot e^{\frac{\Delta}{2}\,y}\,\cdot\,
\frac{A^{\frac{5}{6}}_1\,\cdot\,A^{\frac{5}{6}}_2}{
A^{\frac{1}{3}}_1\,+\,A^{\frac{1}{3}}_2} \,\cdot\,\exp^{ -
\frac{b^2_t}{2(R^2_1 + R^2_2}}\,\,.
\eeq

Using the general formulae of \eq{O1} - \eq{O5} and \eq{AU2} we can 
calculate the asymptotic behaviour of the total and elastic cross sections
in the kinematic region where the opacity $\Omega$ is small ( $\Omega
\,\leq
1 $ ).  These calculations lead to
\begin{eqnarray}
\sigma_{tot}^{asymp} (y)\,\,&=&\,\,2\,\int \,d^2 b_t \,\Omega_{asymp}
(y,b_t) \,\, \longrightarrow \,\, A^{\frac{5}{6}}_1 \,\times\,
A^{\frac{5}{6}}_2\,\times\,e^{\frac{\Delta}{2}\,y}\,\,, 
\label{AU3}\\
\sigma_{el}^{asymp} (y)\,\,&=&\,\,\int \,d^2 b_t \,\Omega^2_{asymp}
(y,b_t) \,\, \longrightarrow \,\, \frac{ A^{\frac{5}{3}}_1
\,\times\A^{\frac{5}{3}}_2}{ A^{\frac{2}{3}}_1
\,+\,A^{\frac{2}{3}}_2}\,\times e^{\Delta\,y}\,\,.
\label{AU4}
\end{eqnarray}

We will discuss below how to calculate processes of  diffraction
dissociation in our approach, but for the sake of completeness we present
here the result of the asymptotic behaviour of the diffractive cross
section at
ultra high energy. It turns out that this cross section is
\beq \label{AU5}
\sigma^{SD}_{asymp}\,\,\,\propto\,\,\,\frac{\Delta\,\,g_0}{\gamma}
\,\times\,\Le A_1\,\,+\,\,A_2 \Ra\,\times\,e^{\frac{\Delta}{2}\,y}\,\,.
\eeq

It should be stressed that \eq{AU5} is quite different from what we
expect if the asymptote is a real Pomeron exchange with a
different intercept. Let us recall that in the single Pomeron exchange
model
$$
\sigma^{SD}_{asymp}\,\,\,\propto\,\,\,\sigma^{asymp}_{el}
\,\,\propto\,\,e^{2\,\Delta_P\,y}\,\,.$$

\begin{figure}[htbp]
\begin{center}
\epsfig{file=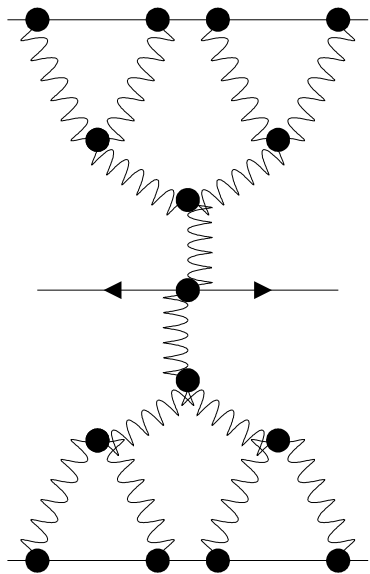,width=10cm}
\end{center}
\caption{\it The Mueller diagram for the inclusive cross section in our
approach for a nucleus-nucleus collision.}
\label{fig23}
\end{figure}

The formula for the inclusive cross section in the kinematic region of
sufficiently small $\Omega$ is also quite different from the single Pomeron
exchange (see Fig.(\ref{fig23}) ).  Indeed, from Fig.(\ref{fig23}) one can
see that the single inclusive cross section  results from the sum of "fan"
( or better to say "tree fan" ) diagrams. Therefore, the density of
produced
particles  in a unit of rapidity 
$$ \rho \equiv \frac{d
\sigma^{asymp}_{incl}/dy_c}{\sigma^{asymp}_{tot}}
$$ 
has an asymptotic limit
\beq \label{AU6}
\rho\,\,\,\propto\,\,\,A^{-\frac{1}{6}}_1 \cdot
A^{-\frac{1}{6}}_2 \cdot
e^{-\,\Delta_P\,y}\,\,\,\longrightarrow\,\,\,0\,\,.
\eeq
Therefore, we can conclude, that the Pomeron interaction with both nuclei
leads to a small number of particles produced in the central rapidity
region. This is the most striking difference of the resulting effective
Pomeron from the input Pomeron, which has a uniform distribution of
particles in rapidity.

It is also  interesting to note,
that in our amplitude the second term  has a stronger energy dependence,
in spite of the fact, that each diagram in this term has a weaker
energy dependence, than each diagram of the same order
in the first term. This means,
that the sum of the diagrams with the weak energy
dependence grows with energy faster, than the
sum of the diagrams with a stronger energy dependence.

\subsection{What energy is asymptotic?}

As has been mentioned, the three terms in \eq{AGA11} have different
asymptotic behaviour. At ultra high energy the last two terms dominate and they
provide the leading asymptotic behaviour 
that we have discussed in the previous section.
To estimate the value of energy that we can consider to be ultra high,
so that we can safely
apply our asymptotic formulae of \eq{AU1} - \eq{AU6},
we  calculate the
three terms of \eq{AGA11} separately, namely,
\begin{eqnarray}
\sigma_{1}&=&2\int d^{2}bd^{2}b'\cdot
\Phi_{1}\Le y, b, b'\Ra\,\,,
\label{WE1}\\
\sigma_{2}&=&2\int d^{2}bd^{2}b'\cdot
\Phi_{2}\Le y, b, b'\Ra\,\,,
\label{WE2}\\
\sigma_{3}&=&2\int d^{2}bd^{2}b'\cdot
\Phi_{3}\Le y, b, b'\Ra\,\,,
\label{WE3}
\end{eqnarray}
where the total cross section at small $\Omega$ is equal to
\beq \label{WE4}
\sigma_{tot}=\sigma_{1}+\sigma_{2}-\sigma_{3}\,\,.
\eeq

\begin{figure}[htbp]
\begin{center}
\begin{tabular}{c c}
\epsfig{file=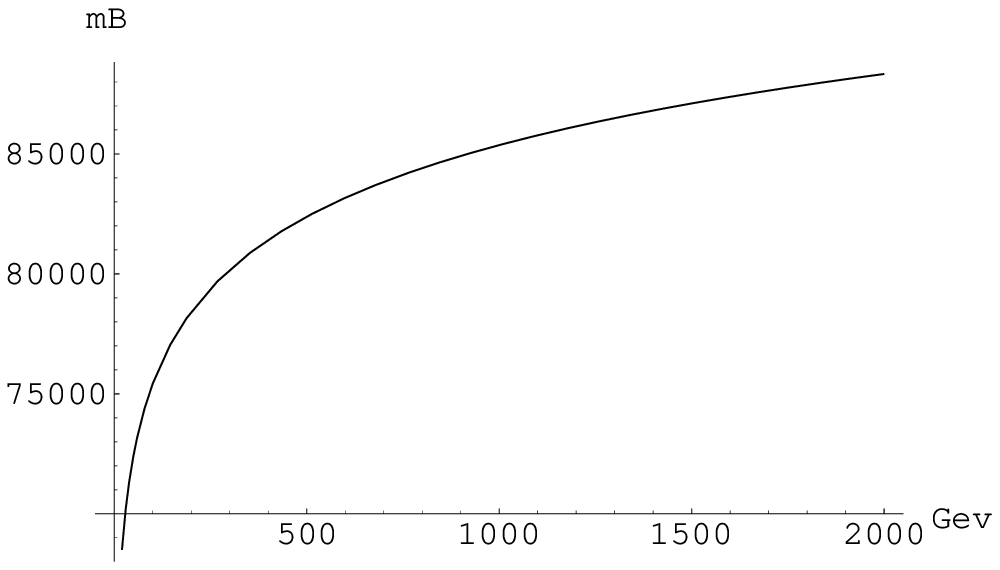,width=8cm} &
\epsfig{file= 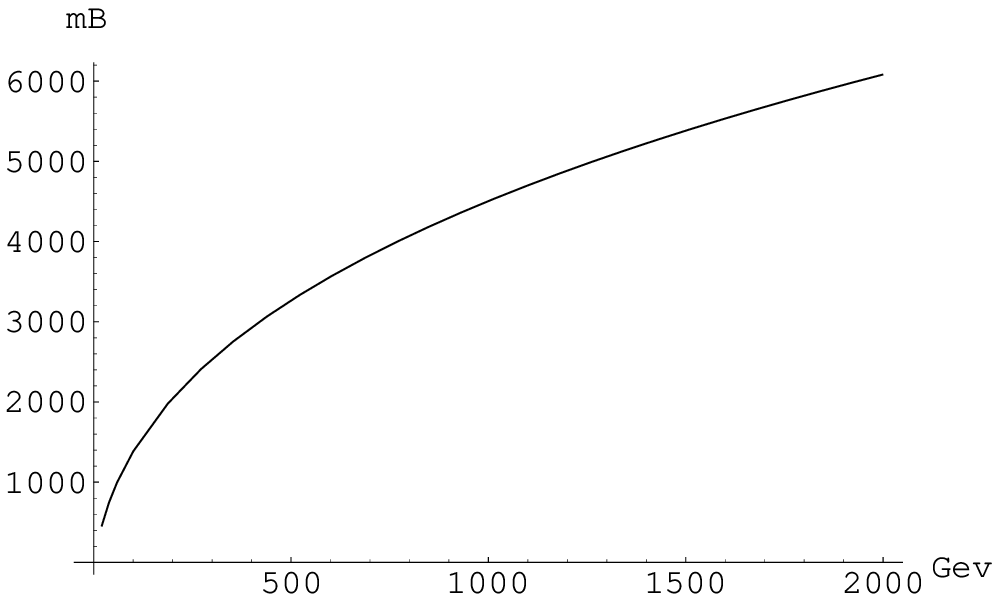,width=8cm}\\
Fig.26-a & Fig.26-b \\
\epsfig{file=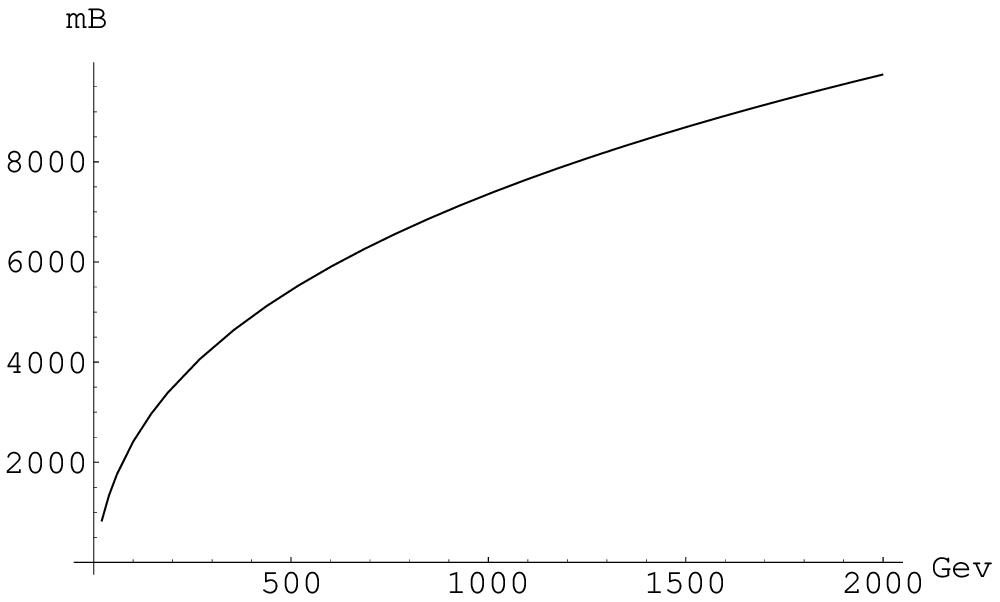,width=8cm} &
\epsfig{file=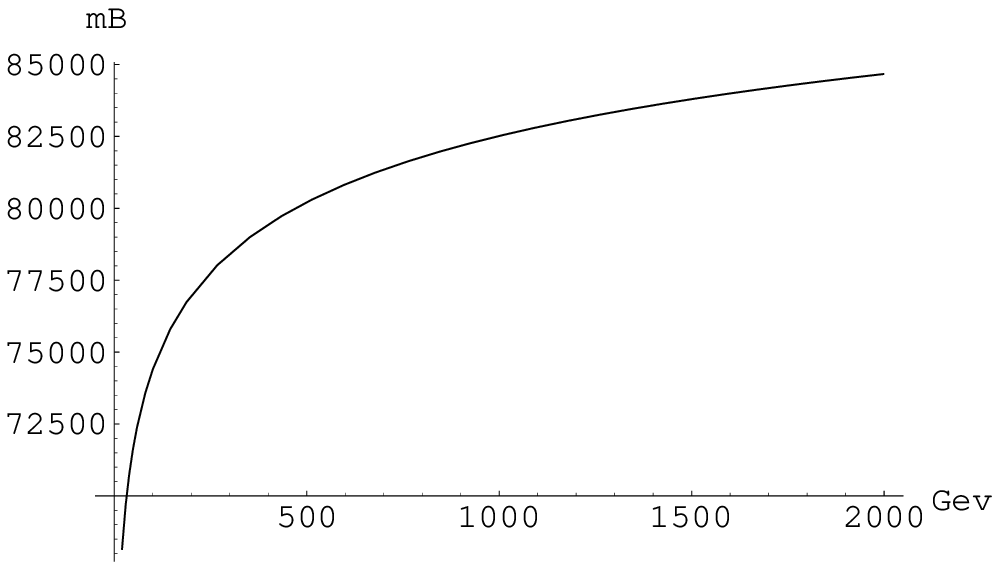,width=8cm}\\
Fig.26-c & Fig.26-d
\end{tabular}
\end{center}
\caption{\it The Energy dependence of the first ( Fig.26-a ),the  second
( Fig.26-b )
and  the third ( Fig.26-c ) terms of \protect\eq{WE4} as well as the value
of the total cross section ( Fig.26-d ).}
\label{fig24}
\end{figure}

The result of the calculation is given in Fig.(\ref{fig24}) where the
parameters of section 2.6 were used for the proton-proton interaction.
One can see from Fig.(\ref{fig24})  that in 
a range of energy, up to the
Tevatron energies, the first term is much larger than the other two.
Therefore, we can conclude:
\begin{itemize}
\item\,\,\,
That  ultra high energies, where our
asymptotic answer is valid, are  above of the Tevatron energies,
and we will only be able to see
the characteristic features of the real asymptote at the LHC.

\item\,\,\, We can take into account  only the first term for all
numerical estimates for the RHIC energies as well as for all energies
below
the Tevatron energy. This is the reason
for only taking the first term in the
following numerical estimates.
\end{itemize}

\subsection{Total and elastic cross sections}
\begin{figure}[htbp]
\begin{center}
\begin{tabular}{c c}
{\bf $\mathbf{\sigma_{tot}}$ for $\mathbf{A_{1}=20}$,$\mathbf{A_{2}=100}$}
 &{\bf $\mathbf{\sigma_{el}}$ for
$\mathbf{A_{1}=20}$,$\mathbf{A_{2}=100}$}\\
\epsfig{file=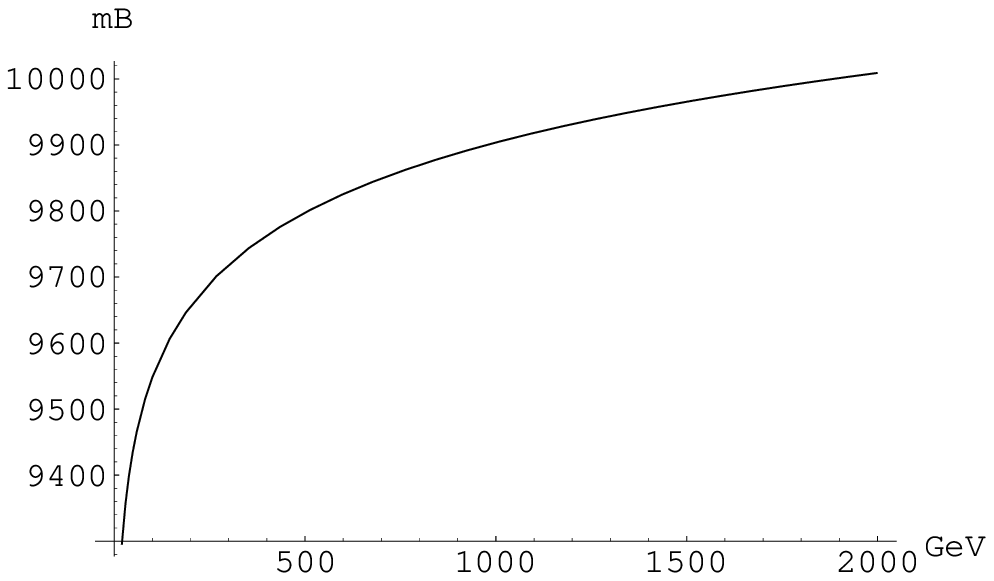,width=8cm} &
\epsfig{file=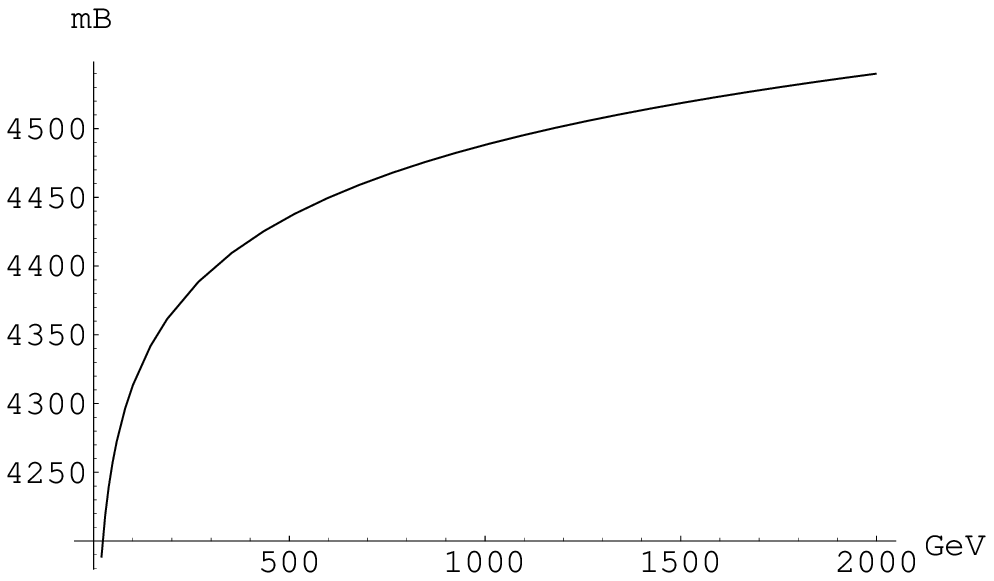,width=8cm}\\
{\bf $\mathbf{\sigma_{tot}}$ for
$\mathbf{A_{1}=200}$,$\mathbf{A_{2}=100}$}
&
{\bf $\mathbf{\sigma_{el}}$ for
$\mathbf{A_{1}=200}$,$\mathbf{A_{2}=100}$}\\
\epsfig{file=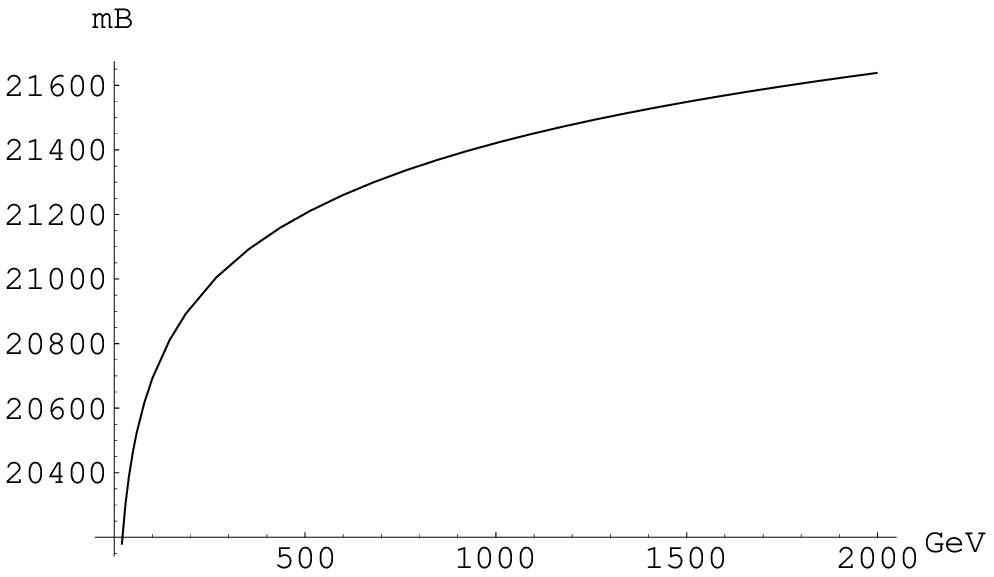,width=8cm}
&
\epsfig{file=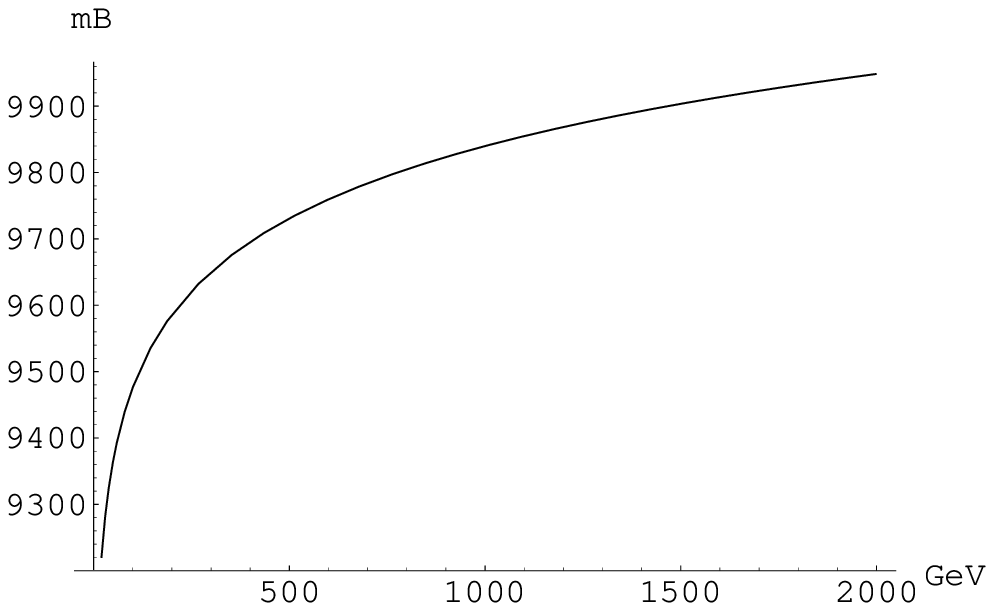,width=8cm}\\
{\bf $\mathbf{\sigma_{tot}}$ for
$\mathbf{A_{1}=200}$,$\mathbf{A_{2}=300}$}
&
{\bf $\mathbf{\sigma_{el}}$ for
$\mathbf{A_{1}=200}$,$\mathbf{A_{2}=300}$}\\
\epsfig{file=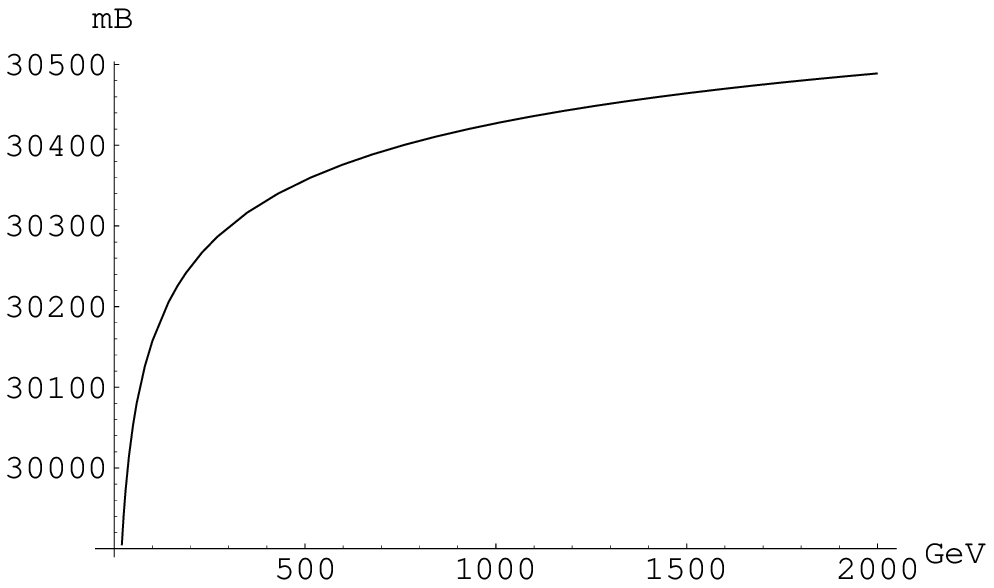,width=8cm} &
\epsfig{file=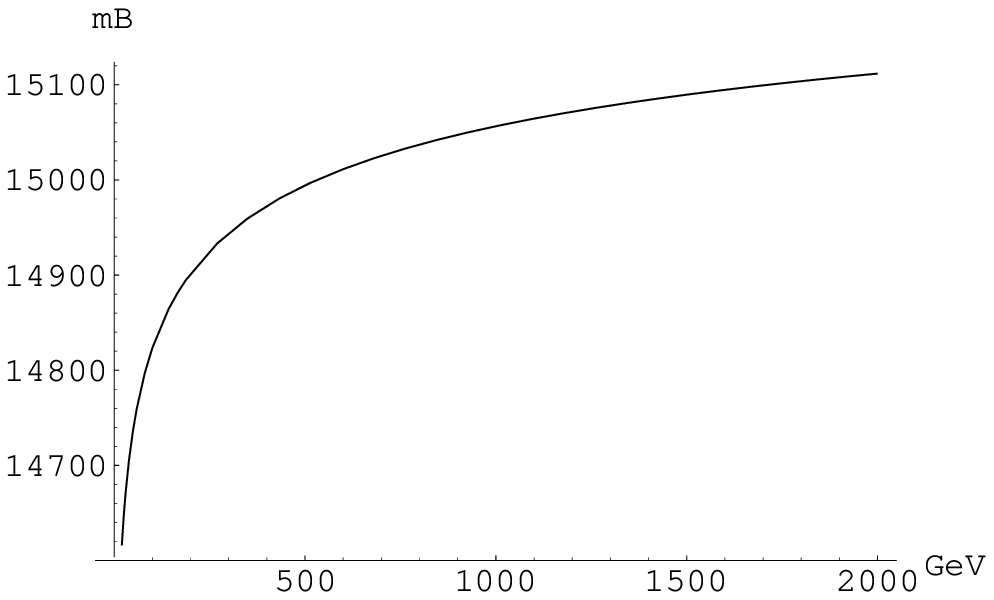,width=8cm}
\end{tabular}
\end{center}
\caption{\it Total and elastic $A_{1}-A_{2}$ cross sections versus energy.}
\label{fig25}
\end{figure}

To calculate total and elastic cross sections we have to use the general
formulae of \eq{O1} and \eq{O2} with the opacity $\Omega(s,b_t)$ defined
in \eq{AU2} through the function $F(y,b,b')$ given in \eq{AGA11}.
The results of these calculation are shown in Fig.(\ref{fig25}).

\subsection{Diffractive dissociation processes}

We start the discussion of the diffractive cross section for 
nucleus-nucleus interaction by rewriting the formula for the amplitude
in terms of the function $S(y,b_t)$ which describes the sum of
the  "fan" diagrams
( see \eq{SFD} ). We introduce two such functions: $S_{\uw}$ and
$S_{\dw}$ which sum the "fan" diagrams with up and down looking "fans",
respectively. Using these functions we rewrite our amplitude as
\beq \label{AD1}
F\Le y, b, b'\Ra=g_{1}\cdot g_{2}\cdot e^{\Delta y}\cdot
S_{\uw}\cdot S_{\dw}\cdot\sum_{n=0}^{n=\infty}\Le
1-S_{\dw}\Ra^{n} \Le 1-S_{\uw}\Ra^{n}\,\,.
\eeq
This is a very convenient representation of the amplitude, because,
we can use the AGK cutting rules \cite{AGK} and reduce the problem of
finding the diffractive cross section to the problem of the diffractive
cross section for hadron-nucleus interaction that has been solved in
section 2.3.
To see this, we rewrite the amplitude in the form:

\beq \label{AD2}
F\Le y, b, b'\Ra=g_{1}\cdot g_{2}\cdot e^{\Delta y}\cdot
S_{\uw}\cdot\sum_{n=0}^{n=\infty}\Le
1-S_{\uw}\Ra^{n} \sum_{m=0}^{m=n}
C_{n}^{m}\Le -1\Ra^{m}S_{\dw}^{m+1}\,\,.
\eeq
This sum can be  rewritten in terms of the amplitude for  single
diffraction dissociation.
We calculate the diffraction dissociation of the nucleus $A_2$.  In this
case we notice that the cut of $S_{\dw}$
leads to a multiparticle production process or to the diffractive
dissociation
of the nucleus $A_1$ and , therefore, it does not contribute to the
process of interest. Finally, we obtain

\begin{eqnarray}
F^{SD}_{\uw}\Le y, b, b'\Ra &=& g_{1}\cdot g_{2}\cdot e^{\Delta y}\cdot
S_{\uw}\cdot\sum_{n=0}^{n=\infty}\Le
1-S_{\uw}\Ra^{n} \sum_{m=0}^{m=n}C_{n}^{m}\Le -1\Ra^{m}\cdot
\nonumber \\
& \cdot & \Le\sum_{k=1}^{k=m+1}\cdot \Le 2S_{\dw}\Ra^{m-k+1}\Le
-S_{\dw}^{SD}\Ra^{k}
\frac{\Le m+1\Ra !}{k!\Le m-k+1\Ra}\cdot\Le -1\Ra^{m}\Ra\,\,.
\label{AD3}
\end{eqnarray}
The function $S_{\dw}^{SD}$ differs from $D(y,y_1,b_t)$ of \eq{DF11} only
by
some factors, namely,
\beq \label{AD5}
S^{SD}_{\dw}\Le y, y_{1}\Ra =
2\frac{g_{1}\cdot G_{3P}\cdot
e^{\Delta y_{1}}}
{\Le g_{1}\cdot\gamma\cdot
\Le 2e^{\Delta y}-e^{\Delta y_{1}}-1\Ra +1\Ra ^{2}}\,\,.
\eeq
After simple algebra, we reduce \eq{AD5} to
\begin{eqnarray}
F^{SD}_{\dw}\Le y, b, b'\Ra &=& g_{1}\cdot g_{2}\cdot e^{\Delta y}\cdot
S_{\uw}\cdot\sum_{n=0}^{n=\infty}\Le
1-S_{\uw}\Ra \sum_{m=0}^{m=n}C_{n}^{m}\Le -1\Ra^{m}\cdot \nonumber \\
& \cdot& \Le\sum_{k=1}^{k=m+1}\cdot \Le 2S_{\dw}\Ra^{m-k+1}\Le
-S_{\dw}^{SD}\Ra^{k}
\frac{\Le m+1\Ra !}{k!\Le m-k+1\Ra}\cdot\Le -1\Ra^{m}
+\Le 2S_{\dw}\Ra^{m+1}-\Le 2S_{\dw}\Ra^{m+1}\Ra \nonumber \\
& =&
-\frac{g_{1}\Le b-b'\Ra
\cdot g_{2}\Le b'\Ra\cdot e^{\Delta y}\cdot
\Le 2S_{\dw}-S_{\dw}^{SD}\Ra S_{\uw}}
{1-\Le 1-S_{\uw}\Ra\Le 1+2S_{\dw}-S_{\dw}^{SD}\Ra} \nonumber \\
 & + &  \frac{g_{1}\Le b-b'\Ra\cdot g_{2}\Le b'\Ra\cdot e^{\Delta y}\cdot
S_{\dw}\cdot S_{\uw}}
{1-\Le 1-S_{\uw}\Ra\Le 1-2S_{\dw}\Ra} \,\,. \label{AD6}
\end{eqnarray}

Using \eq{AGA10} and \eq{AD6} , we can obtain a simple expression for
the cross section of the   single diffractive production:
\beq \label{AD7}
\sigma^{SD}=\int d^{2}b_t
\cdot\Le F^{SD}_{\uw}\Le y, b_t\Ra+
F^{SD}_{\dw}\Le y, b_t\Ra\Ra e^{-\Omega\Le y, b_t\Ra}\,\,.
\eeq

\begin{figure}
\begin{center}
\begin{tabular}{c c}
{\bf $\mathbf{\sigma^{DD}}$ for $\mathbf{A_{1}=20}$,$\mathbf{A_{2}=100}$}
 &{\bf $\mathbf{\sigma^{DD}}$ for
$\mathbf{A_{1}=200}$,$\mathbf{A_{2}=100}$ }\\
\epsfig{file=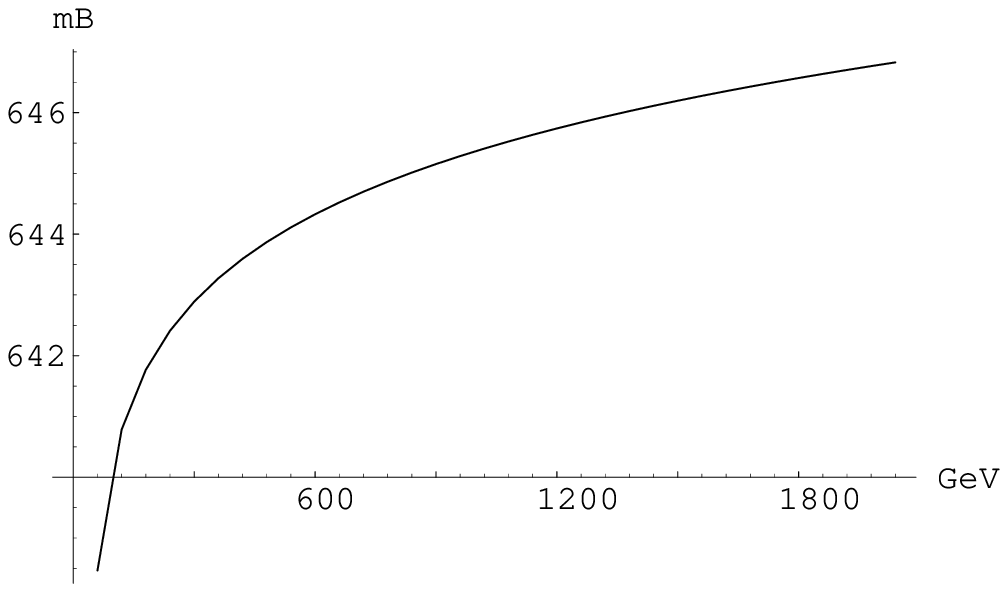,width=8cm} &
\epsfig{file=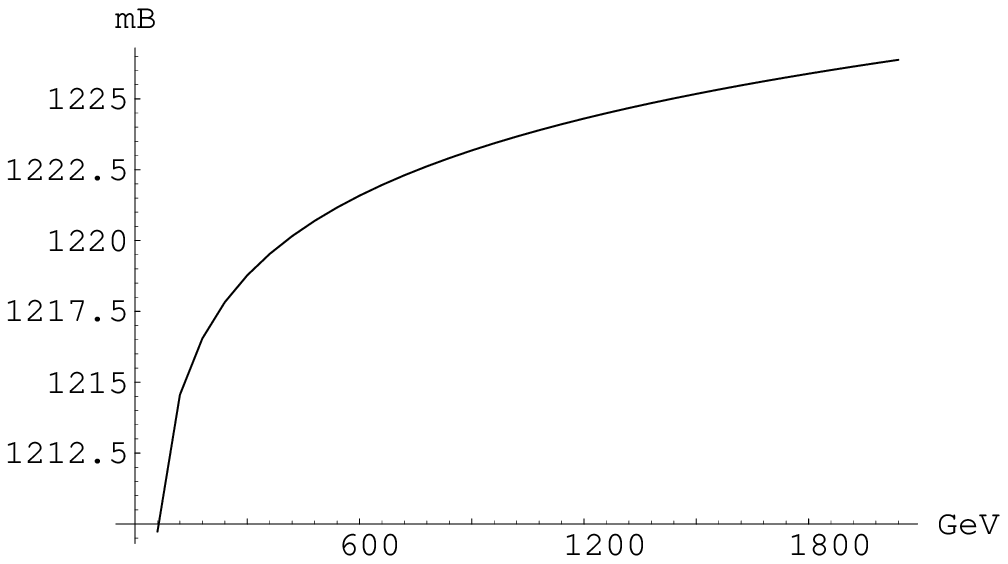,width=8cm}\\
{\bf $\mathbf{\sigma^{DD}}$ for
$\mathbf{A_{1}=200}$,$\mathbf{A_{2}=300}$ }
&
\\
\epsfig{file=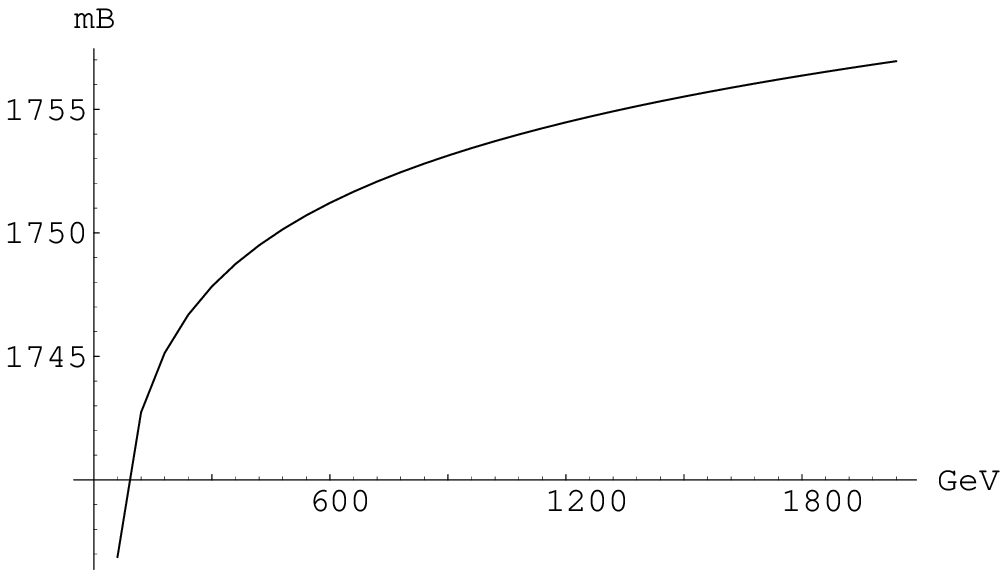,width=8cm}
&
\end{tabular}
\end{center}
\caption{\it The total diffractive 
dissociaton $A_{1}-A_{2}$ cross sections as a function of energy.}
\label{fig26}
\end{figure}

\begin{figure}
\begin{center}
\begin{tabular}{c c}
{\bf $\mathbf{\sigma^{SD}}$ for
$\mathbf{A_{1}=20}$,$\mathbf{A_{2}=100}$}
&{\bf $\mathbf{\sigma^{SD}}$ for $\mathbf{A_{1}=200}$,$\mathbf{A_{2}=100}$}\\
\epsfig{file=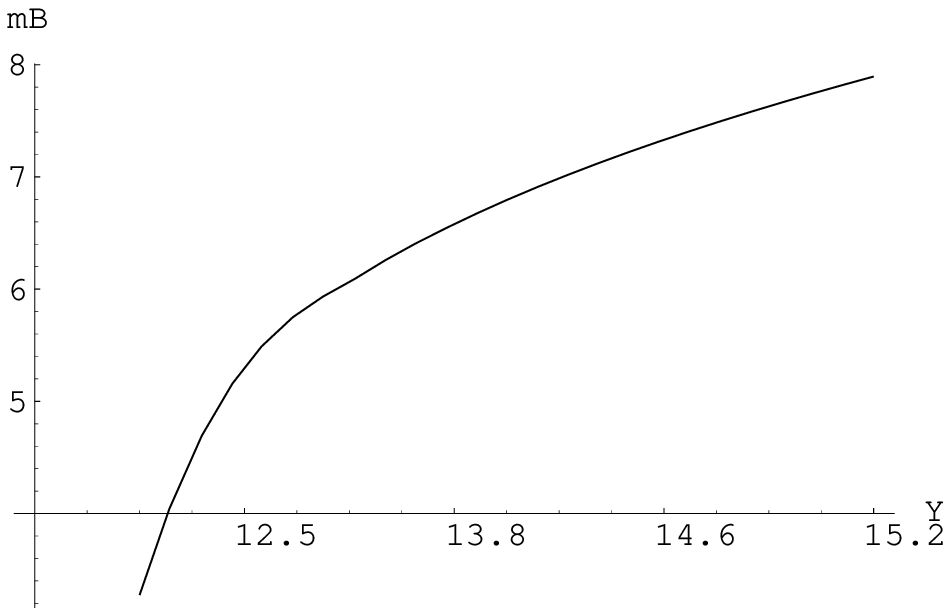,width=8cm} &
\epsfig{file=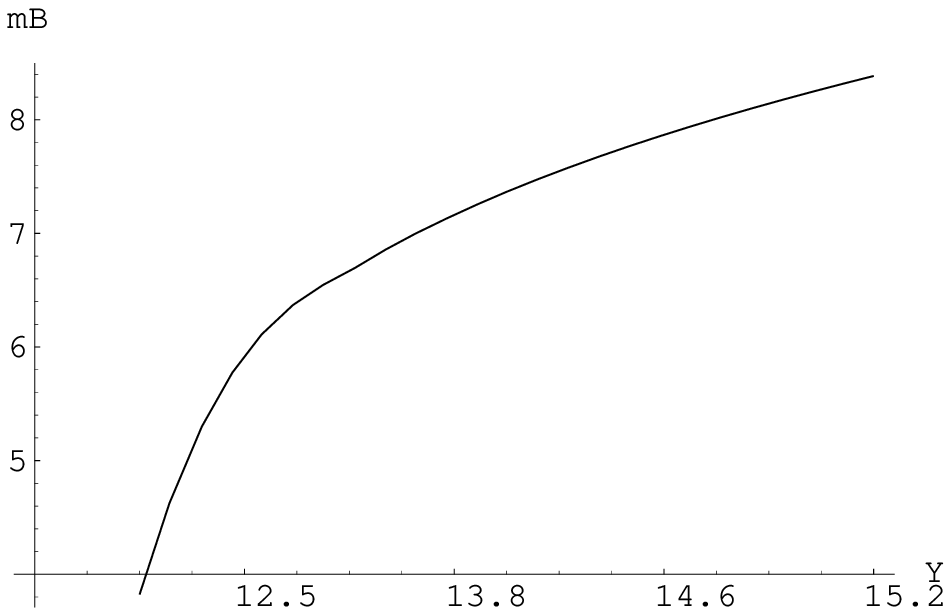,width=8cm}\\
{\bf $\mathbf{\sigma^{SD}}$ for
$\mathbf{A_{1}=200}$,$\mathbf{A_{2}=300}$}
&
\\
\epsfig{file=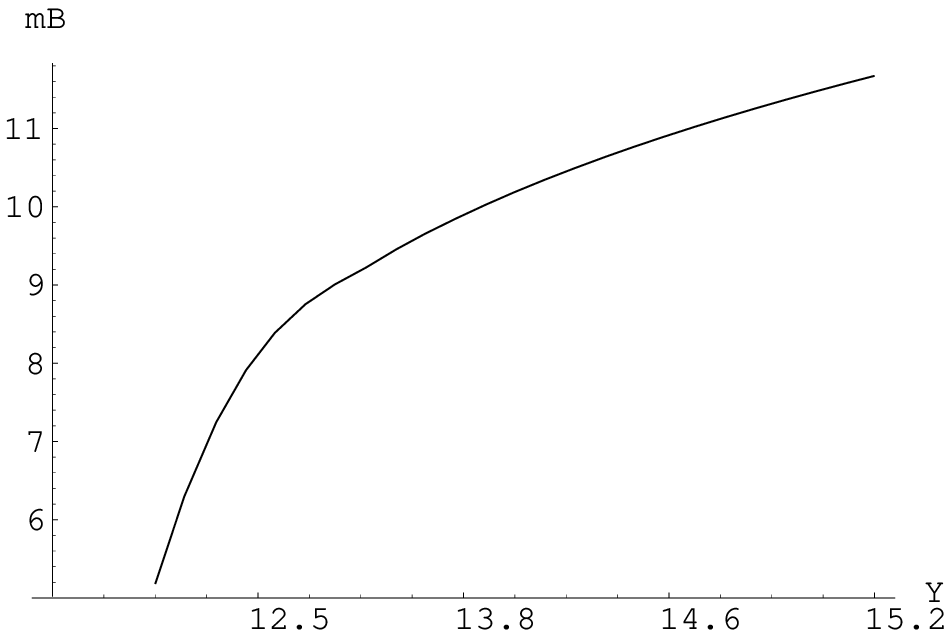,width=8cm}
&
\end{tabular}
\end{center}
\caption{\it The single diffractive dissociaton $A_{1}-A_{2}$ cross sections
as a function of rapidity gap y for a fixed 
rapidity $Y=\ln S/S_{0}$ for the $\sqrt{S}=2000$ GeV
and $\sqrt{S_{0}}=1$ GeV.}
\label{fig26_a}
\end{figure}

We can also obtain a simple formula for the total cross section of
diffractive dissociation directly using the unitarity constraint of
\eq{DF8} which leads to
\begin{eqnarray}
F^{DD} \Le y,b,b' \Ra &=& \,2\,\frac{g_1\cdot g_2 \cdot e^{\Delta\,y}}{
\Le\,g_1 \,+\, g_2  \Ra\,\gamma\,\Le e^{\Delta\,y}\, - \,1 \Ra \,\,+\,\,1}
\nonumber \\
 & - & 2\,\frac{g_1\cdot  g_2 \cdot e^{\Delta\,y}}{ 2\,
\Le\,g_1 \,+\, g_2  \Ra\,\gamma\,\Le e^{\Delta\,y}\, - \,1 \Ra \,\,+\,\,1}
\,\,. \label{AD8}
\end{eqnarray}
\eq{AD8} leads to the total cross section of diffraction dissociation:
\beq 
\sigma^{DD}\,\,\,=\,\,\,\int\,d^2 b_t\,\,F^{DD}\,\,e^{
-\Omega(y,b_t)}\,\,.
\eeq
The calculations were performed for  $A_{1}-A_{2}$
interaction
with  ($A_{1}=20$) - ($A_{2}=100$),
($A_{1}=200$) - ($A_{2}=100$),
($A_{1}=200$) - ($A_{2}=300$).
The results for the total diffractive 
dissociation cross sections are shown in
Fig.\ref{fig26}.
The single diffractive dissociation in Fig.\ref{fig26_a},
is calculated as a function
of rapidity  gap y
for a fixed rapidity $Y=\ln S/S_{0}$ for the $\sqrt{S}=2000$ GeV
and $\sqrt{S_{0}}=1 GeV$.
All parameters were taken to be the same as for the elastic and total
cross sections calculations.

\subsection{Survival probability of Large Rapidity Gaps}

We can calculate the survival probability of the LRG processes using the
expression for the scattering amplitude in the form of \eq{AD1} which we
rewrite as 
\beq \label{ASP1}
F\Le y, b, b'\Ra\,\,=\,\,g_{1}\cdot g_{2}\cdot e^{\Delta y}\cdot
\sum_{n=0}^{n=\infty}\sum_{m=0}^{m=n}
C^{n}_{m}\Le - S_{\dw}\Ra^{m+1}
\sum_{k=0}^{k=n}
C^{n}_{k}\Le -S_{\uw}\Ra^{k+1}\,\,.
\eeq
Substituting the function $S^{SP}_{\dw}$ instead of one of $S_{\dw}$ in
\eq{ASP1}  we obtain the survival probability for a process of di-jet
production which can be accompanied by hadrons with rapidities
which are  smaller
than $y_2$. The function $S^{SP}_{\dw}$ differs from $L(y,y_1,y_2,b_t)$, found
in \eq{SP9}, by some factors,
\begin{eqnarray}
S_{\dw}^{SP} & = &  \label{ASP2}\\
      & \sigma_{jet}& \frac{
e^{-\Delta\Le y_{2}-y_{1}\Ra }}
{\Le g_{1}\cdot\gamma\cdot\Le
e^{\Delta y_{1}}-1\Ra +1 \Ra}\cdot
\frac{\Le g_{1}\cdot\gamma\cdot\Le
2e^{\Delta y_{2}}-e^{\Delta y_{1}}-
1\Ra +1 \Ra^{2}}
{\Le g_{1}\cdot\gamma\cdot\Le 2e^{\Delta y}-
e^{\Delta y_{1}}-
1\Ra +1 \Ra^{2}}\,\,.\nonumber
\end{eqnarray}

Replacing one of the $S_{\dw}$ by $S_{\dw}^{SP}$  and performing
the explicit summation over all the indices, we obtain:

\begin{eqnarray}
F^{SP}_{\dw}\Le y, b, b'\Ra & = & g_{1}\cdot g_{2}\cdot e^{\Delta y}\cdot
\frac{S_{\dw}^{SP}\cdot\Le S_{\uw}S_{\dw}\Ra}
{S_{\uw}+S_{\dw}-S_{\uw}S_{\dw}} \label{ASP3} \\
 &=& g_{1}\cdot g_{2}\cdot e^{\Delta y}\cdot
S_{\dw}^{SP}\cdot
\frac{\Le g_{1}\gamma\Le e^{\Delta y}
-1\Ra +1\Ra \Le g_{2}\gamma\Le e^{\Delta y}
-1\Ra +1\Ra}{ \Le \Le g_1 \,+\,g_2 \Ra \,\gamma\,\Le e^{\Delta
y}\,-\,1\,\Ra
\,+\,1 \Ra^2}.
\label{ASP4}
\end{eqnarray}
The total cross section for the two particle irreducible diagrams is a sum
of two terms $F^{SP} = F^{SP}_{\dw}\,\,+\,\,F^{SP}_{\uw}$ which is equal
to
\begin{eqnarray}
F^{SP}\Le y, b, b'\Ra &=& g_{1}\cdot g_{2}\cdot e^{\Delta y}\cdot
\Le S_{\dw}^{SP}+S_{\uw}^{SP}\Ra \nonumber \\
 &\cdot& \frac{\Le g_{1}\gamma\Le e^{\Delta y}
-1\Ra +1\Ra \Le g_{2}\gamma\Le e^{\Delta y}
-1\Ra +1\Ra}
{\Le\Le g_{1}+g_{2}\Ra \gamma\Le e^{\Delta y}
-1\Ra +1\Ra^{2}} \,\,, \label{ASP5}
\end{eqnarray}
where  $S_{\dw}^{SP}$ is given by \eq{ASP2} and
\begin{eqnarray}
S_{\uw}^{SP} &= & \label{ASP6} \\
 & &\sigma_{jet}\frac{
e^{-\Delta\Le y_{2}-y_{1}\Ra }}
{\Le g_{2}\cdot\gamma\cdot\Le
e^{\Delta\Le y-y_{2}\Ra}-1\Ra +1 \Ra}\cdot
\frac{\Le g_{2}\cdot\gamma\cdot\Le
2e^{\Delta\Le y-y_{1}\Ra}-e^{\Delta\Le y-y_{2}\Ra}-
1\Ra +1 \Ra^{2}}
{\Le g_{2}\cdot\gamma\cdot\Le 2e^{\Delta y}-
e^{\Delta\Le y-y_{2}\Ra}-
1\Ra +1 \Ra^{2}}\,\,. \nonumber
\end{eqnarray}
For the calculation of the survival probability we also need
the expression for the inclusive process in the case of 
nucleus-nucleus interactions.

\begin{figure}[htbp]
\begin{center}
\epsfig{file=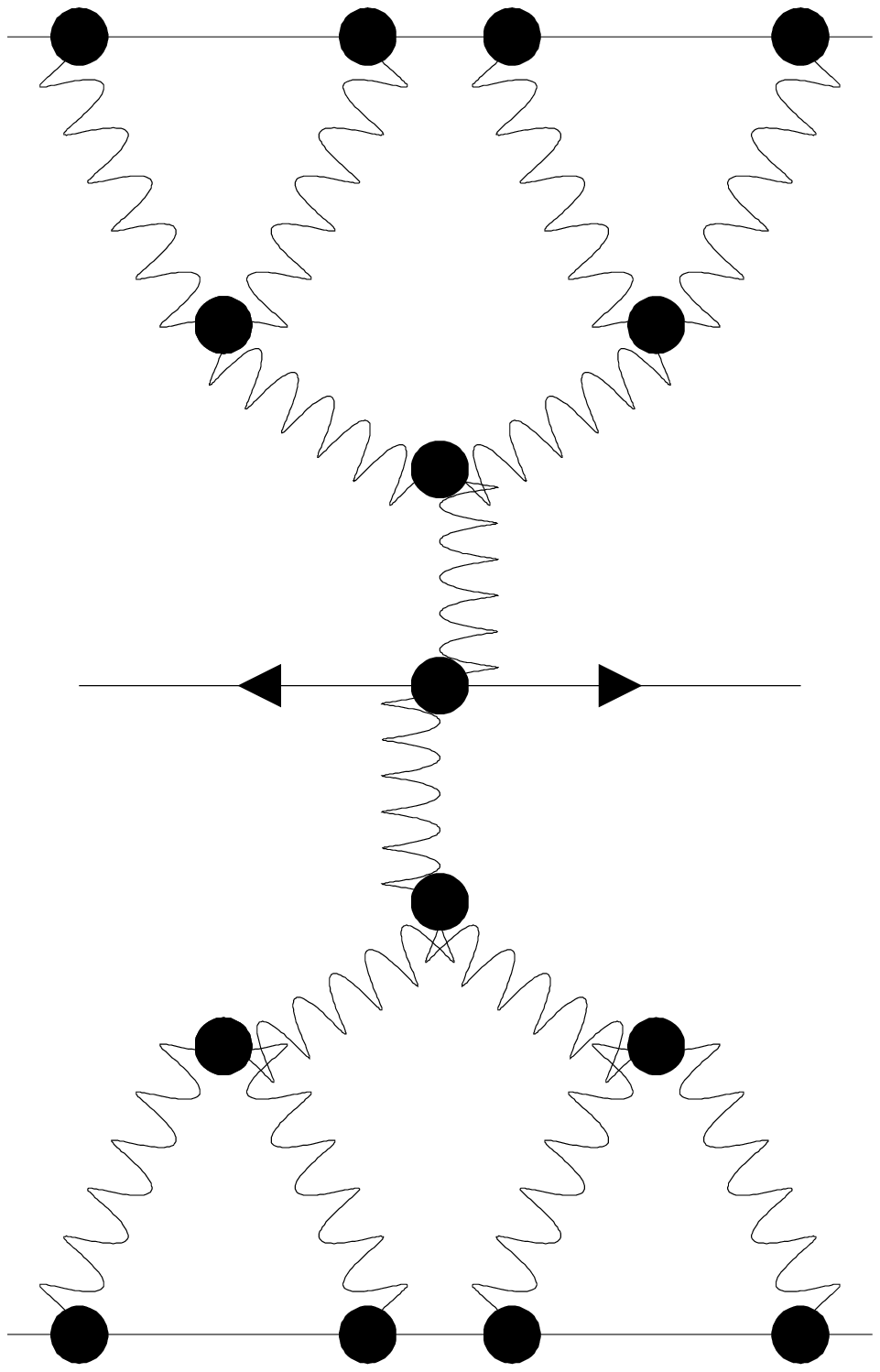,width=10cm}
\end{center}
\caption{\it The Mueller diagram for the inclusive process for a  A-A
interaction.}
\label{fig27}
\end{figure}
\begin{figure}
\begin{center}
\begin{tabular}{c c}
{\bf $\mathbf{<|S^2|>}$ for $\mathbf{A_{1}=20}$,$\mathbf{A_{2}=100}$}
 &{\bf $\mathbf{<|S^2|>}$ for
$\mathbf{A_{1}=200}$,$\mathbf{A_{2}=100}$}\\
\epsfig{file=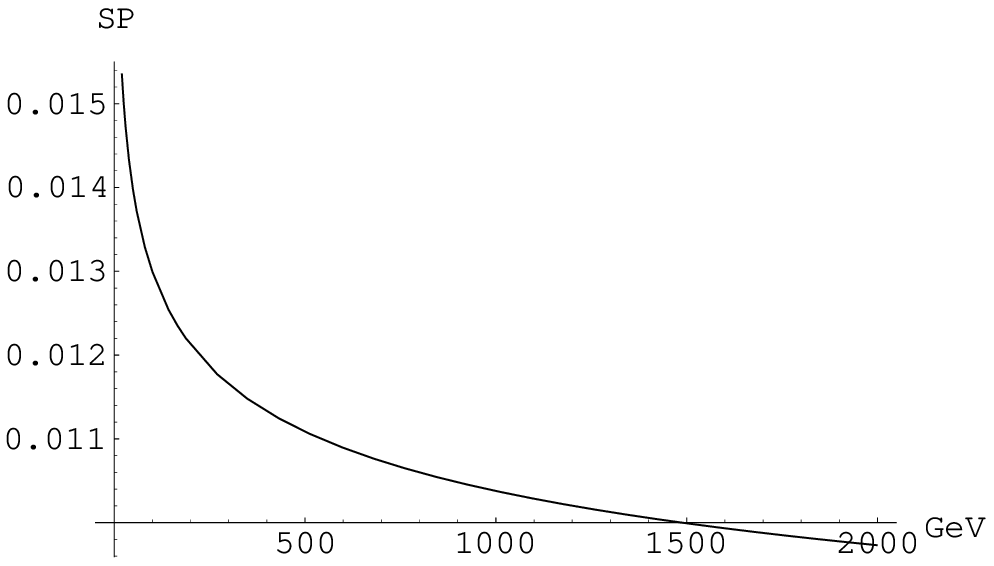,width=8cm} &
\epsfig{file=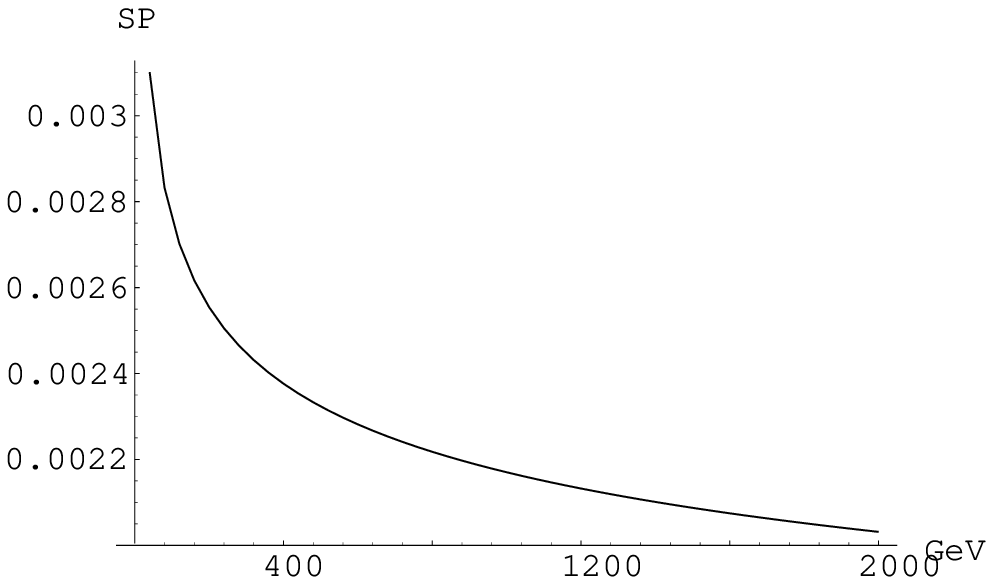,width=8cm}\\
{\bf $\mathbf{<|S^2|>}$ for
$\mathbf{A_{1}=200}$,$\mathbf{A_{2}=300}$}
&
\\
\epsfig{file=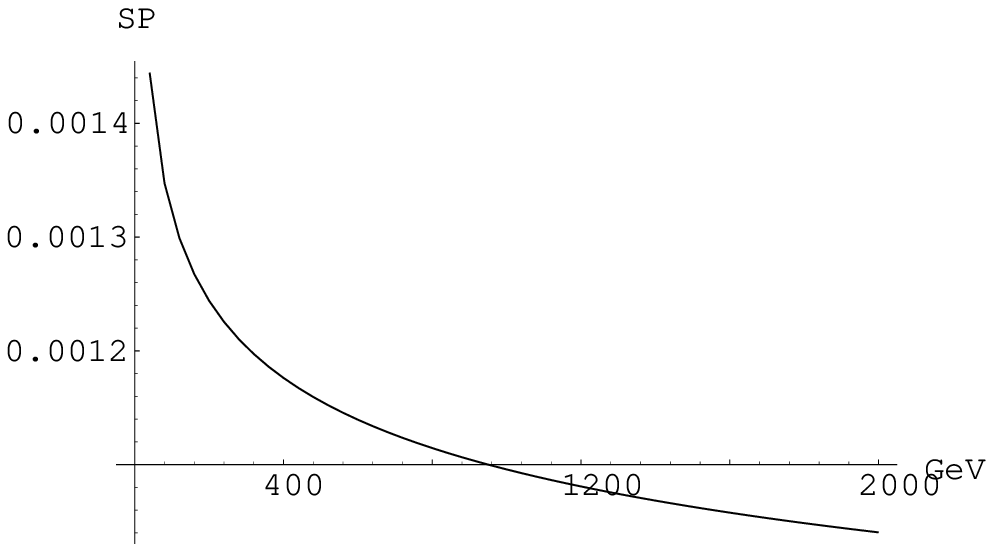,width=8cm}
&
\end{tabular}
\end{center}
\caption{\it The Survival probability for
di-jet production in the rapidity interval $10-12.5$ versus energy
in $A_{1}-A_{2}$ interaction.}
\label{fig57}
\end{figure}

The Mueller diagram for the inclusive cross section is given in
Fig.\ref{fig27} and the analytic expression is
\beq \label{ASP7}
F^{incl}\Le y, b, b'\Ra =\sigma_{jet}\cdot
\frac{g_{1}
\cdot g_{2}\cdot e^{\Delta\Le y-y_{2}+y_{1}\Ra}}
{\Le g_{1}\gamma\Le e^{\Delta y_{1}}
-1\Ra +1\Ra\cdot\Le g_{2}\gamma\Le e^{\Delta\Le y-y_{2}\Ra}
-1\Ra +1\Ra}\,\,.
\eeq
Finally, from \eq{SP4} we obtain
\beq \label{ASP8}
<|S\Le y, y_1, y_2 \Ra|^{2} > \,\,=\,\,\frac{\int d^{2}b\cdot F^{SP}\Le y,
b\Ra\cdot
e^{-\Omega\Le y, b\Ra}}
{\int d^{2}bd^{2}b'\cdot F^{incl}\Le y, b, b'\Ra}\,\,,
\eeq
where $\Omega$ is defined by \eq{AGA10}.
Using \eq{ASP8}  we performed calculations 
with  ($A_{1}=20$) - ($A_{2}=100$),
($A_{1}=200$) - ($A_{2}=100$),($A_{1}=200$) - ($A_{2}=300$),
for
di-jet production in the rapidity interval 10-12.5.
The result of the calculation is given in
Fig.\ref{fig57} as a survival probability versus energy.

One can see in Fig.\ref{fig57} that the value of the survival probability
turns out to be rather small.

\subsection{Inclusive production}

In this subsection we repeat the calculation, that has been performed for
hadron-nucleus collision in section 2.5.

\subsubsection{Single inclusive cross section}

\begin{figure}
\begin{center}
\begin{tabular}{c c}
{\bf $\mathbf{N}$ for $\mathbf{A_{1}=20}$,$\mathbf{A_{2}=100}$}
 &{\bf $\mathbf{N}$ for
$\mathbf{A_{1}=200}$,$\mathbf{A_{2}=100}$}\\
\epsfig{file=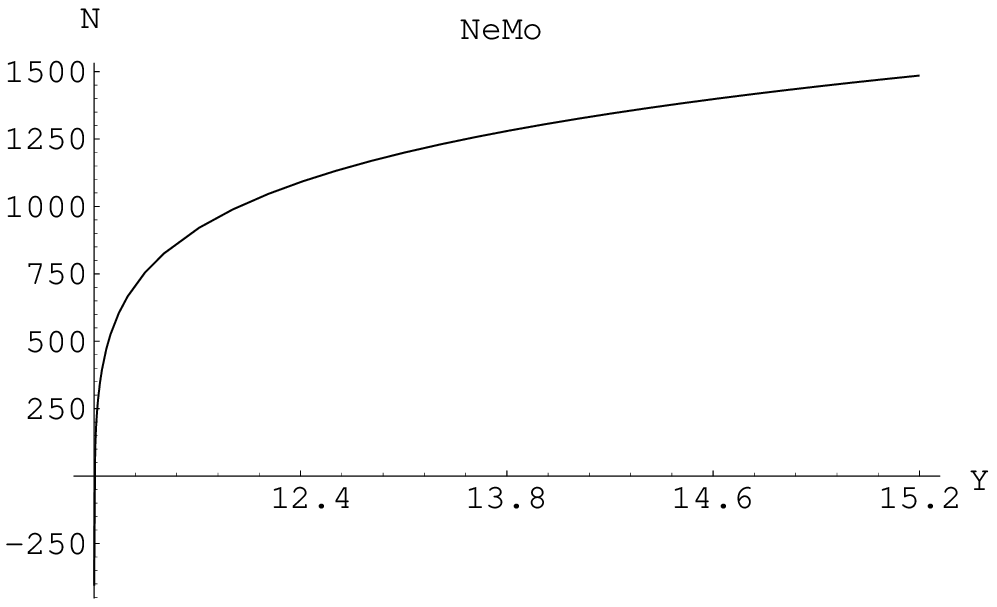,width=8cm} &
\epsfig{file=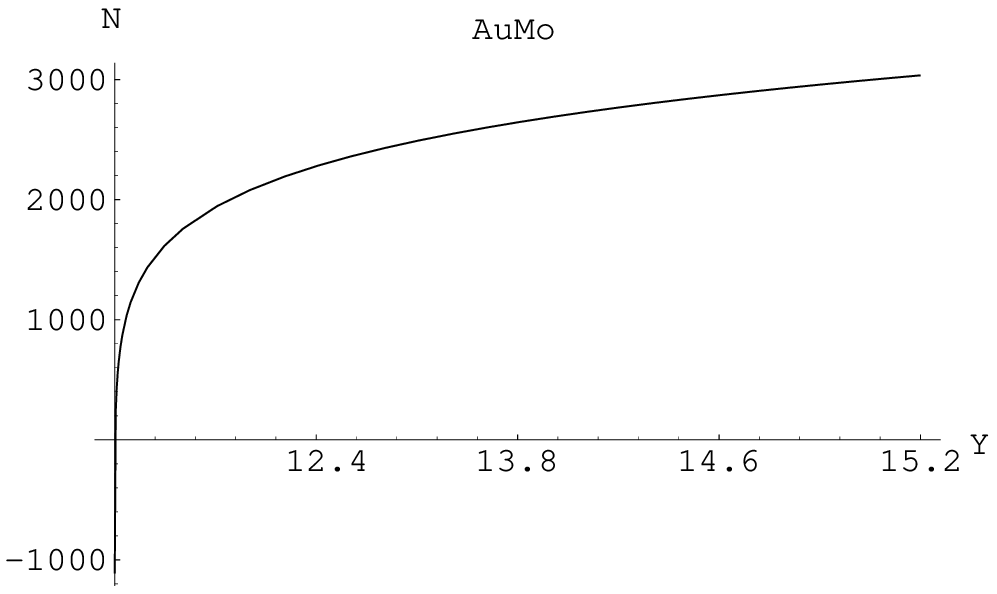,width=8cm}\\
{\bf $\mathbf{N}$ for
$\mathbf{A_{1}=200}$,$\mathbf{A_{2}=300}$}
&
\\
\epsfig{file=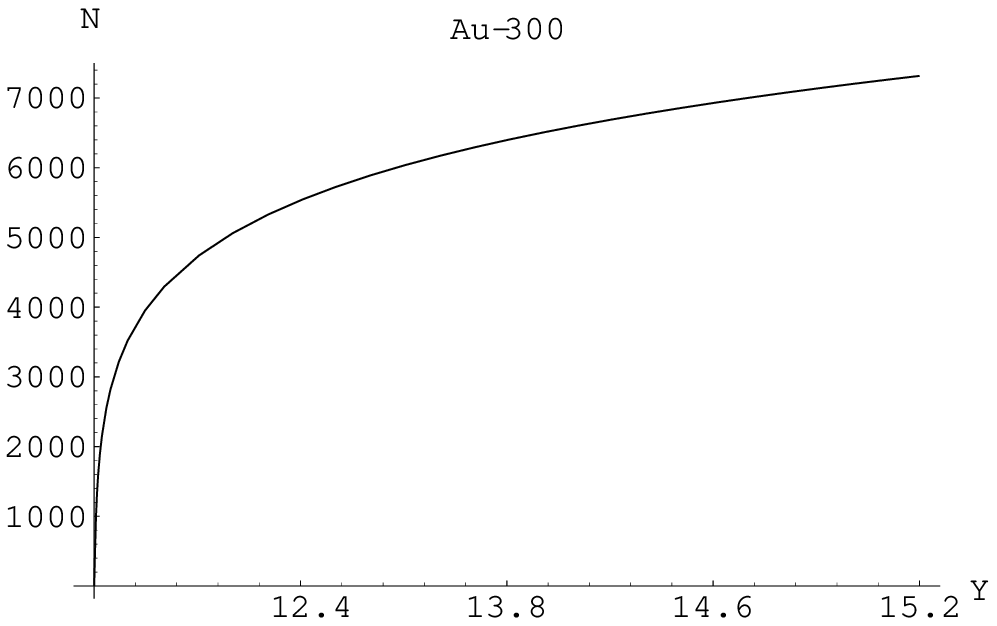,width=8cm}
&
\end{tabular}
\end{center}
\caption{\it The multiplicity of produced hadrons for 
Ne-Mo, Mo-Au, Au-A=300 
in $A_{1}-A_{2}$ interaction.}
\label{FF1}
\end{figure}

The Mueller diagram is shown in Fig.~\ref{fig27}. This diagram leads to a
simple formula
\begin{eqnarray}
\frac{d \sigma (A_1 + A_2 )}{d y_c} &= &\int \,\,d^2 b \,\,d^2 b'
\,\,a^P_P\,\,S_{A_1}\Le y - y_c, b \Ra \,\cdot\,S_{A_2 }\Le y_c, b' \Ra
\label{AIN1} \\
 &=&  a^P_P\,\,\int \,\,d^2 b \,\,d^2 b'
\,\,\frac{g_{P-A_1}(b)\,g_{P-A_2}(b')\,\, e^{\Delta \,\,y}}{ \Le
\,\kappa_{A_1} \Le y - y_c, b \Ra \,\,+\,\,1\,\Ra \,\Le \,\kappa_{A_2} \Le
y_c, b \Ra \,\,+\,\,1\,\Ra}\,\,.\nonumber
\end{eqnarray}
Integrating this equation over $y_{c}$ and dividing by the
$\sigma_{tot}$ we obtain the multiplicity N of the produced hadrons
in A-A interaction, where the results are
shown in  Fig.\ref{FF1}
for the cases of interactions of Ne-Mo, Mo-Au, Au-A=300.

It is interesting to note that the density of produced hadrons in the
central rapidity region  at high
energy has a very simple relation to the density in nucleon-nucleon
collision:

\beq \label{AIN2}
\frac{\frac{d \sigma (A_1 + A_2 )}{d y_c}}{\sigma_{tot}( A_1 + A_2)}\,
\,\longrightarrow|_{y \gg 1} \,\,\frac{g^2_{P-N}}{\gamma}\,\frac{
A^{\frac{2}{3}}_2\,\cdot\,
\ln A^{\frac{1}{3}}_{1}\cdot\ln A_{2}^{\frac{1}{3}}}
{\Le 
1\,\,+\,\,\Le\frac{A_{1}}{A_{2}}\Ra^{\frac{1}{3}}\Ra\,}
\,\,\frac{\frac{d \sigma (N + N )}{d
y_c}}{\sigma_{tot}( N + N)}\,\,.
\eeq
where we assumed
\beq
R_{A}^{2}\approx R_{N}^{2}\cdot A^{\frac{2}{3}}
\eeq
and
\beq
A_{2}>A_{1}.
\eeq

\subsubsection{Rapidity correlations}
\begin{figure}[htbp]
\begin{center}
\epsfig{file=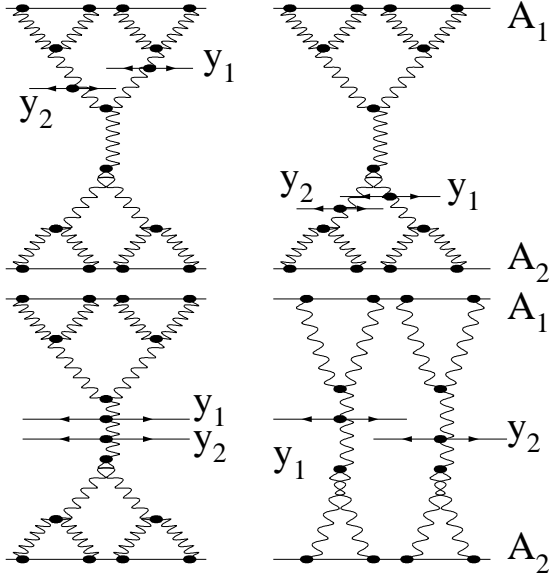,width=10cm}
\end{center}
\caption{\it The Mueller diagram for the double inclusive process in an
$A_{1}-A_{2}$ interaction.}
\label{fig28}
\end{figure}

Fig.\ref{fig28} shows the Mueller diagrams for the double inclusive
cross section which can be written analytically as
\begin{eqnarray}
\frac{d^2 \sigma ( A_1 + A_2 )}{d y_1 \,dy_2} &=& ( a^P_P )^2 \int \,\,d^2
b \,\,d^2 b'\cdot \label{AIN3} \\
& & \cdot e^{ \Delta ( y_2 - y_1 )} \,S_{A_1}\Le y - y_2 , b \Ra \,S_{A_2}\Le
y_1 ,
b' \Ra   \label{AIN4} \\
 &+& S_{A_1}\Le y - y_2, b \Ra \,\,
S_{A_1}\Le y_2, b' \Ra \,\,
S_{A_2} \Le y-y_1, b \Ra\,
\int \,\,d^2 b'' \,\,
S_{A_2} \Le y_1, b'' \Ra\,
\label{AIN5} \\
& + & G_{3P}\,\int^y_{y_2}\,d y' e^{\Delta ( 2y' - y_1 - y_2  )}\,\,
S_{A_1}\Le y-y', b \Ra\,\,
S_{A_2}\Le y_2, b' \Ra\,\,S_{A_2} \Le y_1, b' \Ra \label{AIN6}\\
& + & G_{3P}\,\int^{y_1}_{0}\,d y' e^{\Delta (  y_1 + y_2 - 2y'  )}
S_{A_2}\Le y', b' \Ra
S_{A_1}\Le y - y_2, b \Ra S_{A_1} \Le y - y_1, b \Ra\,,
\label{AIN7}
\end{eqnarray}

where each term of \eq{AIN4} - \eq{AIN7} corresponds to
Fig. \ref{fig28}-1 - Fig. \ref{fig28}-4.
The asymptotic behaviour for the double inclusive cross section
is:
\begin{eqnarray}
\frac{d^2 \sigma ( A_1 + A_2 )}{d y_1 \,dy_2} &=& 
\frac{( a^P_P )^2 }{\gamma^2}\Le R^2_{A_{1}}R^2_{A_{2}}\Ra
\,\,\ln^{2} A_{1}^{\frac{1}{3}}\,\,\ln^{2} A_{2}^{\frac{1}{3}}\,\,.
\end{eqnarray}

\section{Conclusions}
In this paper we found a natural generalization of the Glauber approach to
hadron-nucleus and nucleus-nucleus collisions.
It has been known for three decades \cite{GG,GG1} that such a
generalization
requires finding a theoretical way to take the triple Pomeron
interaction into account or, in other words, 
to include the diffractive dissociation processes.
In the framework of the Pomeron approach, this problem was solved here
using two main physical ideas, or more precisely two small
phenomenological parameters:
\begin{enumerate}
\item\,\, Typical distances
for the Pomeron structure which are sufficiently small,
  justify using the pQCD approach
for the evaluation of different types of the Pomeron interaction.

\item\,\,\, A small value of the triple Pomeron vertex ($ G_{3P}/g_{P-N}
\,\ll\,1 $) which allows us to use the following set of parameters in
summing the Pomeron diagrams:
$$
\gamma \,\,=\,\,
\frac{G_{3P}}{\Delta}\,\,\ll\,\,1\,\,,\,\,\,\,\,\,\,\,\,\,\,\,
 \gamma\,\times A^{\frac{1}{3}}\,\,\approx\,\,1 \,\,.
$$

\end{enumerate}

Based on these parameters we were able to formulate the selection rules
for the Pomeron diagrams and sum them. This lead to a significant
generalization of the
oversimplified eikonal type model for shadowing corrections without
losing the advantages of such models.
In this approach we calculated
diffractive dissociation processes and survival probability
of the large rapidity gap processes, and calculated
the shadowing (screening) corrections for a large class of the "soft"
interaction processes at high energies.

The physics of our generalization is very simple. In the parton model,
the Glauber (eikonal) approach takes into account only the interaction of the
fastest parton with the target. In our approach, we consider
the interactions of all partons both with the target and with the projectile, if
the last is a heavy nucleus. We are successful in
finding a simple  closed expression for
such types of interaction making use of the fact that the
nucleus is a dense parton system.

Our main results are:

\begin{enumerate}

\item\,\,\,We have generalized the Glauber approach
for the nucleus-nucleus interaction at high energies.
\item\,\,\, We show that the asymptote of the two particle irreducible
diagrams leads to the exchange of an effective Pomeron with an intercept which is
two times smaller than the intercept of the input Pomeron.
\item\,\, We show that the asymptotic behaviour for nucleus-nucleus
collision starts at sufficiently high energies ( higher than the Tevatron ).
\item\,\,A systematic approach to diffractive dissociation has been
developed both for hadron-nucleus and nucleus-nucleus collisions.
\item\,\, Our approach leads to saturation of the rapidity density of
produced hadrons
\beq \label{C1}
\frac{\frac{d \sigma (A_1 + A_2 )}{d y_c}}{\sigma_{tot}( A_1 + A_2)}\,
\,\longrightarrow|_{y \gg 1} \,\,\frac{g^2_{P-N}}{\gamma}\,
\frac{ A^{\frac{2}{3}}_2 \cdot 
\ln A^{\frac{1}{3}}_{1}\cdot\ln A_{2}^{\frac{1}{3}}}
{\Le 
1\,\,+\,\,\Le\frac{A_{1}}{A_{2}}\Ra^{\frac{1}{3}}\Ra\,}
\,\,\frac{\frac{d \sigma (N + N )}{d
y_c}}{\sigma_{tot}( N + N)}\,\,.
\eeq

This  prediction is quite different from the Glauber-Gribov result given
in \eq{GG2}.  In particular, \eq{C1} predicts  that the multiplicity
for nucleus - nucleus collisions  $< N >(A_1 - A_2 ) = A^{\frac{2}{3}}_1
< N >(N - N )$ where $A_1 < A_2$, while \eq{GG2} leads to $< N >(A_1 - A_2
) = A_1 \cdot A^{\frac{2}{3}}_2 $. We predict the same dependence also for
central collisions.

On the other hand, the prediction of \eq{C1} is quite different from the
predictions of the high density QCD approach \cite{THEHDQCD}. Indeed, the
high density QCD approach leads to a formula which is very similar to
\eq{C1}, but due to the integration over the transverse momentum of the emitted
partons
the extra factor $Q^2_s$ appears in front of the equation. In particular,
it turns out that $< N >(A_1 - A_2 ) \,\propto A^{\frac{2}{3}}_1 \cdot
Q^2_s(s,A) \cdot < N >(N - N ) $.  $Q^2_s(s,A)$ is the saturation 
scale which increases with $A$
and energy squared  $s$. We expect that $Q^2_s(s,A)$ grows with $A$ (  at
least
$Q^2_s \,\propto A^{\frac{1}{3}}$ \cite{THEHDQCD} but it could 
even be
proportional to $A^{\frac{2}{3}}$ \cite{KOV} ), and ,therefore,  high
density QCD
predicts $< N >(A_1 - A_2 ) \,\propto \, A^{\frac{2}{3}}_1 \times \left(
A^{\frac{1}{3}}_2 \,\div\,A^{\frac{2}{3}}_2 \right)$, where $A_2 > A_1$.

\item\,\, The survival probability of the LRG processes has been calculated
both for the hadron-nucleus and the nucleus-nucleus collision.

\item\,\, The theoretical solution to the nucleus-nucleus interaction  can
serve as guiding light for a theoretical approach to the
dense parton system.

\item\,\, The numerical estimates
which we have made, serve as
a basis for run of the mill
physics estimates. Hence,
results found at the next generation
of accelerators such as RHIC and LHC, should be compared to
our estimates to determine if there are any unusual features.

\end{enumerate}

\begin{figure}[htbp]
\begin{tabular}{l l}
\epsfig{file=diag3.eps,width=8cm,height=7.5cm} &
\epsfig{file=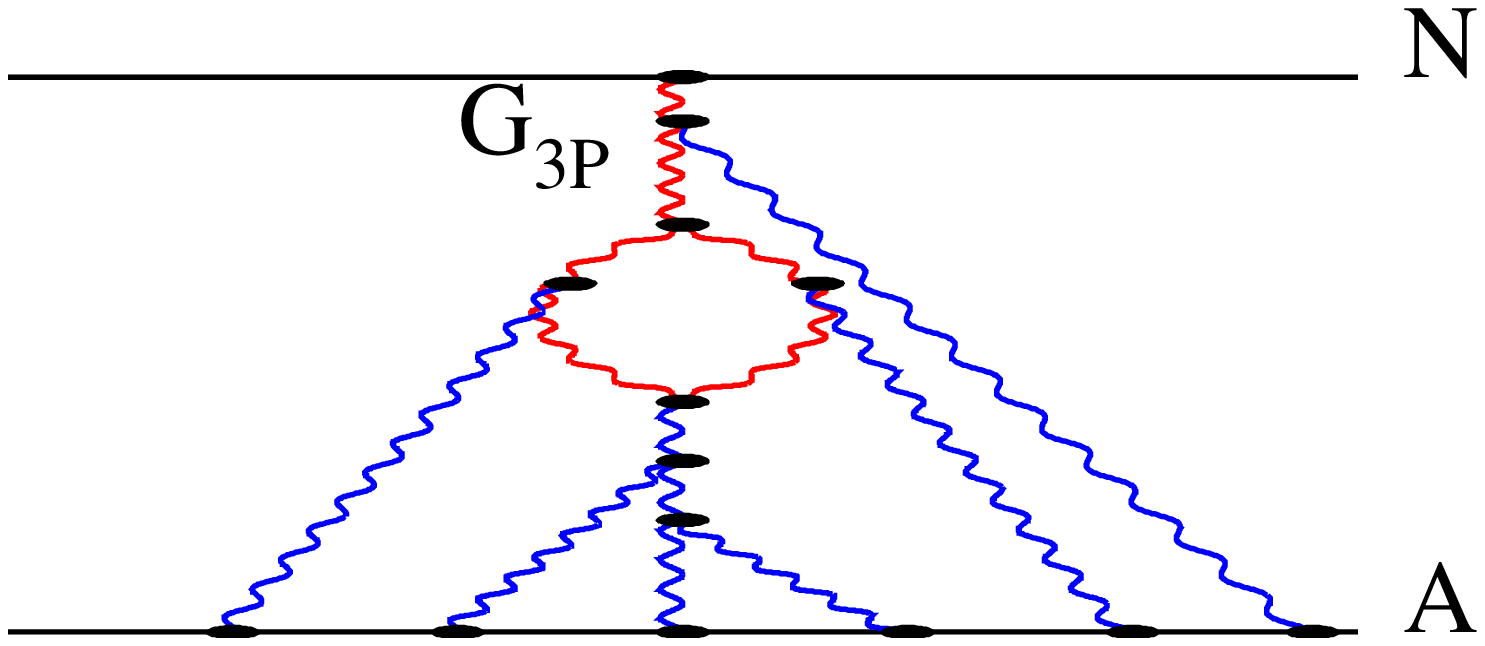,width=7cm,height=7.5cm}\\
``Fan" diagrams & First enhanced diagram\\
\end{tabular}
\caption{\it  The first correction to our approach.}
\label{acur}
\end{figure}

The most important corrections to our approach stem from the
enhanced diagram  ( see Fig.\ref{acur} ).    

Direct calculation of the ratio\footnote{We thank K. Tuchin for help in
calculating \eq{C2}.} 
\beq \label{C2} 
\frac{Enhanced\,\,\,\, diagram}{``Fan"\,\,\,\, diagram}
\,\,\,\, \Longrightarrow|_{s \rightarrow \infty}
\frac{G^2_{3P}}{2 \,\alpha'_P} \,\,y^2
\eeq
shows  that in our approach we have achieved reasonable accuracy only for
sufficiently
small values of the triple Pomeron vertex. In pQCD (see \eq{VE1} and \eq{VE2}
) one can see that the ratio of \eq{C2} is proportional to $\alpha^2_S 
 y^2$, and, therefore, it appears that we can neglect the contribution
of the enhanced diagram in the wide region of energy $y = \ln s <
1/\alpha_S $. However, the numerical parameters of our fit 
(see section 2.6
) gives $G^2_{3P}$/$2 \,\alpha'_P $ $\,\,\approx\,\,1$.

\section*{Acknowledgments}

The authors would like to acknowledge helpful and encouraging
discussions with  Alexey Kaidalov, Dima Kharzeev,  Larry McLerran,  Al
Mueller, Rob Pisarski,  
Misha Ryskin,  Kirill
Tuchin, Raju
Venugopalan, and Heribert Weigert.

This work was carried out while E.L. was on sabbatical leave at
BNL. E.L. wants to thank the nuclear theory group at BNL for their
hospitality and support during that time. U.M. wishes to thank 
the nuclear theory group at BNL for their hospitality during his
visit there.

This research has been supported in part by the Israel Science
Foundation, founded by the Israeli Academy of Science and Humanities and  
by the
United States--Israel BSF Grant $\#$ 9800276. This manuscript has been
authorized under Contract No. DE-AC02-98CH10886 with the
U.S. Department of Energy.

\end{document}